\documentclass[aps,amsmath,amssymb,reprint,showkeys,superscriptaddress]{revtex4-2}

\usepackage[dvipsnames]{xcolor}
\usepackage{graphicx}
\usepackage{hyperref}
\usepackage[utf8]{inputenc}
\usepackage[T1]{fontenc}
\usepackage{physics}
\usepackage{mathtools}
\usepackage{siunitx}
\usepackage{bm}
\usepackage{dcolumn}
\usepackage{appendix}
\usepackage{lipsum}
\usepackage{multirow}
\usepackage{chemformula}

\newcommand{\ver}[1]{\hat{\vb {#1}}}
\newcommand{\mY}{\mathcal{Y}}

\usepackage[normalem]{ulem}

\newcommand{\Hope}[1]{\textcolor{red}{#1}}

\begin{document}
\title{A general formalism for machine-learning models based on multipolar-spherical harmonics}

\author{M. Domina}
\affiliation{School of Physics, AMBER and CRANN Institute, Trinity College, Dublin 2, Ireland}
\affiliation{Laboratory of Computational Science and Modeling, Institut des Mat\'eriaux, \'Ecole Polytechnique F\'ed\'erale de Lausanne, 1015 Lausanne, Switzerland}
\author{S.~Sanvito}
\affiliation{School of Physics, AMBER and CRANN Institute, Trinity College, Dublin 2, Ireland}
\date{\today}

\begin{abstract}
The formulation of descriptors of the local chemical environment, enabling the
construction of machine-learning models, is usually obtained by studying the 
properties of the expansion coefficients of a neighborhood density. In this work, 
we show that all the transformation properties of the descriptors and their 
behavior under rotation, inversion and complex conjugation, are derived from 
the choice of the basis over which the density is expanded. Furthermore, crucially
they are independent from the explicit mathematical form of the neighborhood 
density. In particular, we show that all the descriptors investigated, can be 
obtained by an expansion in multipolar spherical harmonics, which constitutes 
the core of this work, and which is introduced and analyzed in great detail. 
By exploiting the orthogonality and the transformation rules of the multipolar
spherical harmonics, we show that several formulations are simplified, such 
as the one needed to obtain the $\lambda-$SOAP kernel and its properties. We 
close this work by applying our framework to several multi-body descriptors
available in literature, providing an in-depth analysis of their main properties, 
as made clear from the vantage viewpoint of a basis-centered approach.
\end{abstract}

\maketitle

\section{Introduction}

In recent years there has been a paradigm shift in the computational study of
materials, ignited by the rapid establishment of machine-learning potentials (MLPs) 
as essential tools for performing molecular dynamics and evaluating thermodynamical
properties of molecules and solids \cite{Review1,Review2,Review3,Review4}. The 
ambitious goal of MLPs is to reach the same state-of-the-art accuracy of 
computationally expensive \emph{ab initio} methods, such as density functional theory
(DFT) \cite{HK,KS}, at a fraction of their cost. In practice, one first aims to 
perform a limited number of highly accurate calculations for selected structures 
and configurations. This is usually done at the DFT or quantum-chemistry level. 
The resulting dataset, is then used to \emph{train} a carefully selected and 
optimized MLP, which will be able to predict quantities of interest on previously
unseen configurations, ranging from different atomic environments to large-sized
systems, inaccessible to \emph{ab-initio} calculations 
\cite{GAP17,GAP20,GAP_Si,ACE_C}. 

An MLP is mainly made of three parts. The first, called \emph{descriptors} (or 
fingerprints), corresponds to the choice of an adequate mathematical representation 
for the system at hand. The \emph{targets} are the quantity that the MLP will 
predict, while the mathematical relation between the descriptors and the 
targets  constitute the optimized interpolative model. 
Clearly, the choice of the fingerprints is crucial to determine the descriptive
capability of the model and its ability to properly account for the aimed targets. 
This results in a careful selection of the symmetries of the descriptors, which 
have to mirror the ones of the targets, under translation, rotation, inversion 
(parity) and atoms permutations.
In recent years, there has been a large effort aimed to define descriptors capable 
of addressing the most disparate targets, ranging from scalar quantities, namely 
energies \cite{GAP,SOAP,MTP,Performance_Assesment}, with a focus on multi-specie
systems \cite{SOAP_multi,EME-SNAP}, or with the inclusion of the spin degree of 
freedom \cite{ACE_vec,Domina,Suzuki}, to vectorial and tensorial quantities
\cite{Glielmo,SAGPR,Lunghi2019a,Lunghi2019b,HNguyen}, or even to electronic 
densities \cite{density_symmetry_functions,SALTED,SNAPGrid,JLCDM,JLCDM1}, and 
Hamiltonians \cite{ACE_Hamilt}. 
Also the choice of the ML models has been thoroughly explored. These include 
neural networks (NN) \cite{symmfunc,ANI,DeepMD,m3gnet}, kernel-based methods
\cite{GP4ML,SOAP,CeriottiNMR}, linear models \cite{ACE,SNAP,JLP} and 
message-passing schemes \cite{MACE,NigamUnified}.

For the majority of these schemes, the construction of the descriptors proceeds 
from the definition of a local atomic density, which describes the atomic 
neighborhood surrounding a central atom. This is usually taken as the fundamental
quantity, from which all the descriptors are derived. The spotlight in the study 
of the atomic density is taken by its expansion coefficients with respect to a 
suitable basis. In particular, since the density represents an atom-centered
environment, the choice of basis falls into selecting a set of radial functions, 
for the radial expansion, and in spherical harmonics, for the angular one. 
Crucially, the evaluation of the expansion coefficients scales linearly with respect 
to the number of atoms in the neighborhood, making it computationally convenient. 
Because of the central role of the atomic density, the properties of its expansion
coefficients have been thoroughly studied. The investigation of such 
\emph{atomic basis} is usually performed by means of the explicit expression of 
this coefficients \cite{ACE_dusson,BatatiaE3} or by means of a 
Dirac-bracket-derived formalism \cite{Willat,NICE}. 

The first approach consists in taking appropriate contractions of products of 
the expansion coefficients, and directly investigate their explicit expressions. 
This way of proceeding, however, does not fully exploit the fact that such objects 
can be obtained from products of atomic densities. In fact, all the properties 
should be obtainable with no knowledge of the explicit form of the expansion
coefficients, as this is the fundamental rationale behind the use of an 
atomic density. This is, indeed, the idea behind the second approach, where the 
product of atomic densities takes the central role. In that case, all the properties 
of interests are obtained by studying the projection of these products onto the 
basis of interests, which is usually a coupled multi-body basis. On the one hand, 
this approach enforces and exploits the idea of the atomic density as the main 
quantity behind the design of MLPs descriptors. On the other hand, as it mostly 
study the properties of the projections (namely expansion coefficients with respect 
to the multi-body basis), it does not fully exploit the fact that most properties 
are inherited from the multi-body basis itself. Moreover, the derivations usually
follow a bottom-up approach, obtaining properties for the density product from 
 the expansion coefficients of single-density ones. 

This work aims to investigate a third route. We take elements from both the previous 
methods and show that, not only the choice of the basis represents the most fundamental 
step in the design of ML descriptors, but also that a top-down approach can be followed, 
with a direct focus on the expansion over a multi-body basis. The proposed approach 
can be seen as a middle ground between the two strategies discussed above. On the one 
hand, it exploits the expansion of products of atomic densities (as the second formalism). 
On the other hand, it also follows a position representation for the basis (as done in the 
first). In particular, we will show that all the properties of interests can be derived from 
a formalism based on \emph{multipolar-spherical harmonics} (MultiSHs)~\cite{Angular}, 
which are the many-body generalization of the spherical harmonics. These have been 
already exploited in astrophysical studies \cite{Hajian,Joshi} (we also refer to Ref.~\cite{Cahn} 
for a general and in-depth treatment of the isotropic case). Indeed, a crucial result of this 
paper is that all the derivations and properties of the most well-known ML descriptors are 
directly inherited from the choice of MultiSHs as a multi-body basis, and do require just 
an handful of symmetries of the atomic density. 

Therefore, this work has two explicit goals. The first consists in deriving and discussing the 
MultiSHs-expansion-based formalism. We will show that it is possible to derive all the 
most well-known ML fingerprints, namely the powerspectrum, bispectrum and Smooth 
Overlap of Atomic Positions (SOAP) kernels, as a straightforward application of the core 
properties of the MultiSHs basis (orthogonality, behavior under rotation and under inversion), 
without any explicit study of the expansion coefficients. This will disentangle the role of the 
density and of the basis, clarifying what are the crucial symmetries of the atomic density. 
The second goal consists in providing a general recipe to derive linear ML models (for both 
scalar and tensor fields), which can be used to perform deep analytical studies of their properties 
and to establish connections across formalisms. In particular, we will use the case of the Atomic 
Cluster Expansion (ACE)~\cite{ACE} and the Spectral Neighbor Analysis Potential (SNAP) as 
examples to show that exploiting a MultiSHs based formalism offers the right tools for in-depth 
investigations. 

This work is structured as follows. In the next section, Section~\ref{sec:Methods}, we will introduce 
some descriptors for state-of-the-art ML models. Then, we will begin the discussion of the MultiSHs 
formalism, which will be tackled in a hierarchical way. We will first introduce the two- and three-points 
generalization of the spherical harmonics, namely the bipolar- and tripolar-spherical harmonics. These
form the backbone for the full MultiSHs formalism, which will be introduced in Section~\ref{sec:multipolar_spherical}.
Then, we will proceed with providing a first example of the usefulness of the MultiSHs-based approach: 
by exploiting only the orthogonality and the rotational covariance of the MultiSHs, we will be able to provide 
a compact and straightforward evaluation of the general-body $\lambda$-SOAP kernel~\cite{SAGPR}. 
We will conclude the section by showing that the crucial properties of the expansion coefficients, namely 
their transformation under rotation, inversion and conjugation, are inherited by analogous properties of the 
basis, and by only using the fact that we are expanding a real and scalar function. In particular, we will show
that the explicit form of the expansion coefficients is irrelevant for this discussion.

In Section~\ref{sec:powerspectrum_bispectrum_SOAP}, by further exploiting the fact that the 
expanded function is a \emph{product} of atomic densities, we will show that the MultiSHs formalism 
leads to the usual definition of the powerspectrum and bispectrum components as the rotationally invariant 
terms of the MultiSHs expansion. We will also show how to evaluate the explicit expression of the 
SOAP kernel~\cite{SOAP}. In Section~\ref{sec:on_the_atomic_density}, we will finally discuss the explicit 
form of the expansion coefficients, in order to connect the formalism of the MultiSHs to a multi-body expansion: 
this will be the only instance throughout this work in which the explicit form of these coefficients has any 
relevance. Thus, in Section~\ref{sec:reproducing_model}, we will exploit this connection to define a 
general linear models based on the MultiSHs expansion. In doing so, we will re-derive the ACE expansion, 
as the general basis for scalar quantities. We will also investigate the SNAP formalism, demonstrating 
how the MultiSHs framework provides a useful tool to an in-depth investigation of the potential at hand, 
in particular showing how the compactness of the SNAP model is also related to an incomplete expansion 
of four-body potentials. We will conclude this section by providing the formal connection between the 
MultiSHs formalism and internal-coordinates based expansions, by explicitly investigating the case of the 
Moments Tensor Potential (MTP) and our Jacobi-Legendre Potential (JLP). In the last section, 
\ref{sec:tensor_model_linear}, we will generalize the recipe for constructing linear model targeting scalar
quantities, to the tensorial case. Not only we will be able to provide a general recipe to construct the covariant 
components of a tensor, but we will also exploit the inherent property of the MultiSHs concerning the coupling 
scheme of the angular momenta, to show how to severely reduce the number of independent expansion 
coefficients. Crucially, the MultiSHs will make manifest that this reduction holds only when expanding 
symmetric functions.

\section{Methods}\label{sec:Methods}

\subsection{Introduction}
The main aim of this work is to introduce the MultiSHs and to show how they can be used to immediately 
generate all the main ML models for both scalar and tensorial quantities. Here, we will consider only the 
short-ranged contribution to the physical quantities of interests. For example, when targeting the potential 
energy surface (PES), the total energy is assumed to be partitioned as the sum of a long- and a short-range
contribution,
\begin{equation}
    E = E_{\text{short}} + E_{\text{long}}\:,
\end{equation}
and then only $E_{\text{short}}$ will be considered. Moreover, we always assume the further partition 
of $E_{\text{short}}$ in local-atomic contributions. This implies the decomposition
\begin{equation}
    E_{\text{short}} = \sum_{i}^{\text{atoms}} E_{i}\:,\nonumber
\end{equation}
where $E_i$ are the local terms, which are defined for each atom of the system. We will adopt the same 
hypothesis also for the case of a general tensor $\bm T$ (as done in Ref.~\cite{SAGPR}), and that one can
write
\begin{equation}
        \bm T \simeq\sum^{\text{atoms}}_i \bm T_i\:,
\end{equation}
for some local contribution, $\bm T_i$. Note that, in the case of intensive quantities, the left-hand side must be divided by the number of atoms in order to recover the mean: in the following, we will always imply this distinction.

An important object in the designing of ML models is the atomic density, $\rho_i$, which has the form 
\begin{equation}\label{eq:intro_atomic_density}
    \rho_i(\vb r) = \sum_{j}^\text{atoms} h_{Z_jZ_i}(\vb r - \vb r_{ji})\:.
\end{equation}
Here the $h_{Z_jZ_i}$'s are localization functions, usually Gaussians or Dirac-deltas. The vector 
$\vb r_{ji}:= \vb r_j-\vb r_i$, is the relative vector between the positions of the $i$-th atom, taken as 
the center of the reference frame, and its $j$-th neighbor. Thus, the atomic density $\rho_i$ is a mathematical 
representation of the atomic environment around the $i$-th atom. We observe that the localization functions 
can be defined as parametrically dependent on the atomic species, $Z_i$ and $Z_j$, of the two atoms 
involved. Taking the Dirac-deltas as an example, the localization functions can be written as
\begin{equation}\label{eq:Dirac_density_case}
    h(\vb r - \vb r_{ji}) = f_{\text{cut}}(r_{ji}) \delta(\vb r- \vb r_{ji})\:,
\end{equation}
where the cut-off function, $f_c(\text{cut})$, smoothly vanishes when the distance $r_{ji}$ between the 
two atoms approaches an optimized distance, called the cut-off radius, $r_\text{cut}$. Its role is to enforce 
the short-ranged nature of the mathematical description in a continuous way~\cite{symmfunc}.

A crucial role in the design of a description for $\rho_i(\vb r)$ is covered by its expansion coefficients over 
the product of a radial basis with spherical harmonics. Namely, we adopt the expansion
\begin{equation}
    \rho_i(\vb r) = \sum_{nlm} c_{i,nlm} R_{nl} (r) Y_l^m(\ver r)\:,
\end{equation}
with the expansion coefficients, $c_{i,nlm}$, given by
\begin{equation}
    c_{i,nlm} = \int \dd \vb r\, \rho_i(\vb r) R_{nl}(r)Y^{m*}_l (\ver r)\:.
\end{equation}
We remark that this expression holds only when the radial basis is real and ortho-normal. For example, in 
the case of the Dirac's-delta localization functions of Eq.~\eqref{eq:Dirac_density_case} (also called 
``sharp''-case~\cite{Review4}), the coefficients take the form 
\begin{equation}\label{eq:expansion_coefficients_density}
    c_{i,nlm} = \sum_{j}^{\text{atoms}} f_c(r_{ji})R_{nl}(r_{ji})Y^{m*}_l(\ver r_{ji})\:.
\end{equation}
The expansion coefficients of the atomic density can be considered as the most relevant and delicate
quantity for the construction of physically driven ML models. This is because the cost of their computation 
scales linearly with the number of atoms in the atomic environment, a property that becomes crucial when 
considering relatively large number of neighboring atoms. 

In this work we will explicitly address two ML methods: the linear one and $\lambda$-SOAP. In both cases, 
we will consider an expansion of the atomic energies in terms of body-order terms such as, for 
example~\cite{ACE,JLP},
\begin{equation}
    E_i = E_i^{(1)} + E_i^{(2)} + E_i^{(3)} +\ldots.
\end{equation}
Here, the $E_i^{(2)}$ and $E_i^{(3)}$ can be written in terms of the powerspectrum and the bispectrum 
components~\cite{SOAP} as
\begin{equation}
\begin{dcases}
        E^{(2)}_i :=  \sum_{n_1n_2l} a_{n_1n_2l} p_{i,n_1n_2l}\:,\\
        E_i^{(3)}:= \sum_{\substack{n_1n_2n_3\\l_1l_2l_3}}  a_{\substack{n_1n_2n_3\\l_1l_2l_3}}  B_{i,\substack{n_1n_2n_3\\l_1l_2l_3}}\:,\\
        \ldots,
\end{dcases}
\end{equation}
with the powerspectrum components defined as
\begin{equation}\label{eq:powerspectrum_def}
    p_{i,n_1n_2l} := \sum_m (-1)^m c_{i,n_1lm}c_{i,n_2l-m},
\end{equation}
and the bispectrum components as 
\begin{equation}\label{eq:bispectrum_components}
    B_{i,\substack{n_1n_2n_3\\l_1l_2l_3}} := \sum_{m_1m_2m_3} c^*_{i,n_3l_3m_3}C^{l_3m_3}_{l_1m_1l_2m_2}c_{i,n_1l_1m_1}c_{i,n_2l_2m_2}\:.
\end{equation}
Here the $C^{lm}_{l_1m_1l_2m_2}$ are Clebsch-Gordan (CG) coefficients \cite{Angular}, while the unknown 
coefficients, $a_{n_1n_2l}$ and $a_{\substack{n_1n_2n_3\\l_1l_2l_3}}$, are obtained by means of some 
optimization method, such the $L^2$ minimization. We will not discuss here such optimizations and we will 
simply assume that a proper solver is in place. The rotational invariance of both powerspectrum and 
bispectrum has been fully established (see, for example, Ref.~\cite{SOAP}), and in this work we will prove 
that such invariance is a straightforward consequence of the MultiSHs formalism introduced here.

Similar expressions can be obtained for the (harmonic-)covariant components of the tensor $\bm T_i$, 
here indicated as $T_{i,\lambda\mu}$. In particular, we can assume the multi-body expansion
\begin{equation}
    T_{i,\lambda\mu} = T_{i,\lambda\mu}^{(1)} +  T_{i,\lambda\mu}^{(2)} + T_{i,\lambda\mu}^{(3)} +\ldots\:,
\end{equation}
with 
\begin{equation}\label{eq:tensor_exp}
    T^{(2)}_{i,\lambda\mu} := \sum_{\substack{n_1l_1\\n_2l_2}} a_{\substack{n_1l_1\\n_2l_2}}\sum_{m_1m_2} C^{\lambda\mu}_{l_1m_1l_2m_2} c^*_{i,n_1l_1m_1}c^*_{i,n_2l_2m_2}\:,
\end{equation}
and 
\begin{align}\label{eq:tensor_expansions_3B}
    T_{i,\lambda\mu}^{(3)} &:= \sum_{\substack{n_1n_2n_3\\l_1l_2l_3L}} a_{\substack{n_1n_2n_3\\l_1l_2l_3L}}\sum_{m_3M}C^{\lambda\mu}_{LMl_3m_3}c^*_{i,n_3l_3m_3}\\
    &\qquad\qquad\times\sum_{m_1m_2}C^{LM}_{l_1m_1l_2m_2} c^*_{i,n_1l_1m_1}c^*_{i,n_2l_2m_2}\:.\nonumber
\end{align}
This is, essentially, the basis employed in Reference~\cite{NICE}. Importantly, these cases reduce to the 
scalar ones when $\lambda=\mu=0$, as can be seen by the example
\begin{align}\label{eq:tensor_exp_powerspectrum}
    T^{(2)}_{i,00} = \sum_{n_1n_2l} a_{n_1n_2l}\dfrac{(-1)^l}{\sqrt{2l+1}}p_{i,n_1n_2l} = E_i^{(2)}\:,
\end{align}
obtained by re-defining the expansion coefficients $a_{n_1n_2l}$ and by using the facts that
 \begin{equation}
    C^{00}_{l_1m_1l_2m_2} = (-1)^{l_1-m_1}\dfrac{\delta_{l_1l_2}\delta_{m_1-m_2}}{\sqrt{2l_1+1}}\:,
\end{equation}
and that the powerspectrum components are real. As for the scalar case, the proof that these expansions 
possess the correct transformation rules will be easily derived by means of the MultiSHs framework. 

Another possible approach is provided by the Gaussian approximation potential (GAP) framework \cite{GAP}, 
which establishes a kernel-based alternative for the fit of the PES and of tensorial components. In general, 
the model is defined as 
\begin{equation}
    E^{(\nu)}_i = \sum_{k=1}^{\text{config}}\alpha_k K^{(\nu)}(\rho_i,\rho^{(k)})\:,
\end{equation}
where $K^{(\nu)}(\rho_i,\rho^{(k)})$ is the kernel, which provides a measure of similarity between the atomic 
environments describing $\rho_i$ and $\rho^{(k)}$. Here, the sum runs over the $k$-configurations used to 
train the model. If we indicate the kernel evaluated between two configurations in the training set with 
$\bm K^{(\nu)}_{kk'} := K^{(\nu)}(\rho^{(k)},\rho^{(k')})$, the expansion coefficients are obtained by the 
analytical formula
\begin{equation}
    \alpha_k = \sum_{k'=1}^\text{config}(\bm K^{(\nu)}+\gamma^2 \mathbb{I})^{-1}_{kk'}E_{k'}\:,
\end{equation}
where $\mathbb{I}$ is the identity, $\gamma$ is an optimized regularization constant and $E_{k'}$ is the energy 
of the $k'$-th configuration in the training set.

Explicitly, the SOAP kernel is defined as \cite{SOAP}
\begin{equation}\label{eq:SOAP_kernel_scalar}
    K^{(\nu)}(\rho,\rho') = \int \dd \hat{R}\, \abs{\int \dd \vb r\, \rho(\vb r)\rho'(\hat{R}\vb r)}^\nu\:,
\end{equation}
where the first integration is an Haar integral performed over all possible rotations, $\hat R$, usually 
parameterized by Euler angles. Here, increasing $\nu$ increases the body order of this expression, 
as we will show in later sections. While an explicit expression for this kernel exists, usually in the form 
of inner products of powerspectrum and bispectrum components, a result of this work will consists in 
providing a simple derivation for the explicit evaluation for any order $\nu$ of interest.

Also in the kernel case, a similar approach can be adopted for the covariant components of a tensor, 
$\bm T$, more explicitly
\begin{equation}
        T^{(\nu)}_{i,\lambda\mu} = \sum_{k=1}^{\text{config}}\alpha_{k,\mu'} (K^{(\nu)}(\rho_i,\rho^{(k)}))^\lambda_{\mu\mu'}\:,
\end{equation}
where again this admits an analytical solution for the coefficients $\alpha_{k,\mu'}$. Here, 
$(K^{(\nu)}(\rho_i,\rho^{(k)}))^\lambda_{\mu\mu'}$ is the covariant generalization of the scalar kernel and, 
as such, it correctly reduces to the scalar case for $\lambda=\mu=0$. 

The covariant, $\lambda$-SOAP, generalized version of the kernel above is provided by (see 
references~\cite{Glielmo_covariant,SAGPR})
\begin{equation}\label{eq:covariant_kernel_def0}
    (K^{(\nu)}(\rho,\rho'))^\lambda_{\mu\mu'} = \int \dd \hat{R}\, \left(\bm D^\dagger(\hat{R})\right)^\lambda_{\mu\mu'} \abs{\int \dd \vb r\, \rho(\vb r)\rho'(\hat{R}\vb r)}^\nu\:,
\end{equation}
where the covariance is enforced by means of the Hermitian conjugate of the Wigner $D$-matrix \cite{Angular}, 
$\bm D^\dagger$. From the fact that $D^0_{00}=1$, we can see how this expression does, indeed, reduce to the 
scalar one for $\lambda=0$. We anticipate that the result of Section~\ref{sec:covariant_kernel_simple_derivation}, 
which can be regarded as the first result of this work, will provide a compact derivation and formula for the covariant 
kernel of Eq.~(\ref{eq:covariant_kernel_def0}). This will be achieved by exploiting just two properties of the MultiSHs 
formalism.

Before proceeding with the introduction of the bipolar spherical harmonics, it is interesting to show that all the 
formulas introduced till this point can be casted in terms of the product of atomic densities, defined as
\begin{equation}
    \rho^{\otimes \nu}(\vb r_1,\ldots,\vb r_\nu) := \rho(\vb r_1)\cdot\ldots\cdot\rho(\vb r_\nu)\:,
\end{equation}
which is arguably the core quantity in the construction of models and descriptors, as already observed 
previously~\cite{Review4,Willat}. In order to provide two meaningful examples, we can first observe that the 
product of $\nu$ expansion coefficients $c_{i,nlm}$ is given by
\begin{align}
    \prod_{\alpha=1}^{\nu}c_{i,n_\alpha l_\alpha m_\alpha}=\int \dd \vb r_1 &\ldots \dd \vb r_\nu\, \rho^{\otimes \nu}(\vb r_1,\ldots, \vb r_\nu)\nonumber\\
    &\times\left[\prod_{\alpha = 1}^\nu R_{n_\alpha l_\alpha}(r_\alpha)Y_{l_\alpha}^{m_\alpha}(\ver r_\alpha)\right].\nonumber
\end{align}
This implies that any expression like the one in Eq.~(\ref{eq:tensor_expansions_3B}) can be evaluated 
from $\rho^{\otimes \nu}$. Analogously, the expression for the covariant kernel can be written as
\begin{equation}\label{eq:covariant_kernel_def}
\begin{split}
      (K^{(\nu)}(\rho,\rho'))^\lambda_{\mu\mu'} =  \int \dd \vb r_1\ldots \dd \vb r_\nu \rho^{\otimes n}(\vb r_1,\ldots,\vb r_\nu)\\
      \times\int \dd \hat{R}\, D^{\lambda*}_{\mu'\mu}(\hat{R})\,\rho'^{\otimes n}(\hat{R}\vb r_1,\ldots, \hat{R} \vb r_\nu)\:,
\end{split}
\end{equation}
which again is written in terms of $\rho^{\otimes \nu}$ only.

Having identified $\rho^{\otimes \nu}$ as the principal object of our investigation, the remaining of this section will be 
devoted to the definition of a suitable basis for a function that depends, simultaneously, on the $\nu$ vectors 
$\vb r_1,\ldots,\vb r_\nu$. We want to remark that the actual form of the $\rho^{\otimes \nu}$ is not crucial. Indeed, 
most of the properties of interests will be derived only by means of the properties of the basis, encoded in 
Eqs.~\eqref{eq:Multi_ortho}, \eqref{eq:Multi_rotation} and \eqref{eq:radial_ortho}, together with the fact that 
$\rho^{\otimes \nu}$ is real, scalar, and is constructed from a product of functions. The specific form of the local 
atomic density, as given in Eq.~\eqref{eq:intro_atomic_density}, will be first discussed as far as 
Section~\ref{sec:on_the_atomic_density}, and used only as a bridge towards a compact and general recipe to 
define linear model for PES and tensorial quantities. 

We will now introduce the formalism based on the MultiSHs in a systematic way, starting from the simpler cases 
of the bipolar- and tripolar-spherical harmonics. In this way, we will develop the basic tools and form the intuition 
that will be then used in the general formulation of the formalism of Section~\ref{sec:multipolar_spherical}.

\subsection{Bipolar spherical harmonics}\label{sec:Bipo}

Let us start the discussion on the MultiSHs formalism by looking at the simpler case of the bipolar spherical 
harmonics (BipoSHs). Firstly, we will provide the formal definition and then we will show the core properties 
of these objects.

A bipolar spherical harmonic is defined as
\begin{equation}\label{eq:def_BipoSH}
    \mathcal{Y}_{l_1l_2}^{\lambda\mu}(\ver r_1, \ver r_2) := \sum_{m_1m_2} C^{\lambda\mu}_{l_1m_1l_2m_2} Y_{l_1}^{m_1}(\ver r_1)Y_{l_2}^{m_2}(\ver r_2)\:,
\end{equation}
namely it is obtained by contracting two spherical harmonics with a set of CG coefficients. Within standard 
angular-momentum theory \cite{Angular,Weissbluth}, the BipoSHs can be seen as the position representation 
of the coupling of two angular momenta, $l_1$ and $l_2$, namely
\begin{equation}\nonumber
    \mathcal{Y}_{l_1l_2}^{\lambda\mu}(\ver r_1, \ver r_2) \equiv \braket{\ver r_1,\ver r_2}{l_1l_2\lambda\mu}\:,
\end{equation}
where the coupled basis is defined as
\begin{equation}\nonumber
     \ket{l_1l_2\lambda\mu} = \sum_{m_1m_2}C^{\lambda\mu}_{l_1m_1l_2m_2} \ket{l_1m_1l_2m_2}\:.
\end{equation}

\begin{figure*}
    \centering
    \includegraphics[width=\textwidth]{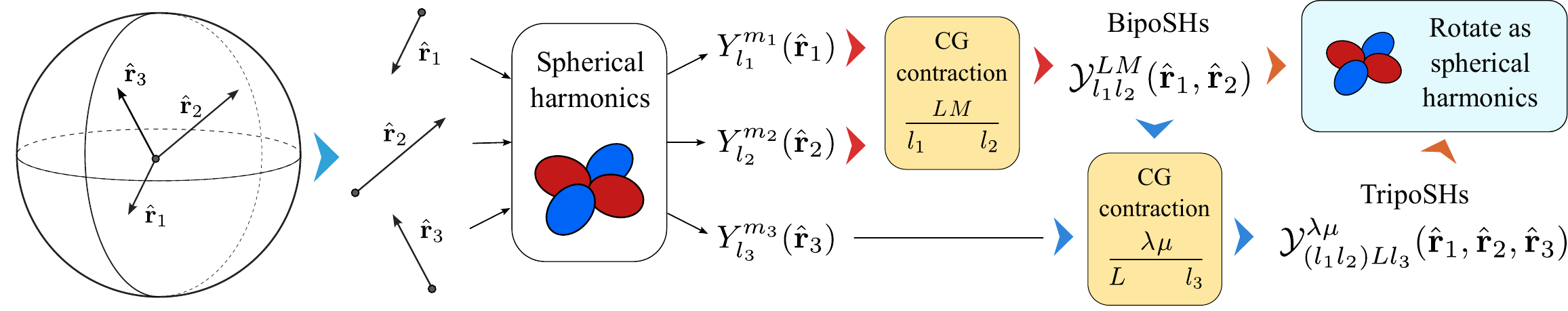}
    \caption{A pictorial view of all the main ingredients associated to the BipoSHs and TripoSHs. The BipoSHs constitute a 
    basis set for the space of two-point functions defined on the surfaces of a sphere. In particular, they can 
    be constructed from the product of two spherical harmonics projected onto the space of total angular momentum 
    $(LM)$. For this reason, the main property of the BipoSHs is that, under a global rotation, they 
    transform as the single spherical harmonic $Y^M_L$. Analogously, the TripoSHs constitute a basis for the space of 
    three-point functions of the sphere, they are constructed by a further projection into the space of angular momentum 
    $(\lambda\mu)$, and transform under rotation as $Y^\mu_\lambda$. We can appreciate how the construction of higher 
    order of multi-point spherical harmonics can be performed iteratively and can be easily generalized to any order.}
    \label{fig:1}
\end{figure*}

Introducing the BipoSHs enables one to immediately exploit their core properties. Firstly and foremost, the 
BipoSHs constitute a complete orthonormal basis, namely
\begin{equation}
    \int \dd \ver r_1 \dd \ver r_2\, \mathcal{Y}_{l_1l_2}^{\lambda\mu}(\ver r_1, \ver r_2)\mathcal{Y}_{l'_1l'_2}^{\lambda'\mu' *}(\ver r_1, \ver r_2) = \delta_{l_1l'_1}\delta_{l_2l'_2} \delta_{\lambda\lambda'}\delta_{\mu\mu'}\:.
\end{equation}
Thus, a general two points function depending on the unit vectors, $\ver r_1$ and $\ver r_2$, that admits an expansion 
in terms of spherical harmonics, can be expanded as
\begin{equation}
    f(\ver r_1,\ver r_2) = \sum_{l_1l_2\lambda\mu} u^{\lambda\mu}_{l_1l_2}\mathcal{Y}_{l_1l_2}^{\lambda\mu}(\ver r_1,\ver r_2)\:,
\end{equation}
with the expansion coefficients given by
\begin{align}
    u^{\lambda\mu}_{l_1l_2} = \int \dd \ver r_1 \dd\ver r_2\,  f(\ver r_1,\ver r_2)\mathcal{Y}_{l_1l_2}^{\lambda\mu*}(\ver r_1,\ver r_2)\:.
\end{align}
As shown in Appendix~\ref{Appendix_sph_BipoSHs}, this formulation can be obtained from the standard expansion in 
terms of two spherical harmonics through the orthogonality of the CG coefficients, namely~\cite{Angular}
\begin{equation}\label{eq:orthogonality_CG}
    \sum_{\lambda\mu} C^{\lambda\mu}_{l_1m_1l_2m_2}C^{\lambda\mu}_{l_1m'_1l_2m'_2} = \delta_{m_1m'_1}\delta_{m_2m'_2}\:.
\end{equation}
This, incidentally, also shows that the CG coefficients constitute the (orthogonal) matrix of the change of basis between 
the two representations. As we will see in this section, such property will be crucial in constructing the MultiSHs formalism, 
since the matrix of the change of basis from a spherical harmonics representation to MultiSHS, can be always represented 
in terms of contractions of products of CG coefficients.

The second crucial property of the BipoSHs, which is the main reason behind the development of this formalism, is their 
behavior under the simultaneous rotation, $\hat{R}$, of its arguments. Explicitly, the following relation holds  
\begin{equation}\label{eq:rotation_BipoSH}
     \mathcal{Y}_{l_1l_2}^{\lambda\mu}(\hat{R}\ver r_1, \hat{R}\ver r_2) = \sum_{\mu'} D^{\lambda*}_{\mu\mu'}(\hat{R})\mathcal{Y}_{l_1l_2}^{\lambda\mu'}(\ver r_1, \ver r_2)\:,
\end{equation}
where, again, we have used the Wigner $D$-matrices. By comparing this behavior with the transformation of a single 
spherical harmonics under rotation,
\begin{equation}\label{eq:transf_rules_sph_harm}
     Y_\lambda^{\mu}(\hat{R}\ver r) = \sum_{\mu'} D^{\lambda*}_{\mu\mu'}(\hat{R})  Y_\lambda^{\mu'}(\ver r)\:,
\end{equation}
we can see how the two transformations are analogous, namely the BipoSHs transforms under rotations as a single 
spherical harmonics. A pictorial representation of the BipoSHs, obtained by using this property, is shown in Fig.~\ref{fig:1}. 
It is worth noticing that Eq.~(\ref{eq:rotation_BipoSH}) can be obtained by using Eq.~(\ref{eq:orthogonality_CG}) and the 
so-called CG series, which states the relation between the CG coefficients and the Wigner $D$-matrices. Explicitly, the 
CG series is given by~\cite{Angular},
\begin{align}\label{eq:CG_series}
    &\sum_{m_1m_2} C^{\lambda\mu}_{l_1m_1l_2m_2}D^{l_1*}_{m_1m_1'}(\hat{R})D^{l_2*}_{m_2m_2'}(\hat{R})\\
    &\qquad\qquad\qquad\qquad\qquad=C^{\lambda\mu'}_{l_1m_1'l_2m_2'} D^{\lambda*}_{m'_1m'_2}(\hat{R})\:,\nonumber
\end{align}
a relation that shows how two rotations (represented by the Wigner $D$-matrices) are decomposed into a single 
one by means of the CG coefficients. As we will show in the following  paragraphs, it is possible to iteratively use 
this property to obtain a generalization of Eq.~\eqref{eq:transf_rules_sph_harm} to the MultiSHs case.

Given the rotation rule of Eq.~\eqref{eq:rotation_BipoSH}, the expansion of a general function $f$ in terms 
of BipoSH results in a decomposition over the irreducibles of the rotations group with respect to a \emph{global} 
rotation of all the inputs of the function. Indeed, a rotation $\hat{R}$
results in
\begin{equation}
\begin{split}
     f(\ver r_1,\ver r_2)&\xrightarrow{\hat{R}}f(\hat{R}\ver r_1,\hat{R}\ver r_2) \\&= \sum_{\lambda\mu'}
     \left(\sum_{\mu}D^{\lambda*}_{\mu\mu'}(\hat{R}) u_{l_1l_2}^{\lambda\mu}\right)\mY^{\lambda\mu'}_{l_1l_2}(\ver r_1, \ver r_2)\:,\nonumber
\end{split}
\end{equation}
namely the coefficients of the expansion inherit the transformation rule
\begin{equation}\label{eq:transformation_rule_2B_coeff_BipoSH}
    u^{\lambda\mu}_{l_1l_2} \xrightarrow{\hat{R}} \sum_{\mu}D^{\lambda*}_{\mu\mu'}(\hat{R}) u_{l_1l_2}^{\lambda\mu}\:.
\end{equation}
In other words, also the coefficients belong to the sub-space of angular momentum $\lambda$. At this point, it is crucial 
to notice that we have derived the transformation properties of the expansion coefficient \emph{regardless} of the specific 
form of the function $f$. This is an example of the general fact that investigating the transformation properties of the basis 
is enough to fully determine the behavior of its coefficients.

Another example of the fact that investigating the properties of the basis can be extremely informative stems from the 
fact that the arguments can be interchanged by means of the following transformation
\begin{equation}\label{eq:symmetry_BipoSH}
    \mathcal{Y}_{l_1l_2}^{\lambda\mu}(\ver r_1,\ver r_2) = (-1)^{l_1+l_2-\lambda}\mathcal{Y}_{l_2l_1}^{\lambda\mu}(\ver r_2,\ver r_1),
\end{equation}
which can be directly derived from the analogous symmetry of the CG coefficients. This property has consequences over 
the expansion of a generic symmetric function $f$. Indeed, if $f(\ver r_1,\ver r_2) = f(\ver r_2,\ver r_1)$, then the coefficients 
inherit the same transformation rules of the BipoSH, namely
\begin{equation}\label{eq:symmetry_coeff}
    u_{l_1l_2}^{\lambda\mu} = (-1)^{l_1+l_2-\lambda} u_{l_2l_1}^{\lambda\mu}\:,
\end{equation}
as can be easily proven by following the same derivation of Eq.~\eqref{eq:transformation_rule_2B_coeff_BipoSH}. Again, it 
is worth noticing that we have been able to recover this property without any information on the explicit form of the coefficients 
$u^{\lambda\mu}_{l_1l_2}$. Moreover, Eq.~\eqref{eq:symmetry_coeff} implicitly states that any expression writeable as a 
linear combination of the coefficients $u_{l_1l_2}^{\lambda\mu}$, can be obtained by considering only the $l_1\geq l_2$ 
terms. This property, appropriately generalized in the very last section of this work, is extremely useful in removing 
redundancies from linear ML models.

\subsubsection{Graphical Representation}

We conclude this section by showing that the BipoSHs admit a graphical representation by means of the 
definition
\begin{equation}
\includegraphics[width=.22\textwidth]{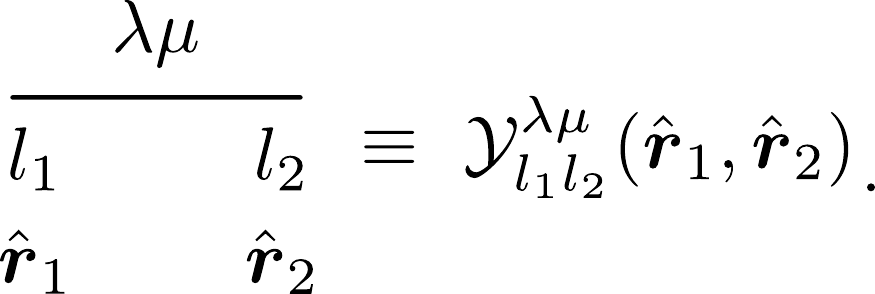}
\end{equation}
Comparing this representation with the definition of Eq.~\eqref{eq:def_BipoSH}, we can notice how the 
horizontal line represents the CG coefficients connecting the channels of angular momentum $l_1$ and $l_2$. 
With this representation, Eq.~\eqref{eq:symmetry_BipoSH} can be graphically written as
\begin{equation}\label{eq:symm_graph}
    \includegraphics[width=.30\textwidth]{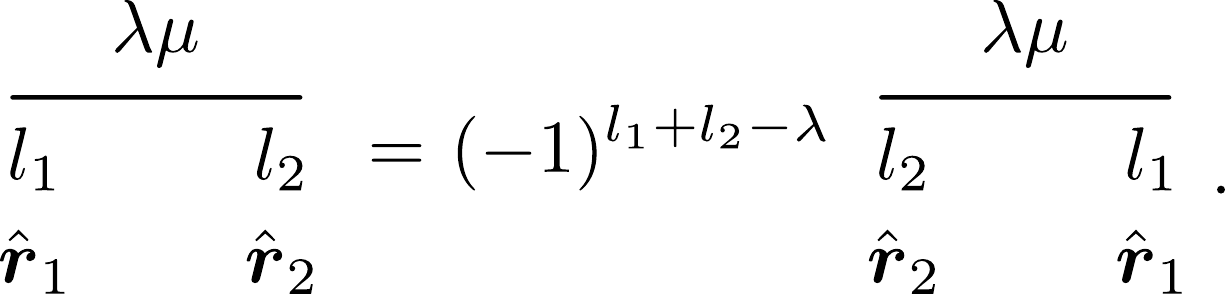}
\end{equation}
This shows that the graph is oriented, from left to right, and that changing the order gives rise to a phase change. 
This graphical representation is trivial for the case of the BipoSH, but it will become a useful tool in the following 
sections, when we will deal with different choices of coupling schemes.

\subsection{Tripolar spherical harmonics}\label{sec:Tripo}
Following the same approach of the previous section, we start here with the definition of the tripolar-spherical 
harmonics (TripoSHs), followed by a discussion of their most important properties and of their use as a basis. 

The TripoSHs are the generalization of the BipoSHs for the case of three-points functions. Indeed, they can be 
defined as \begin{align}\label{eq:TripoDef}
&\mY^{\lambda\mu}_{(l_1l_2)Ll_3}(\ver r_1,\ver r_2,\ver r_3)\\
&:= \sum_{\substack{m_1m_2\\m_3M}} C^{\lambda\mu}_{LMl_3m_3}C^{LM}_{l_1m_1l_2m_2} Y_{l_1}^{m_1}(\ver r_1)Y_{l_2}^{m_2}(\ver r_2)Y_{l_3}^{m_3}(\ver r_3)\nonumber\:,
\end{align}
and can be represented graphically as
\begin{equation}\label{eq:3B_graph_1}
    \includegraphics[width=.35\textwidth]{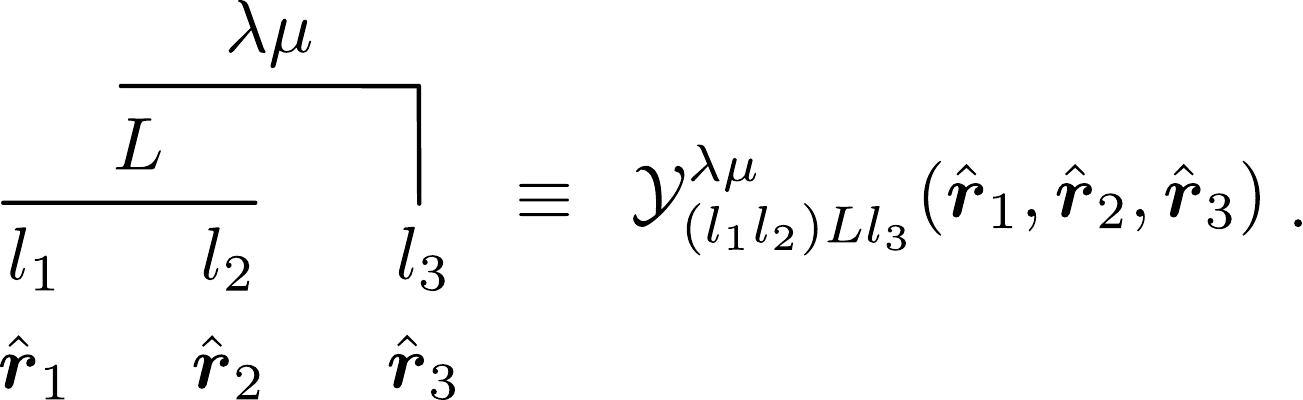}
\end{equation}
A pictorial representation of the TripoSHs is shown in Fig.~\eqref{fig:1}

The TripoSHs form an orthonormal basis for the three-points functions defined on the surface of a sphere. 
Explicitly, their orthogonality reads
\begin{align}
    \int\dd \ver r_1\dd\ver r_2\dd\ver r_3 \,\mY^{\lambda\mu}_{(l_1l_2)Ll_3}(\ver r_1,\ver r_2,\ver r_3)\mY^{\lambda'\mu'*}_{(l'_1l'_2)L'l'_3}(\ver r_1,\ver r_2,\ver r_3)\nonumber \\
    = \delta_{\lambda\lambda'}\delta_{\mu\mu'}\delta_{l_1l'_1}\delta_{l_2l'_2}\delta_{l_3l'_3}\delta_{LL'}\:.\nonumber
\end{align}
Since the TripoSHs form a basis, when a function $f(\ver r_1,\ver r_2,\ver r_3)$ admits an expansion in terms of 
spherical harmonics, then it can be also expanded as
\begin{equation}\label{eq:completeness_TripoSH}
    f(\ver r_1,\ver r_2,\ver r_3) = \sum_{\substack{l_1l_2\\l_3L}}\sum_{\lambda\mu}u^{\lambda\mu}_{(l_1l_2)Ll_3}\mY^{\lambda\mu}_{(l_1l_2)Ll_3}(\ver r_1,\ver r_2,\ver r_3)\:,
\end{equation}
with the expansion coefficients given by
\begin{equation}
    u^{\lambda\mu}_{(l_1l_2)Ll_3} =  \int\dd \ver r_1\dd\ver r_2\dd\ver r_3 \, f(\ver r_1,\ver r_2,\ver r_3)\mY^{\lambda\mu*}_{(l_1l_2)Ll_3}(\ver r_1,\ver r_2,\ver r_3)\:.
\end{equation}

Similarly to the BipoSHs of Eq.~\eqref{eq:rotation_BipoSH}, the defining property of the TripoSHs is their behavior 
under rotation, which is analogous to the rotation of a single spherical harmonics of order $\lambda\mu$. Explicitly, 
we have
\begin{equation}\label{eq:rotation_TripoSH}
\begin{split}
    &\mY^{\lambda\mu}_{(l_1l_2)Ll_3}(\hat{R}\ver r_1,\hat{R}\ver r_2,\hat{R}\ver r_3)\\ 
    &\qquad\qquad= \sum_{\mu'}D^{\lambda*}_{\mu\mu'}(\hat{R})\mY^{\lambda\mu'}_{(l_1l_2)Ll_3}(\ver r_1,\ver r_2,\ver r_3)\:.
\end{split}
\end{equation}

From their definition and graphical representation, it follows that a TripoSH can be constructed by contracting a 
BipoSHs with a spherical harmonic, namely
\begin{equation}\label{eq:recursion_TripoSHs}
    \mY^{\lambda\mu}_{(l_1l_2)Ll_3}(\ver r_1,\ver r_2,\ver r_3)= \sum_{m_3M} C^{\lambda\mu}_{LMl_3m_3}\mY^{LM}_{l_1l_2}(\ver r_1,\ver r_2)Y_{l_3}^{m_3}(\ver r_3)\:.
\end{equation}
The importance of this recursion relation is twofold. On the one hand, once the BipoSHs have been already 
evaluated and stored, the TripoSHs can be constructed in a simple way. On the other hand, it allows to prove 
Eq.~\eqref{eq:rotation_TripoSH}, by using the CG series of Eq.~\eqref{eq:CG_series}. Indeed, since rotating 
the arguments of BipoSHs introduces only one Wigner $D$-matrix, we are again in the situation in which we 
are contracting two Wigner $D$-matrices and one CG coefficients, namely we can directly apply the CG series. 
Not only this proves Eq.~\eqref{eq:rotation_TripoSH}, but it also shows a general property. Namely, sequentially 
contracting spherical harmonics with CG-coefficients projects the expression into the space of the required global 
angular momentum.
This statement will be crucial 
for the construction of the general MultiSHs formalism. Before proceeding, let us briefly discuss the choice of 
the coupling scheme in the definition of the TripoSHs.

\subsubsection{Change of coupling for symmetric functions}

An important difference between the BipoSHs and the TripoSHs, is that the definition of the latter requires 
the selection of a coupling scheme. Indeed, in the example above, we have that the channel $l_1$ and $l_2$ 
are first coupled into the $L$ channel, which is then coupled to the channel $l_3$ and finally projected into the 
space $(\lambda,\mu)$. This is represented by the string $((l_1l_2)Ll_3)$. However, not only it is possible to 
define other coupling schemes by means of different strings or graphs, but these also form a separate set 
of TripoSHs. For example, the string $(l_1(l_2l_3)L)$ produces the TripoSHs
\begin{equation}\label{eq:3Bgraph_tree2}
    \includegraphics[width=.37\textwidth]{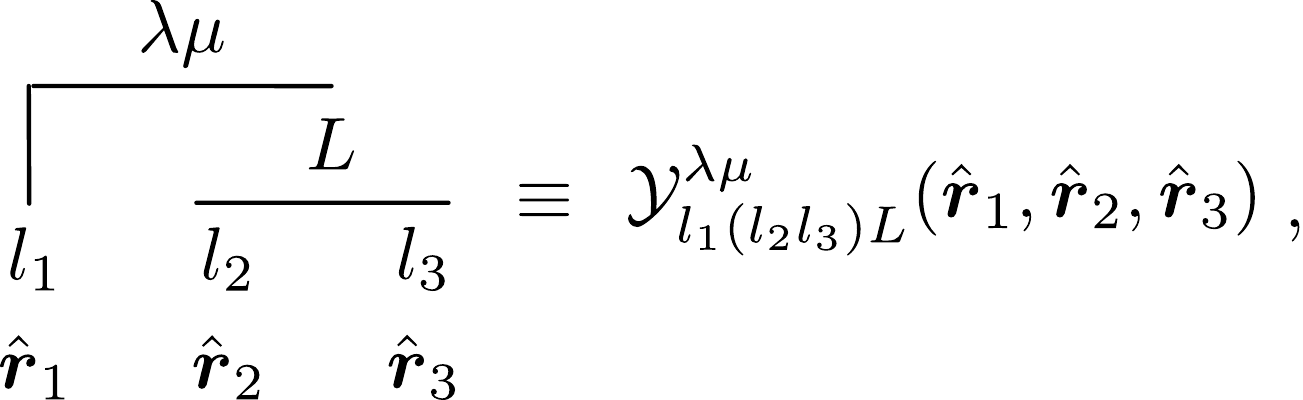}
\end{equation}
which corresponds to the analytical expression
\begin{align}\label{eq:TripoDef2}
&\mY^{\lambda\mu}_{l_1(l_2l_3)L}(\ver r_1,\ver r_2,\ver r_3)\\
&= \sum_{\substack{m_1m_2\\m_3M}} C^{\lambda\mu}_{l_1m_1LM}C^{LM}_{l_2m_2l_3m_3} Y_{l_1}^{m_1}(\ver r_1)Y_{l_2}^{m_2}(\ver r_2)Y_{l_3}^{m_3}(\ver r_3)\nonumber.
\end{align}
It is clear that, by comparing this graph with the one of Eq.~\eqref{eq:3B_graph_1}, the two 
sets of TripoSHs are different. However, since both produce a basis, there must exist a unitary 
transformation connecting the two different couplings. For example, in the cases just discussed, 
we have the following relation 
\begin{align}
    \mY^{\lambda\mu}_{(l_1l_2)Ll_3}(\ver r_1,\ver r_2,\ver r_3) &= \sum_{L'} \sqrt{(2L+1)(2L'+1)}\\&\times W(l_1l_2\lambda l_3;LL')\mY^{\lambda\mu}_{l_1(l_2l_3)L'}(\ver r_1,\ver r_2,\ver r_3)\nonumber,
\end{align}
where the components of the unitary matrix are given by the Racah-$W$ coefficients 
(see reference~\cite{Angular}). Another possible case has already been given in Eq.~\eqref{eq:symm_graph}, 
where the change of ordering can be interpreted as a trivial change of coupling. 

To further stress the importance of the coupling-scheme choice, let us proceed with an example. 
Consider a 3-points function, $f(\ver r_1,\ver r_2,\ver r_3)$, symmetric under any permutation of 
its arguments. This can be expanded as
\begin{equation}
    f(\ver r_2,\ver r_1,\ver r_3) = \sum_{\substack{l_2l_3\\l_1L}}\sum_{\lambda\mu}u^{\lambda\mu}_{(l_2l_3)Ll_1}\mY^{\lambda\mu}_{(l_2l_3)Ll_1}(\ver r_2,\ver r_3,\ver r_1)\:.
\end{equation}
On the one hand, because of the symmetries of the function $f$, this expression must be equal to the expansion of Eq.~\eqref{eq:completeness_TripoSH}. On the other hand we know that there must be a unitary matrix that allows the 
change of coupling
\begin{align}
    &\mY^{\lambda\mu}_{(l_2l_3)Ll_1}(\ver r_2,\ver r_3,\ver r_1) \\
    &\quad= \sum_{L'} \braket{(l_1l_2)L'l_3,\lambda}{(l_2l_3)Ll_1,\lambda} \mathcal{Y}^{\lambda\mu}_{(l_1l_2)L'l_3}(\ver r_1,\ver r_2,\ver r_3)\nonumber,
\end{align}
namely that restores the order of the arguments of the function. Here we have expressed the unitary 
matrix through a Dirac-braket notation, since its explicit expression (in terms of Racah-$W$ coefficients) 
is unessential for our discussion. By inserting this last expression into the expansion of $f$, and comparing 
with Eq.~\eqref{eq:completeness_TripoSH}, leads to the relation
\begin{equation}\label{eq:Tripo_redundancy_reduction}
    u^{\lambda\mu}_{(l_1l_2)Ll_3} = \sum_{L'}\braket{(l_1l_2)Ll_3,\lambda}{(l_2l_3)L'l_1,\lambda}u^{\lambda\mu}_{(l_2l_3)L'l_1}\:,
\end{equation}
for the expansion coefficients. Several other relations similar to this one can be obtained by a change 
of coupling. Crucially, the example shows that, in the case of a symmetric function, the order of the $l$-s 
channels can be always changed by means of an appropriate change of coupling. This fact will be important 
in reducing the redundancies for the general MultiSHs formalism, as shown in the last section of this work. 
Again, we remark that this symmetry has been obtained by exploiting the symmetry of the function $f$, but 
with no knowledge of the specific form of the expansion coefficients $u$.

Before proceeding to a discussion of the general MultiSHs formalism, we point out that, as a matter of notation, 
we will indicate different graphs (or coupling trees) with the letter $\tau$. For example, a possible choice is to 
label the graph of Eq.~\eqref{eq:3B_graph_1} with $\tau_1$ and the one of Eq.~\eqref{eq:3Bgraph_tree2} 
with $\tau_2$.

\subsection{Multipolar spherical harmonics (MultiSHs)}\label{sec:multipolar_spherical}
\subsubsection{Notation and definitions} 
In this section, we will introduce the MultiSHs in their generality. Because of the rapid growth 
in the number of indexes involved in the various expressions, and in order to keep the notation 
more compact and readable, let us first define a few practical short-hands (analogously to the 
ones introduced in reference~\cite{ACE}). 

Firstly, $\nu$-position vectors will be defined by the single vector
\begin{equation}
    \vb x := (\vb r_1,\ldots,\vb r_\nu)\:.
\end{equation}
From this, we can define the collection of lengths and directions as
\begin{equation}
    x := (r_1,\ldots,r_\nu), \qquad \text{and}\qquad\ver x := (\ver r_1,\ldots,\ver r_\nu)\:, 
\end{equation}
respectively. To further reduce the number of indexes, we will always imply the dimensionality 
of these vectors, which will be understood from the context. Similarly, we can also collect all the 
$l$-indexes in a vector $\bm l$, defined as
\begin{equation}
    \bm l:= (l_1,\cdots,l_\nu)\:.
\end{equation}
In contrast, we will indicate the intermediate coupling channels by means of the vector
\begin{equation}
    \bm L:= (L_1,\cdots, L_{\nu-2})\:.
\end{equation}
We remark here that, if the length of the vector $\bm l$ is $\nu\geq2$, then the length of $\bm L$ will always 
be $(\nu-2)$. In general, the specific coupling tree (or, equivalently, the string of the coupling) will be 
labelled by $\tau$. 

With these definitions in place, the MultiSHs will be written as 
\begin{equation}
    \mY^{\lambda\mu}_{\bm l\bm L}(\ver x) \equiv \mY^{\lambda\mu}_{\bm l\bm L,\tau}(\ver x)\:,
\end{equation}
with the coupling tree, $\tau$, usually implied. The explicitly construction of the MultiSHs will be 
presented shortly and investigated in detail. Before that, however, we will introduce the core properties 
that the basis is required to satisfy, and we will show that they already lead to a practical evaluation of 
the general $\nu$-points $\lambda$-SOAP kernel of Eq.~\eqref{eq:covariant_kernel_def}. Then, in the 
explicit construction of the MultiSHs, we will show how these properties are satisfied.

\subsubsection{Core properties of the MultiSHs}

The core properties for the MultiSHs are their orthonormality, 
\begin{align}\label{eq:Multi_ortho}
    \int \dd \ver x \,  \mY^{\lambda\mu}_{\bm l\bm L}(\ver x) \mY^{\lambda'\mu'*}_{\bm l'\bm L'}(\ver x) = \delta_{\bm l \bm l'}\delta_{\bm L\bm L'} \delta_{\lambda\lambda'}\delta_{\mu\mu'}\:,
\end{align}
and the transformation rule for simultaneous rotation of all the arguments,
\begin{align}\label{eq:Multi_rotation}
    \mY^{\lambda\mu}_{\bm l\bm L}(\hat{R}\ver x) = \sum_{\mu'} D^{\lambda*}_{\mu\mu'}(\hat{R})\mY^{\lambda\mu'}_{\bm l\bm L}(\ver x)\:,
\end{align}
which hold for same number of coupled angular momenta and for fixed coupling.
Here, we have defined the Kronecker delta between two vectors as the product of the 
Kronecker deltas of each component, and we have introduced 
$\dd \ver x := \dd \ver r_1\ldots\dd \ver r_\nu$, and $\hat{R}\ver x:=(\hat{R}\ver r_1,\ldots,\hat{R}\ver r_\nu)$. 
These properties, which generalize the analogous ones for the BipoSHs and TripoSHs, together with the 
fact that the MultiSHs constitute a basis (for any fixed coupling scheme), are among the most 
useful expressions for this work.

It is now also necessary to construct a global radial basis, namely an expansion over the vector $x$. 
With the short-hand notation
\begin{equation}
    \bm n := (n_1,\ldots,n_\nu)\:,
\end{equation}
a suitable basis is given by the product of $\nu$ orthonormal-radial basis of single variable, namely
\begin{equation}\label{eq:def_product_radial_basis}
    \mathcal{R}_{\bm n\bm l} (x) := R_{n_1l_1}(r_1)\cdot\ldots\cdot R_{n_\nu l_\nu}(r_\nu)\:.
\end{equation}
Here, the orthogonality of this basis is expressed by means of the integral
\begin{equation}\label{eq:radial_ortho}
    \int \dd x \,r_1^2\cdot\ldots\cdot r_\nu^2\,  \mathcal{R}_{\bm n\bm l} (x) \mathcal{R}_{\bm n'\bm l'} (x) = \delta_{\bm n \bm n'}\delta_{\bm l\bm l'}\:,
\end{equation}
where, again, $\dd x = \dd r_1\ldots \dd r_\nu$.  With this radial basis and the MultiSHs at hand, we can 
now define an expansion for a general $\nu$-points function, $f(\vb x)$, and we can write
\begin{equation}\label{eq:expansion_in_MultiSHs}
    f(\vb x) = \sum_{\bm n \bm l \bm L}\sum_{\lambda\mu} u^{\lambda\mu}_{\bm n \bm l \bm L} \mathcal{R}_{\bm n \bm l }(x) \mY^{\lambda\mu}_{\bm l \bm L}(\ver x)\:,
\end{equation}
where 
\begin{equation}
    u^{\lambda\mu}_{\bm n \bm l \bm L} = \int \dd \vb x  \,\mathcal{R}_{\bm n \bm l }(x) \mY^{\lambda\mu*}_{\bm l \bm L}(\ver x)
    {f(\vb x)}\:.
\end{equation}

Importantly, from these definitions, we are already able to infer the rotational behaviour of the expansion coefficients, 
$u^{\lambda\mu}_{\bm n \bm l \bm L}$. In fact, by generalizing what done in Eq.~\eqref{eq:transformation_rule_2B_coeff_BipoSH} 
and by employing the transformation rule of Eq.~\eqref{eq:Multi_rotation}, we obtain
\begin{equation}
\begin{split}
    f(\vb x)&\xrightarrow{\hat{R}}f(\hat{R}\vb x) =\\&= \sum_{\bm n \bm l \bm L}\sum_{\lambda\mu'}
     \left(\sum_{m}D^{\lambda*}_{\mu\mu'}(\hat{R}) u_{\bm n \bm l \bm L}^{\lambda\mu}\right)\mathcal{R}_{\bm n\bm l}(x)\mY^{\lambda\mu'}_{\bm l \bm L}(\ver x)\:.
\end{split}
\end{equation}
From this we deduce that, if $f(\vb x)$ is a scalar functions, then the coefficients will undergo the 
following transformation under a rotation, $\hat{R}$,
\begin{equation}\label{eq:transformation_rule_coeff_SH_passive}
    u^{\lambda\mu}_{\bm n \bm l \bm L} \xrightarrow{\hat{R}} \sum_{\mu}D^{\lambda*}_{\mu\mu'}(\hat{R}) u_{\bm n\bm l \bm L}^{\lambda\mu}\:.
\end{equation}
Remarkably, this means that we already know the transformation rules of the coefficients under rotation, without having 
properly defined the MultiSH and without any knowledge of the actual form of the scalar function $f(\ver x)$.

With all the definitions in place, we will now show that representing a function over the MultiSHs greatly simplifies 
the calculation of all the properties of interests for ML models. A first example to demonstrate the usefulness of this 
representation is provided by the calculation of the general covariant kernel of Eq.~\eqref{eq:covariant_kernel_def}. 
We wish to remark that what follows will not be connected with the specific form of the expansion coefficients 
$u^{\lambda\mu}_{\bm n \bm l \bm L}$ and, instead, will be derived uniquely by the properties of 
Eqs.~\eqref{eq:Multi_ortho} and ~\eqref{eq:Multi_rotation} of the MultiSHs basis.

\subsubsection{Calculation of the general $\lambda-$SOAP covariant kernel}\label{sec:covariant_kernel_simple_derivation} 

In order to evaluate the explicit form of the general $\lambda$-kernel of Eq.~\eqref{eq:covariant_kernel_def} 
we will expand the densities $\rho^{\otimes n}$ and $(\rho')^{\otimes n}$ in terms of the basis 
$\mathcal{R}_{\bm n \bm l}(x)\mY^{\lambda\mu}_{\bm l\bm L}(\ver x)$. If the expansion coefficients are labelled 
by $u^{\lambda\mu}_{\bm n \bm l \bm L}$ and $v^{\lambda\mu}_{\bm n \bm l \bm L}$, respectively, then, from 
Eq.~\eqref{eq:covariant_kernel_def}, we have
\begin{align}
      &(K^{(\nu)}(\rho,\rho'))^\lambda_{\mu_1\mu_2} =\nonumber\\
      &=  \int \dd \vb x \rho^{\otimes \nu}(\vb x)
      \int \dd \hat{R}\, D^{\lambda*}_{\mu_2\mu_1}(\hat{R})\,(\rho')^{\otimes \nu}(\hat{R}\vb x)=\nonumber\\
      &= \sum_{\substack{\bm n \bm n'\\\bm l \bm l'\bm L\bm L}}\sum_{\substack{\lambda'\mu'\\\lambda''\mu''}}u^{\lambda'\mu'}_{\bm n \bm l \bm L}v^{\lambda''\mu''*}_{\bm n' \bm l' \bm L'}\int\dd x\, r_1^2\cdot\ldots\cdot r_\nu^2\,\mathcal{R}_{\bm n \bm l}(x)\mathcal{R}_{\bm n' \bm l'}(x)\nonumber\\
      &\qquad\qquad\times \int \dd \ver x \,\mY^{\lambda'\mu'}_{\bm l\bm L}(\ver x)\int \dd \hat{R}\, D^{\lambda*}_{\mu_2\mu_1}(\hat{R})\mY^{\lambda''\mu''*}_{\bm l'\bm L'}(\hat{R}\ver x)\:,
\end{align}
where we have conveniently used the fact that the density function $(\rho')^{\otimes \nu}$ is real, when 
expanding its complex conjugate. Now, we can use the rotation rule of Eq.~\eqref{eq:Multi_rotation} to 
explicitly calculated the MultiSHs with rotated arguments, namely we write
\begin{align}
    &\int\dd\ver x\,\mY^{\lambda'\mu'}_{\bm l\bm L}(\ver x)\int \dd \hat{R}\, D^{\lambda*}_{\mu_2\mu_1}(\hat{R})\mY^{\lambda''\mu''*}_{\bm l'\bm L'}(\hat{R}\ver x) =\nonumber\\
    &= \sum_{m''}\int\dd \hat {R}\,D^{\lambda*}_{\mu_2\mu_1}(\hat{R})D^{\lambda''}_{\mu''\mu'''}(\hat{R})\int\dd\ver x\,\mY^{\lambda'\mu'}_{\bm l\bm L}(\ver x)\mY^{\lambda''\mu'''*}_{\bm l'\bm L'}(\ver x) = \nonumber\\
    &= \dfrac{8\pi^2}{2\lambda+1}\delta_{\lambda \lambda'' }\delta_{\lambda'\lambda''}\delta_{\mu_1\mu'}\delta_{\mu_2\mu''}\delta_{\bm l\bm l'}\delta_{\bm L\bm L'},\nonumber
\end{align}
where we have used the orthogonality of the Wigner $D$-matrices~\cite{Angular} and of the 
MultiSHs [see Eq.~\eqref{eq:Multi_ortho}]. If we now use also the orthogonality of the radial basis, 
Eq.~\eqref{eq:radial_ortho}, then the final expression for the kernel is
\begin{align}\label{eq:multi_lambda_kernel}
    (K^{(\nu)}(\rho,\rho'))^\lambda_{\mu_1\mu_2} = \dfrac{8\pi^2}{2\lambda+1}\sum_{\bm n\bm l \bm L}u^{\lambda \mu_1}_{\bm n \bm l \bm L}v^{\lambda \mu_2*}_{\bm n \bm l \bm L}\:,
\end{align}
where, again, we remark that $u^{\lambda\mu}_{\bm n \bm l \bm L}$ and $v^{\lambda\mu}_{\bm n \bm l \bm L}$ 
are the expansion coefficients of $\rho^{\otimes \nu}$ and $(\rho')^{\otimes\nu}$, respectively. 
In Appendix~\ref{appendix_kernel_properties} we will prove that the above expression undergoes the 
correct rotations expected by a covariant kernel.

Importantly, Eq.~\eqref{eq:multi_lambda_kernel} explicitly shows that the kernel is an inner product, as one 
would expect, of the $\bm n$, $\bm l$, and $\bm L$ channels of the expansion coefficients, a result does not 
depend on the explicit form of these functions. For the actual expressions of the expansion coefficients we refer to 
Section~\ref{sec:on_the_atomic_density}. This compact derivation, holding for any order $\nu$, can be regarded 
as the first result of this work.

We close this section by remarking once again that the explicit form of the $\lambda$-SOAP kernel has been 
obtained by means of the properties of Eqs.~\eqref{eq:Multi_ortho}, \eqref{eq:Multi_rotation} and \eqref{eq:radial_ortho} 
only (and the fact that $\rho^{\otimes \nu}$ is real). Indeed, it is remarkable that we still have not discussed 
how to explicitly construct the MultiSHs. The next paragraph will show how such construction can be achieved, 
by exploiting the graphical methods presented in the previous section.

\subsubsection{Construction of the MultiSHs}
Following the same strategy used in Sections~\ref{sec:Bipo} and ~\ref{sec:Tripo}, we can construct the 
MultiSHs as a contraction of the product of $\nu$ spherical harmonics, by mean of the CG coefficients. 
Explicitly, the general form of MultiSHs is given by
\begin{equation}\label{eq:def_general_multiSH_Gamma}
    \mY^{\lambda\mu}_{\bm l\bm L,\tau}(\ver x) = \sum_{\bm m} \Gamma^{\lambda\mu}_{\substack{\bm l \bm m \bm L}}(\tau) Y^{m_1}_{l_1}(\ver r_1)\cdot\ldots\cdot Y^{m_\nu}_{l_\nu}(\ver r_\nu)\:,
\end{equation}
where we have introduced the additional short-hand notation
\begin{equation}
    \bm m := (m_1,\ldots ,m_\nu)\:,
\end{equation}
while the $\Gamma$'s tensors are the so-called generalized CG coefficients (see for example Ref.~\cite{Yutsis}).
Here, the form of the tensor $\Gamma$ depends on the choice of coupling scheme (or tree). 
For example, in the discussion on the BipoSHs in Section~\ref{sec:Bipo} ($\nu=2$), we have 
shown the two set of coupling tensors
\begin{equation}\label{eq:Gamma_n2_tau1}
    \Gamma^{\lambda\mu}_{\substack{l_1m_1l_2m_2}}(\tau_1) = C^{\lambda\mu}_{l_1m_1l_2m_2}\:, 
\end{equation}
and
\begin{equation}
    \Gamma^{\lambda\mu}_{\substack{l_1m_1l_2m_2}}(\tau_2) = C^{\lambda\mu}_{l_2m_2l_1m_1}\:, 
\end{equation}
where the two graphs $\tau_1$ and $\tau_2$ are given by
\begin{equation}\label{eq:graph_bipo_taus}
    \includegraphics[width =.35\textwidth]{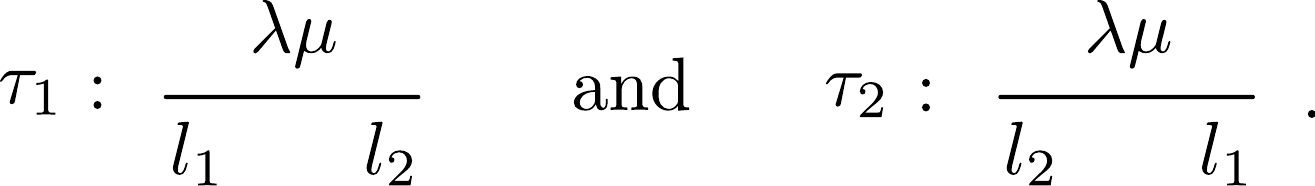}
\end{equation}
It is clear that these coupling trees are the fundamental blocks to define the graphical representation 
of Eq.~\eqref{eq:symm_graph}, which is realized by the combination of such trees together with the 
position representation provided by the vectors $\ver r_1$ and $\ver r_2$. 
From these examples, we can see how an horizontal line in the tree implies the introduction of a CG coefficient 
in the construction of the tensor. Further insight can be acquired from the example provided in Section~\ref{sec:Tripo}, 
where we have introduced the two coupling trees 
\begin{equation}
        \includegraphics[width =.43\textwidth]{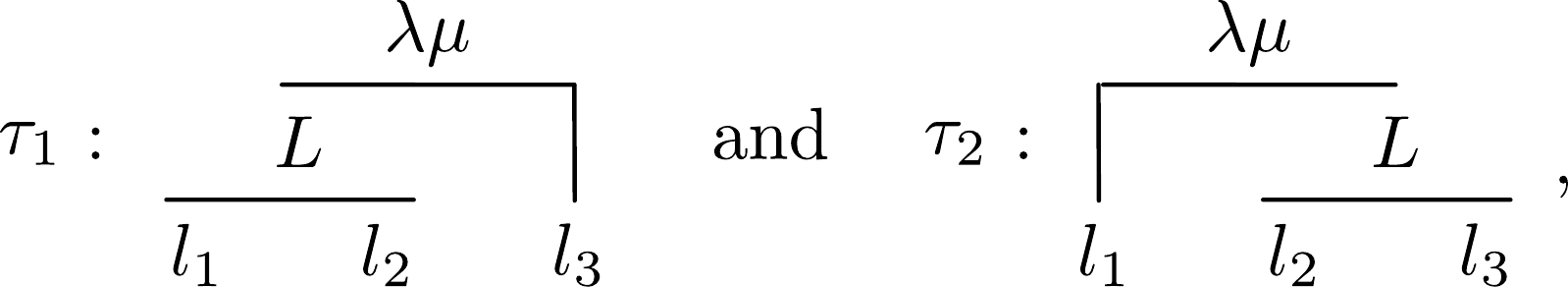}
\end{equation}
contained in Eqs.~\eqref{eq:3B_graph_1} and ~\eqref{eq:3Bgraph_tree2}, respectively. Moreover, as can be seen 
from Eqs.~\eqref{eq:TripoDef} and ~\eqref{eq:TripoDef2}, the corresponding $\Gamma$ tensors are given by
\begin{equation}\label{eq:Gamma_tau_n3}
     \Gamma^{\lambda\mu}_{\substack{l_1l_2l_3\\
     m_1m_2m_3},L}(\tau_1) = \sum_M C^{\lambda\mu}_{LMl_3m_3}C^{LM}_{l_1m_1l_2m_2}\:,
\end{equation}
and
\begin{equation}
     \Gamma^{\lambda\mu}_{\substack{l_1l_2l_3\\
     m_1m_2m_3},L}(\tau_2) = \sum_M C^{LM}_{l_2m_2l_3m_3}C^{\lambda\mu}_{l_1m_1LM}\:.
\end{equation}

These two expressions show that the tensors $\Gamma$ are always constructed from the contraction 
of products of CG coefficients. In particular, the contractions are always performed over all the CG 
coefficients with the only exception being the last projection onto the $(\lambda\mu)$ space. With only 
these rules, once a coupling tree is selected, constructing the corresponding tensor $\Gamma$ (and 
therefore the relative MultiSH), is straightforward. For example, the construction suggested in ACE for 
the $\nu=4$ terms, is given by the following coupling scheme
\begin{equation}
    \includegraphics[width=.25\textwidth]{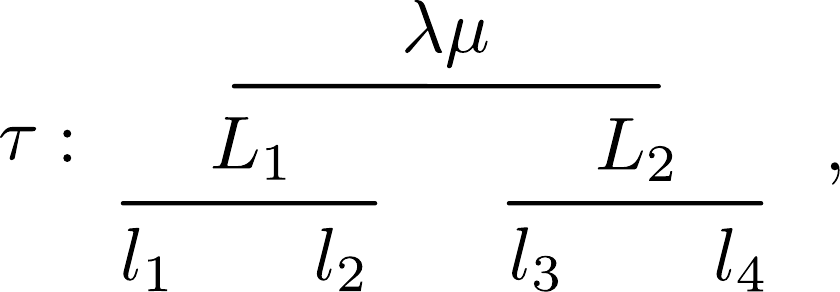}
\end{equation}
which corresponds to the tensor
\begin{equation}\label{eq:Gamma_ACE}
\begin{split}
&\Gamma_{\substack{l_1l_2l_3l_4\\m_1m_2m_3m_4},L_1L_2}^{\lambda\mu}(\tau) \\
&\qquad=\sum_{M_1M_2}C^{\lambda\mu}_{L_1M_1L_2M_2}C^{L_1M_1}_{l_1m_1l_2m_2}C^{L_2M_2}_{l_3m_3l_4m_4}\:.
\end{split}
\end{equation}

Another important example is provided by the generalized CG coefficients used in reference~\cite{BatatiaE3}, 
and described by the tree
\begin{equation}\label{eq:Gamma_generalized_graph}
     \includegraphics[width=.3\textwidth]{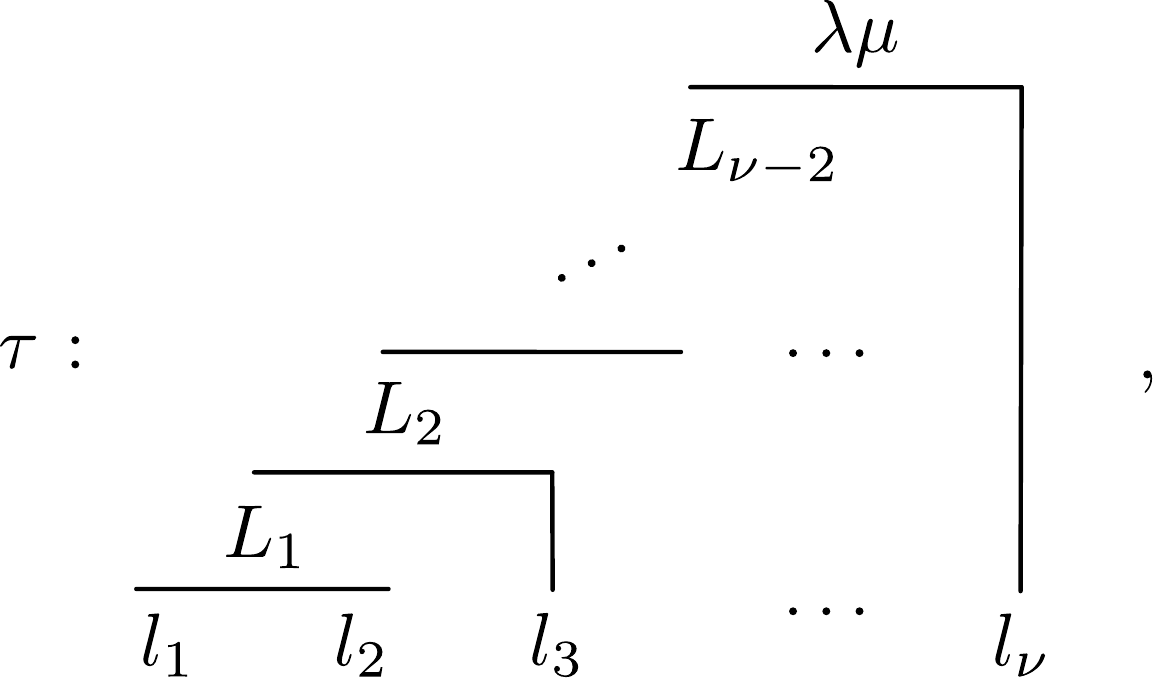}
\end{equation}
and corresponding to the tensor
\begin{equation}\label{eq:Gamma_generalized}
\begin{split}
&\Gamma_{\bm l \bm m \bm L}^{\lambda\mu}(\tau)\\&=\sum_{M_1\ldots M_{\nu-2}}C^{\lambda\mu}_{L_{\nu-2}M_{\nu-2}l_\nu m_\nu}\cdot\ldots\cdot C^{L_2M_2}_{L_1M_1l_3m_3}C^{L_1M_1}_{l_1m_1l_2m_2}\:.
\end{split}
\end{equation}
This example is important, since it allows us to study the property of the tensor $\Gamma$ in a general way. 
In particular, since different coupling-scheme choices are always connected by a unitary transformation, any 
conclusion draw for one choice of the coupling tree can be read in terms of another coupling. This makes 
the above choice ideal, since it allows us to deduce the following recursion relation
\begin{align}\label{eq:recursion_relation_multi}
    &\mY^{\lambda\mu}_{\bm l l_{\nu+1} \bm L L_{\nu-1}, \tau'}(\ver x, \ver r_{\nu+1}) \\
    &= \sum_{m_{\nu+1}M_{\nu-1}}C_{L_{\nu-1}M_{\nu-1}l_{\nu+1}m_{\nu+1}}^{\lambda\mu}\mY^{L_{\nu-1}M_{\nu-1}}_{\bm l \bm L,\tau}(\ver x)Y^{m_{\nu+1}}_{l_{\nu+1}}(\ver r_{\nu+1})\nonumber\:,
\end{align}
which connects the MultiSHs of order $\nu$ with the ones of order $(\nu+1)$. Here, $\tau'$ is a tree 
that contains $\tau$, namely the progressive coupling scheme of the graph in \eqref{eq:Gamma_generalized_graph}. 
This relation is the generalization of the one obtained in Eq.~\eqref{eq:recursion_TripoSHs} for the TripoSHs case. 
On the one hand, it allows for a recursive computation of the MultiSHs of high order, based on those of lower one. 
On the other hand it allows us to perform proof by induction. For example, it can be used to prove that the CG 
series of Eq.~\eqref{eq:CG_series} implies that the tensor $\Gamma$ of Eq.~\eqref{eq:Gamma_generalized} 
satisfies the rotation property of the MultiSHs of Eq.~\eqref{eq:Multi_rotation}. This is obtained by a straightforward 
generalization of the same argument used for the TripoSHs, in Section~\ref{sec:Tripo}.

Another crucial property of the tensor $\Gamma$ is that it is unitary. We have just shown that the recursion relation 
introduced above allows one to seamlessly prove that the tensor $\Gamma$, constructed from contraction of CG 
coefficients, satisfies the rotation condition of Eq.~\eqref{eq:Multi_rotation}. In the same way, the following relation
\begin{equation}\label{eq:contraction_Gamma}
    \sum_{\lambda\mu \bm L}\Gamma^{\lambda\mu}_{\bm l\bm m \bm L}(\tau)\Gamma^{\lambda\mu}_{\bm l\bm m' \bm L}(\tau) = \delta_{\bm m\bm m'}\:,
\end{equation}
can be obtained by repeatedly applying the analogous property of the CG coefficients of Eq.~\eqref{eq:orthogonality_CG} 
(it is the orthogonality relation of the generalized CG coefficients as discussed in detail in Ref.~\cite{Yutsis}).
This, together with the orthogonality of the standard spherical harmonics, allows us to prove the orthogonality condition 
of Eq.~\eqref{eq:Multi_ortho}. Crucially, this property also allows one to move from a representation in terms of product of 
spherical harmonics to one in terms of MultiSHs, as we will now show, also justifying the use 
of the MultiSHs as a basis. 

\subsubsection{General formula for the expansion coefficients}\label{sec:general_properties_exp_coeff}
We will now discuss how to move from a representation in terms of spherical harmonics to one based on the 
MultiSHs formalism. In doing so, we will also provide the corresponding relations for the expansion coefficients 
in the two representations. Let us expand a general function $f(\vb x)$ over a product of radial-basis functions 
and spherical harmonics, namely
\begin{equation}\label{eq:f_expanded_sh}
    f(\vb x) = \sum_{\bm n\bm l \bm m} f_{\bm n\bm l \bm m} \mathcal{R}_{\bm n\bm l}(x) Y^{m_1}_{l_1}(\ver r_1)\cdot\ldots\cdot Y^{m_\nu}_{l_\nu}(\ver r_\nu)\:.
\end{equation}
Let us now change the basis into the MultiSHs formalism, in a way completely analogous to what done in 
Section~\ref{sec:Bipo}. This is achieved by introducing $\sum_{\bm m'}\delta_{\bm m\bm m'}$, and by changing 
all the spherical harmonics indexes from $\bm m$ to $\bm m'$, in order to have them separated from the 
expansion coefficients $f_{\bm n\bm l \bm m}$. We can then exploit Eq.~\eqref{eq:contraction_Gamma} to 
introduce the contraction of two $\Gamma^{\lambda\mu}_{\bm l \bm m \bm L}$ tensors in place of the 
$\delta_{\bm m \bm m'}$. Since the contraction between these tensors and the spherical harmonics produces 
the MultiSHs [See Eq.~\eqref{eq:def_general_multiSH_Gamma}], after factorizing, we obtain 
\begin{equation}\label{eq:f_expanded_multi_Gamma}
    f(\vb x) = \sum_{\bm n\bm l\bm L}\sum_{\lambda\mu} \left[\sum_{\bm m} \Gamma^{\lambda\mu}_{\bm l\bm m \bm L}(\tau)f_{\bm n\bm l \bm m}\right] \mathcal{R}_{\bm n\bm l}(x)  \mY^{\lambda\mu}_{\bm l\bm L,\tau}(\ver x)\:.
\end{equation}
By comparing this expression with that in Eq.~\eqref{eq:expansion_in_MultiSHs}, we are able to make 
the identification
\begin{equation}\label{eq:recipe_u_f}
    u^{\lambda\mu}_{\bm n\bm l\bm L}= \sum_{\bm m}\Gamma^{\lambda\mu}_{\bm l\bm m \bm L}(\tau) f_{\bm n\bm l \bm m}\:,
\end{equation}
which relates the expansion coefficients in terms of the standard spherical harmonics, $f_{\bm n\bm l\bm m}$ with 
those over the MultiSHs, $u^{\lambda\mu}_{\bm n\bm l\bm L}$. Remarkably, the relation is obtained through the 
\emph{same contraction} that defines the MultiSHs in the first place [compare with Eq.~\eqref{eq:def_general_multiSH_Gamma}], 
and gives a general recipe to obtain the coefficients $u^{\lambda\mu}_{\bm n\bm l\bm L}$, once the coefficients 
$f_{\bm n\bm l\bm m}$ are known.

\subsubsection{Parity, Conjugation and Scalar Terms}

In this foundational section, we will derive the remaining properties of the coefficients 
$u^{\lambda\mu}_{\bm n\bm l\bm L}$. Following the same approach used to analyse 
their transformation under rotation [see Eq.~\eqref{eq:transformation_rule_coeff_SH_passive}],
we will be able to derive their behavior under inversion and conjugation. All that follows 
will be deduced directly from the MultiSHs, with no prior knowledge of the function $f(\vb x)$ 
itself and we will be strictly tied to the transformation character of \emph{proper} and \emph{pseudo} tensors.
 
Let us begin from the parity of the MultiSHs (transformation under reflection). The transformation 
rules of a proper tensor belonging to the space of angular momentum $\lambda$ can be inferred 
by the transformation rule of a single spherical harmonics. Explicitly we have \cite{Angular}
\begin{equation}\label{eq:parity_standard_SH}
    Y_l^m(-\ver r) = (-1)^l Y_l^m(\ver r)\:.
\end{equation}
This encodes the fact, for example, that proper scalars ($\lambda=0$) remain unchanged under inversion, 
while proper vectors ($\lambda = 1$) change sign. On the contrary, a pseudoscalar would have an opposite 
behavior, such as the chirality (a pseudoscalar, $\lambda =0$, that changes sign under parity transformation) 
or the cross-product, which is invariant under inversion despite rotating as a vector ($\lambda = 1$). 

Moving to the analogous transformation for the MultiSHs, Eq.~\eqref{eq:parity_standard_SH} trivially implies that 
\begin{equation}\label{eq:inversion_MultiSHs}
    \hat{P}\mY^{\lambda\mu}_{\bm l \bm L }(\ver x) = \mY^{\lambda\mu}_{\bm l \bm L}(-\ver x) = (-1)^{\Sigma(\bm l)+\lambda} \Big((-1)^\lambda \mY^{\lambda\mu}_{\bm l \bm L}(\ver x)\Big)\:,
\end{equation}
where $\hat{P}$ is the parity operator. Here, we introduced the symbols
\begin{equation}\label{eq:sum_Sigma}
    \Sigma(\bm l) := l_1+\ldots+l_\nu\:,
\end{equation}
for the sum of the $l$-s intermediate channels. We casted have Eq.~\eqref{eq:inversion_MultiSHs} in a way that allows 
for a direct comparison with Eq.~\eqref{eq:parity_standard_SH}. The latter has a sign change depending on the parity 
of the angular momentum channel $l$. A similar inversion is present for the MultiSHs with respect to the angular channel 
$\lambda$, but an additional phase factor, with respect to the parity of the sum $\Sigma(\bm l) + \lambda$, is present. 
Intuitively, this shows that whenever the sum is even the MultiSHs behave like a proper tensor (in agreement with the 
corresponding transformation of the spherical harmonics). Instead, an odd sum causes an additional sign change, 
pointing to a pseudotensorial transformation. We refer to Appendix~\ref{appendix:parity_MultiSH} for a rigorous proof 
of the nature of these terms under parity.

If we now apply the parity to the MultiSHs when used to expand a (proper) scalar function $f(\ver x)$,  
its expansion coefficients must satisfy the same transformation of Eq.~\eqref{eq:inversion_MultiSHs}, namely
\begin{equation}\label{eq:Multi_coefficients_parity}
    u^{\lambda\mu}_{\bm n\bm l \bm L}\xrightarrow{\hat{P}} (-1)^{\Sigma(\bm l)+\lambda}\Big((-1)^\lambda u^{\lambda\mu}_{\bm n\bm l \bm L}\Big)\:.
\end{equation}
This, combined with Eq.~\eqref{eq:transformation_rule_coeff_SH_passive} for the transformation under a general rotation, gives 
the full behavior of the coefficients for any transformation of the $O(3)$ group, without any reference to their actual analytical 
form.

Another important property is the behavior of the coefficients under complex-conjugation. By using the conjugation relation
for the single spherical harmonics
\begin{equation}\nonumber
    Y^{m*}_l(\ver r) = (-1)^m Y^{-m}_l (\ver r),
\end{equation}
and the relation $C^{lm}_{l_1m_1l_2m_2} = (-1)^{l_1+l_2-l} C^{l-m}_{l_1-m_1l_2-m_2},$ we are able to derive an expression
for the complex conjugate of a MultiSH, namely
\begin{equation}
    \mY^{\lambda\mu*}_{\bm l \bm L} (\ver x) = (-1)^{\Sigma(\bm l)+\lambda}\mY^{\lambda -\mu}_{\bm l \bm L} (\ver x)\:.\nonumber
\end{equation}
Since the function $f(\vb x)$ is real, we obtain that also the coefficients must satisfy the same relation, namely
\begin{equation}\label{eq:Multi_coefficients_conjugation}
    u^{\lambda\mu*}_{\bm n \bm l \bm L} = (-1)^{\Sigma(\bm l)+\lambda}u^{\lambda -\mu}_{\bm n \bm l \bm L}\:.
\end{equation}

As we will see in the following section, coefficients that behave as scalars are at the core of the construction of 
ML models for scalar quantities. These coefficients are obtained by setting $\lambda = \mu = 0$. Indeed, from 
Eq.~\eqref{eq:transformation_rule_coeff_SH_passive}, we know that
\begin{equation}\label{eq:coefficients_rotationally_invariant_multi}
    u^{00}_{\bm n \bm l \bm L} \xrightarrow{\hat{R}} D^{0*}_{00'}(\hat{R}) u_{\bm n\bm l \bm L}^{00} = u_{\bm n\bm l \bm L}^{00}\:,
\end{equation} 
which proves that these coefficients are rotationally invariant. Moreover, by combining Eqs.~\eqref{eq:Multi_coefficients_parity} 
and ~\eqref{eq:Multi_coefficients_conjugation}, we also deduce
\begin{equation}
    u^{00}_{\bm n \bm l \bm L} \xrightarrow{\hat{P}} (-1)^{\Sigma(\bm l)}u^{00}_{\bm n \bm l \bm L} = u^{00*}_{\bm n \bm l \bm L}\:.
\end{equation}
This implies the important result that the rotationally invariant coefficients behave like scalars and are real, when $\Sigma(\bm l)$ 
is even. In contrast, when $\Sigma(\bm l)$ is odd, they behave like pseudoscalar, and are imaginary. 

With these information at hand, we can now calculate a $\lambda$-SOAP kernel, which is covariant with respect 
to any transformation of the group $O(3)$. This is done by following a derivation analogous to the one performed 
in section~\ref{sec:covariant_kernel_simple_derivation}, with the inclusion of a sum over a parity inverted density.
Since the details of the calculation are similar to what already shown, here we only mention that the result is again 
an inner product, restricted to coefficients with $\Sigma(\bm l) + \lambda$ even. For an explicit derivation we refer 
to Appendix~\ref{appendix_kernel_03}.

We conclude this section by noticing that all the properties listed above have been derived by investigating only 
the analogous properties of the MultiSHs basis, and by exploiting the fact that $\rho^{\otimes \nu}(\vb x)$ is a real 
and scalar function. Indeed, remarkably, the explicit form of the expansion coefficients $u^{\lambda\mu}_{\bm n \bm l \bm L}$ 
is still left completely general. This proves that, not only adopting the point of view of the MultiSHs allows one to simplify 
the investigation of all the properties of the expansion coefficients, but it also generalize them, showing how they hold 
for a general representation that is not at all linked with a formulation in terms of atomic densities.

\section{The powerspectrum, the bispectrum and the scalar kernel}\label{sec:powerspectrum_bispectrum_SOAP}

Having derived all the transformation rules for the coefficients, we are now able to investigate what happens 
when we choose a special form for the function $f(\vb x)$. In particular, we are going to investigate the case 
when we have a product function, such as $\rho^{\otimes \nu}$. We will show that this will lead to a straightforward 
formulation of the powerspectrum and the bispectrum components, with an immediate proof of their rotational 
invariance. 
Moreover, we will also derive the form of the SOAP-kernel, as obtained in reference~\cite{SOAP}. We anticipate 
that the only requirement for the derivations presented in this section is that the expanded functions are in the form 
of products, with their explicit form not being relevant. Moreover, this section can be seen as a first application of the 
MultiSHs formalism, with the explicit calculation of the tensors $\Gamma(\tau)$ for various order, $\nu$.

Let us consider again the product $\rho_i^{\otimes \nu}(\vb x)$. By expanding this function in terms of the standard 
spherical harmonics, we obtain the chain of relations
\begin{align}
    &\rho_i^{\otimes \nu}(\vb x) = \prod_{\alpha=1}^\nu \rho_i(\vb r_\alpha)=\nonumber\\&= \sum_{\bm n\bm l \bm m}\left[\prod_{\alpha=1}^\nu c_{i,n_\alpha l_\alpha m_\alpha}R_{n_\alpha l_\alpha}(r_\alpha)Y_{l_\alpha}^{m_\alpha}(\ver r_\alpha)\right]=\\
    &= \sum_{\bm n\bm l \bm m}\left[\prod_{\alpha=1}^\nu c_{i,n_\alpha l_\alpha m_\alpha}\right] \mathcal{R}_{\bm n \bm l}(x)  Y^{m_1}_{l_1}(\ver r_1)\cdot\ldots\cdot Y^{m_\nu}_{l_\nu}(\ver r_\nu)\nonumber\:.
\end{align}
The first equality just remarks that $\rho_i^{\otimes \nu}(\vb x)$ is obtained as a simple product, while in the second 
we have expanded each of the constituent functions, $\rho_i(\vb r_\alpha)$, independently. By comparing this expression 
with that of a generic function $f(\ver x)$, Eq.~\eqref{eq:f_expanded_sh}, we can make the identification 
\begin{equation}
    f_{i,\bm n \bm l \bm m} = \prod_{\alpha=1}^\nu c_{i,n_\alpha l_\alpha m_\alpha}\:.
\end{equation}
Therefore, by using Eq.~\eqref{eq:recipe_u_f} for the connection between the coefficients $ f_{i,\bm n \bm l \bm m}$ 
and the MultiSHs' coefficients, $u^{\lambda\mu}_{i,\bm n \bm l \bm L}$, we can write
\begin{equation}\label{eq:recipe_coefficients_full}
    u^{\lambda \mu}_{i,\bm n \bm l \bm L} = \sum_{\bm m} \Gamma^{\lambda\mu}_{\bm l \bm m \bm L} (\tau) \prod_{\alpha=1}^\nu c_{i,n_\alpha l_\alpha m_\alpha}\:,
\end{equation}    
where, for the sake of readability, we remark again that  
\begin{equation*}
    c_{i,nlm} = \int \dd \vb r\, \rho_i(\vb r) R_{nl}(r)Y_l^m(\ver r)\:.
\end{equation*}
These expressions are among the most useful of this work, since they establish the connection between the 
single-density formalism, incorporated into the coefficients $c_{i,nlm}$, and the MultiSHs one. With these formulas 
in place, we are now in the position to re-derive the powerspectrum and the bispectrum components by considering 
the cases for $\nu =2 $ and $\nu =3$, respectively.

\subsubsection{The powerspectrum components}
As explicitly stated in the definition of Eq.~\eqref{eq:powerspectrum_def}, here reported for completeness
\begin{equation}
    p_{i,n_1n_2l} = \sum_{m}(-1)^m c_{i,n_1lm}c_{i,n_2l-m},\nonumber
\end{equation}
the powerspectrum components are the rotational invariants obtained from an appropriate contraction of 
products of two set of coefficients, $c_{i,nlm}$. Therefore, we expect them to be related to the rotationally
invariant coefficients of a MultiSHs formalism, namely the ones satisfying Eq.~\eqref{eq:coefficients_rotationally_invariant_multi}, 
for $\nu =2$. Indeed, by using Eq.~\eqref{eq:recipe_coefficients_full}, the rotationally invariant coefficients read
\begin{equation}\label{eq:u00_n2}
    u^{00}_{i,n_1n_2l_1l_2} = \sum_{m_1m_2} \Gamma^{00}_{l_1m_1l_2m_2}(\tau) c_{i,n_1l_1m_1}c_{i,n_2l_2m_2}\:.
\end{equation}
By selecting the first graph represented in \eqref{eq:graph_bipo_taus}, which corresponds to the tensor $\Gamma(\tau_1)$ 
given in Eq.~\eqref{eq:Gamma_n2_tau1}, we have the simple identification
\begin{equation}
    \Gamma^{00}_{l_1m_1l_2m_2} = C^{00}_{l_1m_2l_2m_2} = \delta_{l_1l_2}\delta_{m_1-m_2} \dfrac{(-1)^{l_1-m_1}}{\sqrt{2l_1+1}}\:,
\end{equation}
where we have used the explicit expression for the CG coefficients. By substituting this expression into 
Eq.~\eqref{eq:u00_n2}, we obtain
\begin{align}
    u^{00}_{i,n_1n_2l_1l_2} = \delta_{l_1l_2}\dfrac{(-1)^{l_1}}{\sqrt{2l_1+1}}\sum_{m_1} (-1)^m c_{i,n_1l_1m_1}c_{i,n_2l_1 -m_1}\:.
\end{align}
When we compare this expression with the definition of the powerspectrum, we obtain 
\begin{equation}\label{eq:multi_powerspectrum}
    u^{00}_{i,n_1n_2ll'} = \delta_{ll'}\dfrac{(-1)^l}{\sqrt{2l+1}}p_{i,n_1n_2l}\:,
\end{equation}
namely, the powerspectrum components are nothing but the scalar coefficients of the $\nu=2$ case. Thus, not only the MultiSH
formalism leads directly to the expression for the powerspectrum components, but it also implicitly proves their rotational invariance 
(since those are scalar components). Moreover, using Eqs.~\eqref{eq:Multi_coefficients_parity} and \eqref{eq:Multi_coefficients_conjugation}, 
respectively for the parity and the conjugation of the expansion coefficients, we also have the immediate information that the 
powerspectrum components are always scalars and real. In particular, since all this informations are obtained \emph{regardless} 
of the explicit form of the coefficients $c_{i,nlm}$, the same conclusion could be obtained from a more generic product function of 
the form $\rho_{12}(\vb r_1,\vb r_2) := \rho_1(\vb r_1)\rho_2(\vb r_2)$, namely that obtained by the product of two different atomic 
densities. The same will hold true also for the bispectrum components.

\subsubsection{The bispectrum components}
As the powerspectrum components are obtained by imposing rotational invariance to the product of two single-density 
expansion coefficients, the bispectrum components are connected with the scalar components in the $\nu=3$ case. Indeed, 
by following the same procedure that led to the powerspectrum components, we can write the rotationally invariant coefficients 
as
\begin{align}
    &u^{00}_{i,\substack{n_1n_2n_3\\l_1l_2l_3}L} \\
    &= \sum_{m_1m_2m_3} \Gamma^{00}_{\substack{l_1l_2l_3\\m_1m_2m_3},L}(\tau) c_{i,n_1l_1m_1}c_{i,n_2l_2m_2}c_{i,n_3l_3m_3}\:.\nonumber
\end{align}
Now, by taking Eq.~\eqref{eq:Gamma_tau_n3} as our choice for the $\Gamma(\tau)$ tensor and, again, by considering that 
$C^{00}_{LMl_3m_3} = \delta_{Ll_3}\delta_{M -m_3}(-1)^{l_3-m_3}/\sqrt{2l_3+1}$, we obtain
\begin{align}\label{eq:multi_bispectrum0}
    &u^{00}_{i,\substack{n_1n_2n_3\\l_1l_2l_3}L}
      =\delta_{l_3L} \dfrac{(-1)^{l_3}}{\sqrt{2l_3+1}}
      \\
      &\qquad\qquad\times\sum_{m_1m_2m_3} c^*_{i,n_3l_3m_3}C^{l_3m_3}_{l_1m_1l_2m_2} c_{i,n_1l_1m_1}c_{i,n_2l_2m_2}\nonumber\:.
\end{align}
Here we have used the property $c^*_{i,nlm}= (-1)^m c_{i,nl-m}$, which holds whenever $\rho(\vb r)$ is a scalar function. By comparing 
this expression with the definition of the bispectrum components of Eq.~\eqref{eq:bispectrum_components}, we obtain that
\begin{equation}\label{eq:multi_bispectrum}
    u^{00}_{i,\substack{n_1n_2n_3\\l_1l_2l_3}L}
      =\delta_{l_3L} \dfrac{(-1)^{l_3}}{\sqrt{2l_3+1}}B_{i,\substack{n_1n_2n_3\\l_1l_2l_3}}\:,
\end{equation}
namely, the bispectrum components are proportional to the rotationally-invariant coefficients of a MultiSHs expansion. Thus, also in this
case, not only we have derived the bispectrum components with a straightforward evaluation of expansion coefficients of the MultiSHs 
formalism, but we have also implicitly proved their rotational invariance. Moreover, given the parity and conjugation properties of 
Eqs.~\eqref{eq:Multi_coefficients_parity} and \eqref{eq:Multi_coefficients_conjugation}, respectively, we have also obtain the additional 
information that the bispectrum components behave like imaginary pseudoscalars when the sum $l_1+l_2+l_3$ is odd, and they are 
real scalars when the same sum is even.

The same method can be expanded to any body order of choice. For example, the approach proposed for the ACE potential for the 
$\nu = 4$ terms can be obtained from an evaluation of the rotationally invariant coefficients with the $\Gamma$ tensor give by 
Eq.~\eqref{eq:Gamma_ACE}. Furthermore, the results of reference~\cite{BatatiaE3} can be obtained by using the tensor 
$\Gamma(\tau)$ from Eq.~\eqref{eq:Gamma_generalized}, with the invariants that can be derived for any $\nu$ by evaluating 
the $\lambda=\mu=0$ case in Eq.~\eqref{eq:recipe_coefficients_full} (and constraining $\Sigma(\bm l)$ to be even for the scalar-real 
case).

\subsubsection{The SOAP kernel}\label{sec:calculation }
Another example of the application of the MultiSH formalism is provided in the evaluation of the SOAP kernel. By using the 
results of Eq.~\eqref{eq:multi_lambda_kernel} for $\lambda=0$, we can write the scalar (SOAP) kernel as 
\begin{equation}
        K^{(\nu)}(\rho,\rho'):=(K^{(\nu)}(\rho,\rho'))^0_{00} = 8\pi^2\sum_{\bm n\bm l \bm L}u^{00}_{\bm n \bm l \bm L}v^{00*}_{\bm n \bm l \bm L}\:.\nonumber
\end{equation}
We are now in the position to study more in detail this expression and provide more information about this inner product. 
Firstly, by using the conjugation property of Eq.~\eqref{eq:Multi_coefficients_conjugation} on the coefficients $v$, we obtain 
the expression
\begin{equation}\label{eq:SOAP_multi_formalism}
    K^{(\nu)}(\rho,\rho') = 8\pi^2\sum_{\bm n\bm l \bm L}(-1)^{\Sigma(\bm l)}u^{00}_{\bm n \bm l \bm L}v^{00}_{\bm n \bm l \bm L}\:,
\end{equation}
which holds true for any $\nu$, and includes also pseudoscalar components, as one can read from the parity transformation of 
Eq.~\eqref{eq:Multi_coefficients_parity}. From this, and with the results provided by Eqs.~\eqref{eq:multi_powerspectrum} and 
\eqref{eq:multi_bispectrum}, we can immediately deduce that the $\nu=2$ case takes the form
\begin{equation}
    K^{(2)}(\rho,\rho')= \sum_{n_1n_2 l }\dfrac{8\pi^2}{2l+1}p_{n_1n_2l}p'_{n_1n_2l}\:,
\end{equation}
where $p_{n_1n_2l}$ and $p'_{n_1n_2l}$ are the powerspectrum obtained from $\rho$ and $\rho'$ respectively, and 
where we have used the fact that the powerspectrum is real. Analogously, for $\nu=3$, we obtain
\begin{equation}
    K^{(3)}(\rho,\rho')= \sum_{\substack{n_1n_2n_3\\l_1l_2l_3 }}(-1)^{l_1+l_2+l_3}\dfrac{8\pi^2}{2l_3+1}B_{\substack{n_1n_2n_3\\l_1l_2l_3}}B'_{\substack{n_1n_2n_3\\l_1l_2l_3}}\:.
\end{equation}
These expressions reproduce the known results [see Reference~\cite{SOAP} for their derivation in the context of 
Gaussian atomic densities] that the SOAP kernel can be obtained by weighted inner products of the powerspectrum 
and bispectrum components, respectively. However, not only the MultiSH formalism recovers these results without 
even taking into account the specific expression for the density $\rho(\vb r)$, but it also provides a generalization 
to any order $\nu$, by means of Eq.~\eqref{eq:SOAP_multi_formalism}.

Having recovered most of the descriptors employed in state-of-the-art ML models, we are now in the position to discuss 
the models themselves and to derive a full framework to obtain them. Before proceeding, however, it is important to remark 
that the requirement to obtain all the descriptors introduced above is the fact of dealing with a product function,
$\rho^{\otimes \nu}(\vb x) = \rho(\vb r_1)\cdot\ldots\cdot \rho(\vb r_\nu)$, and nothing more. This implies that the 
MultiSH formalism, and the corresponding descriptors, are \emph{independent} of the specific shape of the atom 
density, $\rho_i(\vb r)$. This means that, in order to construct the product $\rho^{\otimes \nu}$, we can use up 
to $\nu$ different atomic-densities. 

\section{On the atomic density}\label{sec:on_the_atomic_density}
In this section, we are finally going to discuss the explicit form of the atomic density $\rho(\vb r)$, and that of the 
expansion coefficients of $\rho^{\otimes \nu}$. We anticipate that such investigation is not required to derive and 
investigate the properties of the expansion coefficients $u^{\lambda\mu}_{\bm n \bm l \bm L}$. Instead, this study 
is necessary only to connect the MultiSH formalism to frameworks based on a multi-body expansion. In particular, 
the connection will be realised by mean of an atomic density defined as a sum of Dirac-delta functions. 

Let us consider again an atomic density of the form of Eq.~\eqref{eq:intro_atomic_density}, here reported for 
compleness
\begin{equation}
    \rho_i(\vb r) = \sum_{j}^{\text{atoms}} h_{Z_jZ_i}(\vb r -\vb r_{ji})\:.\nonumber
\end{equation}
By expanding this expression onto an orthonormal radial basis and spherical harmonics, we obtain the coefficients 
$c_{i,nlm}$ of Eq.~\eqref{eq:expansion_coefficients_density} of the general form 
\begin{equation}
    c_{i,nlm} = \sum_{j} g_{nlm}(\vb r_{ji};Z_j,Z_i)\:,
\end{equation}
where $g_{nlm}$ are the expansion coefficients of the functions $h$, namely
\begin{equation}
    g_{nlm}(\vb r_{ji};Z_j,Z_i) = \int \dd \vb r\, h_{Z_jZ_i}(\vb r-\vb r_{ji}) R_{nl}(r) Y^{m*}_l(\ver r)\:.
\end{equation}
From this expression one can appreciate how the coefficients depend on the atomic positions, 
a dependence indicating that they can be further expanded over a radial basis and spherical harmonics. 
Moreover, since $\rho_i(\vb r)$ is a scalar function, the sum of products $\sum_m g_{nlm}(\vb r_{ji})Y_l^m(\ver r)$ must behave as a 
scalar. Therefore, we have that the only possible expansion for such 
coefficients takes the form
\begin{equation}
    g_{nlm}(\vb r_{ji}) = q_{nl}(r_{ji}) Y^{m*}_l(\ver r_{ji})\:,
\end{equation}
where we have implied the dependence on the atomic species, and where the functions $q_{nl}(r_{ji})$ are of the form
\begin{equation}\label{eq:radial_function_of_coefficients_expansion}
    q_{nl}(r_{ji}) := \dfrac{1}{2l+1}\int \dd \ver r_{ji}\, \sum_m g_{nlm}(\vb r_{ji}) Y^{m}_l(\ver r_{ji})\:.
\end{equation}
This expression is explicitly derived in Appendix~\ref{sec:AppendC} by exploiting the orthogonality of the spherical harmonics and the MultiSHs expansion.

Remarkably, despite the simplicity of this layered expansion, such an expression has important consequence 
when analyzed from MultiSH-formalism point of view. Indeed, we now have that the coefficients of the atomic 
density, $c_{i,nlm}$, take the form
\begin{equation}\nonumber
    c_{i,nlm} = \sum_j q_{nl}(r_{ji}) Y^{m*}_l(\ver r_{ji})\:,
\end{equation}
a feature that can be used in Eq.~\eqref{eq:recipe_coefficients_full} for the expansion coefficients of 
$\rho_i^{\otimes \nu}$. Explicitly, we obtain 
\begin{align}
    u^{\lambda \mu}_{i,\bm n \bm l \bm L} &= \sum_{\bm m} \Gamma^{\lambda\mu}_{\bm l \bm m \bm L} (\tau) \prod_{\alpha=1}^\nu c_{i,n_\alpha l_\alpha m_\alpha}=\nonumber\\
    &= \sum_{j_1\ldots j_\nu} q_{n_1l_1}(r_{j_1})\cdot\ldots\cdot q_{n_\nu l_\nu}(r_{j_\nu})\times\\
    &\qquad\times\sum_{\bm m} \Gamma_{\bm l \bm m \bm L}^{\lambda\mu}(\tau) Y^{m_1*}_{l_1}(\ver r_{j_1 i})\cdot\ldots\cdot Y^{m_\nu *}_{l_\nu}(\ver r_{j_\nu i})\nonumber\:.
\end{align} 
By using the definition MultiSHs of Eq.~\eqref{eq:def_general_multiSH_Gamma} and by exploiting the fact that the tensor 
$\Gamma^{\lambda\mu}_{\bm l \bm m \bm L} (\tau)$ is real, we obtain the following expression for the coefficients 
$u^{\lambda\mu}_{i,\bm n \bm l \bm L}$
\begin{equation}\label{eq:Multi_coefficients_singlebody}
    u^{\lambda\mu}_{i,\bm n \bm l \bm L} = \sum_{\bm j} Q_{\bm n \bm l}(x_{\bm j i})\mY^{\lambda\mu *}_{\bm l \bm L}(\ver x_{\bm ji})\:,
\end{equation}
where, to have a more compact expression, we have introduced the short-hand notation
\begin{equation}
    \bm j := (j_1,\ldots, j_\nu),\nonumber \qquad\text{and}\qquad \vb x_{\bm ji} := (\vb r_{j_1 i},\ldots,\vb r_{j_\nu i})
\end{equation}
and the corresponding lengths and directions, encoded in $x_{\bm j i}$ and $\ver x_{\bm j i}$ respectively. 
We have also defined 
\begin{equation}
    Q_{\bm n \bm l}(x_{\bm j i}): = q_{n_1l_1}(r_{j_1})\cdot\ldots\cdot q_{n_\nu l_\nu}(r_{j_\nu})\:. \nonumber
\end{equation}
The expression in Eq.~\eqref{eq:Multi_coefficients_singlebody} implies that, \emph{regardless of the explicit form} of 
the localization functions $h(\vb r-\vb r_{ji})$, the coefficients of the MultiSH expansion are essentially MultiSHs themselves. 
We remark that, not only we have just derived a simple expression for the expansion coefficients $ u^{\lambda\mu}_{i,\bm n \bm l \bm L}$, 
but we have also determined that the actual choice of the localization function $h$ has consequences only on the radial functions 
$Q_{\bm n \bm l}$, and not on the angular part of the description. Indeed, these are the only terms that explicitly depend on the actual 
localization functions chosen to define the local environment described by $\rho(\vb r)$. Everything else, including the behavior under 
rotation, is fully determined by adopting the general expression of Eq.~\eqref{eq:intro_atomic_density}. 
It is now important to notice how these conclusions agree with our finding on the properties against rotation, parity and conjugation, 
of Eqs.~\eqref{eq:transformation_rule_coeff_SH_passive}, \eqref{eq:Multi_coefficients_parity} and \eqref{eq:Multi_coefficients_conjugation}, 
respectively. 
Moreover, it also agrees with our discussion on the active and passive rotations, encoded in Eq.~\eqref{eq:transformation_rule_coeff_SH_active}. 
Indeed, the coefficients $u^{\lambda \mu}_{\bm n \bm l \bm L}$, being constructed from the complex conjugate of MultiSHs, transform 
in a contravariant way, while their own complex conjugate, $u^{\lambda \mu}_{\bm n \bm l \bm L}$, transforms in a fully covariant way. 
This also justifies the use of such complex conjugates in Eqs.~\eqref{eq:tensor_exp} and ~\eqref{eq:tensor_expansions_3B}. In 
order to target the covariant components of a tensor with a linear model, it is necessary to use the complex conjugate of the coefficients 
$c_{i,nlm}$.

An important example for the radial expansion of the coefficients $u^{\lambda\mu}_{\bm n\bm l \bm L}$ is provided in the case the $h$ 
functions are Dirac-deltas, $h(\vb r-\vb r_{ji}) = \delta(\vb r-\vb r_{ji})$. From these we have
\begin{equation}\label{eq:u_coeff_dirac_delta}
     u^{\lambda\mu}_{i,\bm n \bm l \bm L} \substack{\text{Dirac}-\delta\\=} \sum_{\bm j} \mathcal{R}_{\bm n \bm l}(x_{\bm j i})\mY^{\lambda\mu *}_{\bm l \bm L}(\ver x_{\bm ji})\:.
\end{equation}
This expression plays a fundamental role in connecting the MultiSH formalism to the general case of a multi-body expansion. Indeed, it can be read 
as the basis for an expansion in $(\nu+1)$-body clusters, as done in reference~\cite{BatatiaE3}. We remark again that the choice of the 
localization function has an effect only on the radial part of the expansion coefficients, since the angular part can always be casted in terms 
of MultiSHs.

This section concludes our study of the most important properties of the expansion of the product density, $\rho^{\otimes \nu}$, in terms 
of MultiSHs. In doing so, we have successfully reproduced all the most wide-used descriptors of state-of-the-art ML models, and we have been 
able to obtain a compact and general derivation for a general $\lambda$-SOAP kernel. The next section will be devoted to define a simple 
strategy to produce a linear model based on the MultiSH formalism. For this part, we will explore the connection between the expansion in 
terms of the MultiSHs, and cluster-expansion-based schemes. This will lead to a derivation of the ACE formalism~\cite{ACE} and to 
uncover properties of the SNAP~\cite{SNAP}. We will also explore the connection between the MultiSH formalism and internal-coordinate 
based representations, reproducing the MTP~\cite{MTP} and JL formalism~\cite{JLP}. We will conclude this work with a section on linear 
models for covariant quantities, which will lead to the derivation of the basis used in the NICE formalism~\cite{NICE}.

\section{Reproducing linear models: connection with multi-body expansions}\label{sec:reproducing_model}

\begin{figure*}[ht!]
    \centering
    \includegraphics[width=\textwidth]{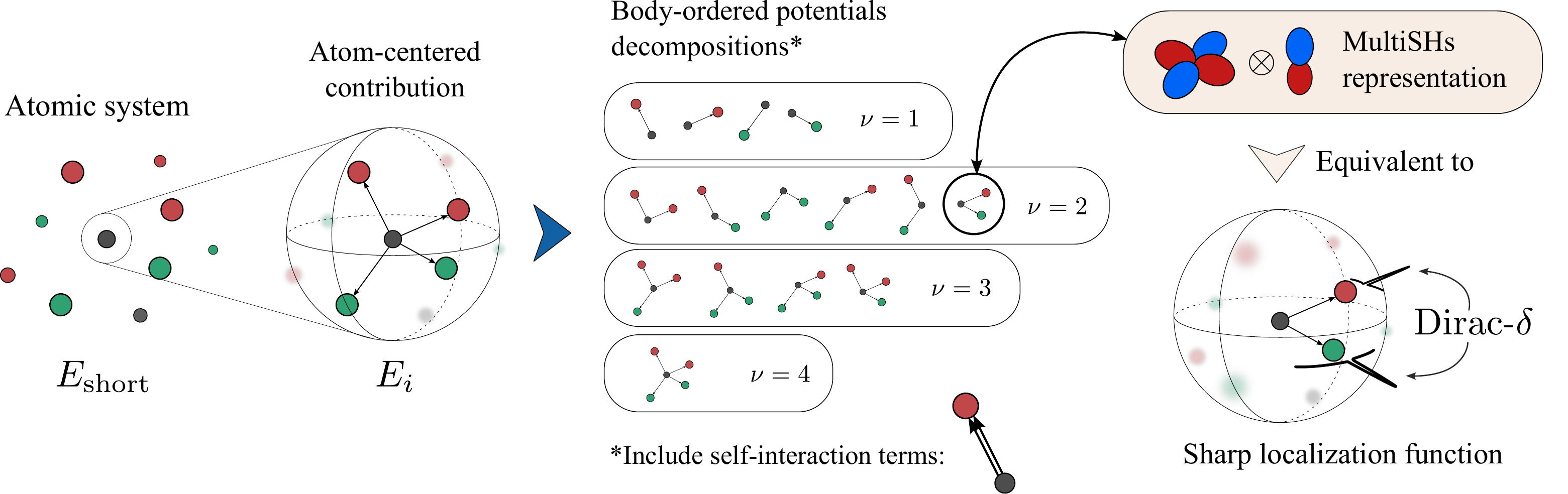}
    \caption{Schematic of a general cluster-expansion formalism. From left to right the figure shows the partition of the energy into 
    local contributions (indicated by $i$) and body orders (indicated by $\nu$). 
    The short-range energy contribution, $E_\text{shorts}$, is partitioned into atom-centered contributions, $E_i$.
    These are further decomposed into $(\nu+1)$-body potentials, $v_i^{(\nu+1)}$, which include also self-interaction 
    contributions in higher body-order terms. Each of the potentials is then expanded in terms of the MultiSHs basis: this is 
    equivalent to using the expansion coefficients $u^{\lambda\mu}_{\bm n \bm l \bm L}$ of a functions $\rho^{\otimes \nu}(\vb x)$ 
    obtained from sharp Dirac-$\delta$ localization functions. The expansion coefficients are given in Eq.~\eqref{eq:u_coeff_dirac_delta}. 
    When looking at the derivation of the properties of the coefficients $u^{\lambda\mu}_{\bm n \bm l \bm L}$, this bridge between 
    the expansion of $\rho^{\otimes \nu}(\vb x)$ and the multi-body formalism is the \emph{only} instance in which the explicit form 
    of the coefficients $u^{\lambda\mu}_{\bm n \bm l \bm L}$ plays a role. All the other properties (covariance, parity and conjugation) 
    are directly inherited from the MultiSH formalism itself and the symmetries of $\rho^{\otimes \nu}(\vb x)$ (a real and scalar function 
    that allows for product factorization).}
    \label{fig:2}
\end{figure*}

The aim of this section is to reproduce some of the most used ML models with the aid of the MultiSH formalism. In doing so we 
will provide a straightforward approach to the construction of linear models, which will show the connection between the MultiSHs 
and multi-body expansions. In the next section, following the same derivation of the JL potential of reference~\cite{JLP}, we will 
derive a general framework for the construction of linear models for scalar quantities. On the one hand, this will allow us to reproduce 
the ACE model. On the other hand, it will lay down the foundations for the investigation of the SNAP. In particular, in pursuing the 
connection between SNAP and the MultiSH formalism, we will show how the SNAP descriptors do not fully cover the intended 
rotationally-invariant space. We will then proceed by exploring the connection between the MultiSH formalism and internal-coordinate 
representations, an analysis that will lead to a derivation of the MTP and the JLP models.

\subsection{Linear models for scalar quantities: ACE}\label{sec:linear_ACE}

We will now introduce a general framework for the construction of linear ML potentials. A summary of the procedure outlined in 
this section is provided in Fig.~\ref{fig:2}. The main hypothesis governing this section is that the short-ranged contribution to the 
energy, $E_\text{short}$, can be expanded in atom-centered contributions, $E_i$. Furthermore, we assume a multi-body expansion 
of each $E_i$ term, such that we have
\begin{equation}
    E_\text{short} = \sum_i^{\text{atoms}} E_i = \sum_i \sum_{\nu=0} \varepsilon^{(\nu+1)}_i\:.
\end{equation}
Here, $(\nu+1)$ indicates the order of the expansion. For example, $\nu=1$ indicates a two-bodies expansion, and $\nu=2$ indicates 
a three-bodies expansion, and so on. Practically, this means that we are assuming that each of the $\varepsilon_i^{(\nu+1)}$ term can 
be expanded as a sum of $(\nu+1)$-body potentials, $v_i^{(\nu+1)}$, such that
\begin{equation}\label{eq:nu_cluster_expansion}
    \varepsilon_i^{(\nu+1)} = \sum_{\bm j} v_i^{(\nu+1)}(\vb x_{\bm j i})\:.
\end{equation}
Importantly, we are implying that the dimensionality of the vector $\bm j$ has to be inferred by the context (the vector $\bm j$ has 
length $\nu$ and, thus, its length depends on the body order). To clarify the above expression, let us consider two examples. The 
two-body expression ($\nu=1$) is given by
\begin{equation}
    \varepsilon_i^{(2)} = \sum_{j}^{\text{atoms}} v^{(2)}_i(\vb r_{ji})\:,
\end{equation}
while the four-body ($\nu=3$) expression is
\begin{equation}
    \varepsilon_i^{(4)} = \sum_{j_1 j_2 j_3}^{\text{atoms}} v^{(4)}_i(\vb r_{j_1i},\vb r_{j_2i},\vb r_{j_3i})\:.
\end{equation}
We remark that a $(\nu+1)$-body order expression is formed by potentials that depend on $\nu$ atomic positions. This is because all the atomic positions are taken with respect to the central atom $i$, which acts as the origin of the local 
reference frame. Thus, when one also considers the central atom, the expression above contains information about $(\nu+1)$ 
atoms at once. Moreover, as discussed in reference~\cite{ACE}, we will not restrict the sum over the atomic labels, $\bm j$. While 
this will introduce ``self-interacting'' terms (when at least two labels refer to the same atom), we can either assume that such terms 
are absorbed in a lower body order, or we can refer to equivalent formulations that show how to produce similar expressions with 
restricted summations, see~\cite{ACE}. 

Now, assuming that the potentials can be expanded over a radial basis and spherical 
harmonics, we are in the condition to perform an expansion in terms of MultiSHs as
\begin{align}
     \varepsilon_i^{(\nu+1)} = \sum_{\lambda\mu}\sum_{\bm n \bm l \bm L} a^{\lambda\mu}_{\nu,\bm n \bm l \bm L} \sum_{\bm j}\mathcal{R}_{\bm n \bm l}(x_{\bm j i})\mY^{\lambda\mu}_{\bm l \bm L}(\ver x_{\bm j i})\:,
\end{align}
where the coefficients $a^{\lambda\mu}_{\nu,\bm n \bm l \bm L}$ are formally obtained through the integral
\begin{equation}
    a^{\lambda\mu}_{\nu,\bm n \bm l \bm L} = \int \dd \vb x_{\bm ji}\, \mathcal{R}_{\bm n \bm l}(x_{\bm j i})\mY^{\lambda\mu*}_{\bm l \bm L}(\ver x_{\bm j i})v_i^{(\nu+1)}(\vb x_{\bm j i})\:.\nonumber
\end{equation}
If we now compare these expressions with Eq.~\eqref{eq:u_coeff_dirac_delta}, we recognize that we are indeed using 
the expansion coefficients of the Dirac-delta $\rho^{\otimes \nu}$ as a multi-body basis. Thus, we can write
\begin{equation}\label{eq:linear_model_MultiSHs_exp}
        \varepsilon_i^{(\nu+1)} = \sum_{\bm j} v_i^{(\nu+1)}(\vb x_{\bm j i}) = \sum_{\lambda\mu}\sum_{\bm n \bm l \bm L} a^{\lambda\mu}_{\nu,\bm n \bm l \bm L} u^{\lambda\mu*}_{i,\nu,\bm n \bm l \bm L}\:,
\end{equation}
where we have explicitly indicated the order $\nu$ as a new index in the expansion coefficients. Finally, we observe 
that the terms $\varepsilon_i^{(\nu+1)}$ are scalars, and therefore invariant under rotations of the atomic positions. 
However, not only we already know how the coefficients $ u^{\lambda\mu*}_{\nu, \bm n \bm l \bm L}$ transforms under 
an active rotation [see Eq.~\eqref{eq:transformation_rule_coeff_SH_active}], but we have already discussed how the 
only rotationally invariant terms are obtained for $\lambda=\mu = 0$ [see Sec.~\ref{sec:general_properties_exp_coeff}]. 
Moreover, as established from the parity transformation of the expansion coefficients, such coefficients act as scalar (and 
are real) only when the sum $\Sigma(\bm l)$ is even [see Eqs.~\eqref{eq:sum_Sigma} and ~\eqref{eq:Multi_coefficients_parity}]. 
Thus, we deduce that the coefficients $a^{\lambda\mu}_{\nu, \bm n \bm l \bm L}$ have to vanish unless $\lambda = 0$, and 
unless the sum $\Sigma(\bm l)$ is an even number. With these constraints, we obtain
\begin{equation}
    \varepsilon_i^{(\nu+1)} = \sideset{}{'}\sum_{\bm n \bm l \bm L}a_{\nu,\bm n \bm l \bm L} u^{00}_{i,\nu,\bm n \bm l \bm L},
\end{equation}
where we have defined $a_{\nu,\bm n \bm l \bm L}:= a^{00}_{\nu,\bm n \bm l \bm L}$, and the prime indicates that the sum 
is constrained to take only cases for which $\Sigma(\bm l)$ is even. If we further define 
\begin{equation}\label{eq:n-body_energy_terms}
    E^{(\nu+1)} := \sum_i \varepsilon_i^{(\nu+1)},\nonumber
\end{equation}
then we obtain the following compact expression of a linear ML model
\begin{align}\label{eq:linear_scalar_model}
    &E_{\text{short}} = \sum_{\nu=0} E^{(\nu+1)}, \\
    &\text{with}\qquad E^{(\nu+1)} = \sideset{}{'}\sum_{\bm n \bm l \bm L}a_{\nu,\bm n \bm l \bm L} \sum_i u^{00}_{i,\nu,\bm n \bm l \bm L},\nonumber
\end{align}
which holds for every body order $(\nu+1)$. 
We can now identify this representation with the analogous one of a general ACE model \cite{ACE,ACE_dusson}. Indeed, we have already proved 
that, for $\nu=2$ and $\nu=3$, the scalar coefficients $u^{00}_{i,\nu,\bm n \bm l \bm L}$ are proportional to the powerspectrum 
and the bispectrum, respectively, as obtained by the ACE formalism. However, writing everything in terms of the MultiSH formalism, 
not only allows us to automatically obtain expressions up to any order, but it also provides access to all the properties derived for 
the expansion coefficients of the MultiSHs, namely Eqs.~\eqref{eq:transformation_rule_coeff_SH_passive}, 
\eqref{eq:transformation_rule_coeff_SH_active} and Section~\ref{sec:general_properties_exp_coeff}.

\subsection{SNAP and 4D Bispectrum}\label{sec:SNAP}
In this section, we are going to show how we can connect the MultiSHs formalism and the linear model described above, with 
the ones employed in SNAP ~\cite{SOAP,SNAP}. This section can be interpreted as an example on how the framework provided 
by the MultiSHs can be used to perform an in-depth investigation of known and well-established descriptors.

The starting point is to map the 3-dimensional vector $\vb r$, onto a 4-dimensional unit vector, $\ver u\in S^3\subset \mathbb{R}^4$ 
(here $S^3$ is the 4-dimensional hypersphere), by mean of a Riemann map. Explicitly, we have that the atomic density is now defined 
as
\begin{equation}\label{eq:SNAP_atomic_density}
    \rho_i(\vb r) \xrightarrow{\vb r \rightarrow \ver u}\rho_i(\ver u) = \sum_{j}^{\text{atoms}} f_{c}(r_{ji})\delta(\ver u-\ver u_{ji})\:,
\end{equation}
where $f_c(r_{ji})$ is the cut-off function, and with the map realized by defining the 4-dimensional unit vector
\begin{equation}\label{eq:Riemann_map}
    \ver u := \mqty(\cos\theta_0\\\sin\theta_0 \ver{r})\in S^3\:.
\end{equation}
Here, the new polar angle, $\theta_0$, is defined in terms of the length of the vector $\vb r$, as $\theta_0 := 4\pi r/(3r_\text{cut})$. 
Now, the new density $\rho(\ver u)$ can be expanded over the hyper-spherical harmonics (HSH)~\cite{HSHs}, which are the 
4-dimensional counterpart of the standard spherical harmonics. Therefore, we obtain
\begin{equation}
    \rho_i(\ver u) = \sum_{l=0,\frac{1}{2},1,\cdots} \dfrac{2l+1}{16\pi^2}\sum_{mm'} c_{i,lmm'} U^l_{mm'}(\ver u)\:,
\end{equation}
where $U^l_{mm'}(\ver u)$ are the HSHs, and where we have introduced the factor $(2l+1)/16\pi^2$, since the HSHs $U^l_{mm'}$ 
are, in general, not normalized. We also remark that the sum over $l$ runs over half-integer values. Here, the expansion coefficients 
are given by
\begin{align}\label{eq:exp_coefficients_SNAP}
     c_{i,lmm'} = \int \dd u\, \rho_i(\ver u)  U^{l*}_{mm'}(\ver u)\:.
\end{align}
At this point, it is important to point out that the HSH can be defined in a multitude of ways from the standard spherical harmonics, as 
shown in reference~\cite{Avery}. We firstly investigate the construction of reference~\cite{SOAP}, which utilizes the so-called 
\emph{parabolic-type} HSH~\cite{HSHs}. There, the 4-dimensional analogous of the bispectrum components is given by
\begin{align}\label{eq:4D_bispectrum_components}
    B_{i,ll_1l_2} := &\sum_{mm'} c^*_{i,lmm'}\\
    &\times\sum_{\substack{m_1m'_1\\m_2m'_2}}H^{lmm'}_{l_1m_1m_1'l_2m_2m_2'} c_{i,lm_1m'_1}c_{i,lm_2m'_2}\nonumber\:,
\end{align}
where the contraction tensor $H$ is defined as the product of two CG coefficients, namely
\begin{equation}\label{eq:tensor_H_SNAP}
    H^{lmm'}_{l_1m_1m_1'l_2m_2m_2'} := C^{lm}_{l_1m_1l_2m_2}C^{lm'}_{l_1m'_1l_2m'_2}\:.
\end{equation}
As proved in reference~\cite{SOAP}, these bispectrum components are invariant under rotations acting on the full 4-dimensional 
space. Finally, the SNAP model introduced in reference~\cite{SNAP} is obtained as a linear model over the bispectrum components 
and targeting the energy $E_\text{short}$. Explicitly
\begin{equation} 
    E_\text{short} \simeq E^{\text{SNAP}} = \sum_{ll_1l_2} a_{ll_1l_2}\sum_{i}^{\text{atoms}} B_{i,ll_1l_2}.
\end{equation}
With the definition of SNAP in place, the MultiSHs formalism will provide the tools to further investigate the properties of this model.

\subsubsection{Hypothesis and approximations behind the SNAP}
We start our discussion by writing an explicit formula for the bispectrum components obtained from the density 
of Eq.~\eqref{eq:SNAP_atomic_density}. In order to keep the expressions as compact as possible, we will employ 
all the short-hands defined in Sections~\ref{sec:multipolar_spherical} and ~\ref{sec:on_the_atomic_density}. 
We also define
\begin{equation}\nonumber
    \hat{\bm{y}}_{\bm j i} := (\ver u_{j_1i}, \ver u_{j_2 i},\ver u_{j_3 i})\:,
\end{equation}
as the collection of three 4-dimensional unit vectors corresponding to three atoms in the local environment of the $i$-th one. 
With these definitions, the bispectrum components can be written as 
\begin{align}\label{eq:4D_bispectrum_explicit_SNAP}
    B_{i,ll_1l_2} \propto \sum_{j_1j_2j_3}^{\text{atoms}}\left[\prod_{\alpha=1}^3 f_c(r_{j_\alpha i})\right]b_{i,ll_1l_2}(\hat{\bm y}_{\bm j i})\:,
\end{align}
with
\begin{align}\label{eq:small_b_SNAP}
    b_{i,l_1l_2l_3}&(\hat{\bm y}_{\bm j i}) := \sum_{mm'} U^{l_1}_{m_1m'_1}(\ver u_{j_1i})\\
    &\times\sum_{\substack{m_1m'_1\\m_2m'_2}}H^{lmm'}_{l_1m_1m_1'l_2m_2m_2'} U^{l_2*}_{m_2m'_2}(\ver u_{j_2i})U^{l_3*}_{m_3m'_3}(\ver u_{j_3 i})\nonumber\:.
\end{align}
The proportionality in Eq.~(\ref{eq:4D_bispectrum_explicit_SNAP}) is resolved by means of coefficients of the 
form of $(2l+1)/16\pi^2$, which are unessential for the present discussion.

We can now follow the same derivation presented in Section~\ref{sec:linear_ACE}, in order to partition the energy 
in multi-body contributions. In particular, we can consider the four-body term ($\nu = 3$) as
\begin{equation}\nonumber
    \varepsilon_i^{(4)} = \sum_{j_1j_2j_3} v_i^{(4)}(\hat{\bm y}_{\bm j i})\:,
\end{equation}
where the terms $v_i^{(4)}$ now depend on three 4-dimensional unit vectors. We can now proceed by \emph{assuming} 
that we can expand the terms $v_i^{(4)}(\hat{\bm y}_{\bm j i})$ as 
\begin{align}\label{eq:intermediate_potential_SNAP}
    v_i^{(4)}(\hat{\bm y}_{\bm j i}) &= \left[\prod_{\alpha=1}^3 f_c(r_{j_\alpha i})\right]\left(\sum_{\bm l} a_{\bm l}b_{i, \bm l}(\hat{\bm y}_{\bm ji})\right)\:,
\end{align}
for some coefficients $a_{\bm l}$. This can be regarded as the main hypothesis behind the SNAP model. 
Finally, inserting this expression back into the definition of $\varepsilon_i^{(4)}$, we obtain
\begin{equation}\nonumber
     \varepsilon_i^{(4)} =\sum_{\bm l} a_{\bm l} B_{i,\bm l}\:,
\end{equation}
where all the unessential factors have been absorbed by re-defining the coefficients $a_{\bm l}$.
By using the same formalism of Eq.~\eqref{eq:n-body_energy_terms}, this implies that we can read the SNAP model 
from the point of view of a multi-body framework as
\begin{equation}\nonumber
    E^{(4)} = \sum_i \varepsilon_i^{(4)} = \sum_{\bm l}a_{\bm l} \sum_i B_{i,\bm l} = E^{\text{SNAP}}.
\end{equation}

This approach to the SNAP model provides two advantages. The first one is that we can clearly see how the SNAP 
is tailored to target four-body contributions to a potential energy surface. The second, and most important observation, 
arises from the expansion assumed in Eq.~\eqref{eq:intermediate_potential_SNAP}. In particular, if we neglect the role 
of the products of cut-off functions, the main hypothesis is that the terms $\bm b_{i,l}$ are able to expand a scalar function 
that depends on the unit vectors $\hat{\bm y}_{\bm j i}$. If we indicate with $f(\hat{\bm y}_{\bm j i})$ such general function, 
this will admit an expansion over products of HSHs as
\begin{equation}\nonumber
    f(\hat{\bm y}_{\bm j i}) = \sum_{\bm l} f_{\bm l\bm m \bm m'} U^{l_1}_{m_1m'_1}(\ver u_{j_1 i})U^{l_2}_{m_2m'_2}(\ver u_{j_2 i})U^{l_3}_{m_3m'_3}(\ver u_{j_3 i})\:.
\end{equation}
Thus, for an expansion in terms of $\bm b_{i,l}$ to be generally valid, there must exist a unitary matrix that allows to write 
the scalar components of this expansion as a linear combination of the $b_{i,\bm l}(\hat{\bm y}_{\bm j i})$. By comparing 
the expansion in terms of HSH with the explicit expression for $\bm b_{i,l}$ provided in Eq.~\eqref{eq:small_b_SNAP}, we 
further notice that the tensor $H$ must be the required unitary matrix. However, as shown in Eq.~\eqref{eq:tensor_H_SNAP}, 
this tensor is not orthogonal with respect to a contraction of the indexes $(lmm')$, and thus is not unitary (in other words, they 
do not satisfy a property, which is analogous to Eq.~\eqref{eq:contraction_Gamma}). This proof by contradiction shows that, 
while an expansion such as the one of Eq.~\eqref{eq:intermediate_potential_SNAP} is surely possible, it is not general, and 
does not span the full rotationally-invariant sub-space. 

On the one hand, obtaining generalized CG-coefficients covering the entirety of the invariant sub-space is beyond the scope 
of this work. On the other hand, the MultiSH formalism provides yet another proof that there are portions of the invariant space 
not explored by the SNAP's bispectrum components. In particular, we will now show how to explicitly connect the SNAP 
formalism with a MultiSH expansion.

\subsubsection{SNAP in terms of MultiSHs}

We will now apply the MultiSHs formalism to an investigation of the SNAP formalism. This section will show how choosing this
basis greatly simplifies the investigations taken in the previous sections.
Let us start again from the atomic density $\rho_i(\ver u)$ of Eq.~\eqref{eq:SNAP_atomic_density} and its expansion in 
terms of HSHs. The following treatment can be greatly simplified, if instead of using the parabolic-type HSHs, $U^l_{mm'}$, 
we use the so-called \emph{spherical}-type HSHs, $Y_{\lambda l m}(\ver u)$ from the work of Meremianin~\cite{HSHs}. 
The two set of HSH, being two basis spanning the same space, are related by the unitary transformations

\begin{equation}\label{eq:change_basis_HSH}
    \begin{dcases}
        U^\lambda_{\mu\mu'}(\ver u) = \sum_{l=0}^{2\lambda}\sum_{m=-l}^l\sqrt{\dfrac{2l+1}{2\lambda+1}} C^{\lambda\mu'}_{\lambda\mu l m}Y_{\lambda lm}(\ver u),\\
        Y_{\lambda l m}(\ver u) = \sqrt{\dfrac{2l+1}{2\lambda+1}} \sum_{\mu\nu = -\lambda}^{\lambda} C^{\lambda\mu'}_{\lambda\mu l m} U^\lambda_{\mu\mu'}(\ver u),
    \end{dcases}
\end{equation}
and so, any result obtained for the spherical-type HSHs, can always be casted in terms of parabolic-type ones. The advantage 
of adopting the spherical-type HSHs can be appreciated from their explicit definition
\begin{equation}
     Y_{\lambda l m}(\ver u) = (-i)^l\sqrt{\dfrac{4\pi}{2\lambda+1}}\chi_{l}^{\lambda}(2\theta_0)Y_{l}^m(\ver r)\:.\nonumber
\end{equation}
Here, $\chi^{\lambda}_l$ are the generalized characters of the $O(3)$ rotation group, see \cite{Angular,HSHs}, and 
$Y^m_{l}(\ver r)$ are the standard spherical harmonics. We remark that, even if this definition neglects the normalization 
of the basis, we will not consider the unessential normalization constants. If we now compare this expression with the 
standard basis, $R_{nl}(r)Y_{ml}(\ver r)$, used throughout this work, we can immediately recognize that this construction 
can be read in terms of the choice of a specific radial basis. More specifically,
\begin{equation}\nonumber
    R_{nl}(r)\xleftrightarrow[r\leftrightarrow \theta_0]{n\leftrightarrow \lambda} (-i)^l\sqrt{\dfrac{4\pi}{2\lambda+1}}\chi_{l}^{\lambda}(2\theta_0)\:,
\end{equation}
where this identification has been induced by the Riemann map (note that we will not investigate the consequences 
of having a complex radial function, but all the derivations and the properties of this paper should be modified accordingly). 
With this identification we can now apply all the tools and results derived for the general MultiSH formalism, in particular, 
the considerations about the rotationally invariant terms. 

For example, from Eqs.~\eqref{eq:multi_bispectrum0} and \eqref{eq:multi_bispectrum}, we can immediately 
write the standard, 3-dimensional, bispectrum components as
\begin{equation}
    B_{i,\substack{\lambda_1\lambda_2\lambda_3\\l_1l_2l_3}}\propto \sum_{m_1m_2m_3} u^*_{i,\lambda_3l_3m_3}C^{l_3m_3}_{l_1m_1l_2m_2} u_{i,\lambda_1l_1m_1}u_{i,\lambda_2l_2m_2}\:,
\end{equation}
where 
\begin{equation}
    u_{i,\lambda l m}\propto \int \dd \ver u\, \rho_i(\ver u) Y^*_{\lambda lm}(\ver u)\nonumber\:.
\end{equation}
Moreover, by writing these bispectrum components in terms of the parabolic HSH and by using
Eq.~\eqref{eq:change_basis_HSH}, we write
\begin{align}
    &B_{i,\substack{\lambda_1\lambda_2\lambda_3\\l_1l_2l_3}}\propto \sum_{\mu_3\mu'_3} c^*_{i,\lambda_3 \mu_3\mu'_3}\sum_{\substack{\mu_1\mu'_1\\\mu_2\mu'_2}}C_{\bm \lambda\bm l \bm \mu \bm \mu'}c_{i,\lambda_1\mu_1\mu'_1}c_{i,\lambda_2\mu_2\mu'_2}\nonumber\:,
\end{align}
where the expansion coefficients, $c_{i,\lambda\mu\mu'}$, are given by Eq.~\eqref{eq:exp_coefficients_SNAP}, and where the coupling tensor is given by
\begin{equation}\nonumber
    C_{\bm \lambda\bm l \bm \mu \bm \mu'} := C^{\lambda_3\mu'_3}_{\lambda_3\mu_3l_3m_3}C^{l_3m_3}_{l_1m_1l_2m_2}C^{\lambda_3\mu'_3}_{\lambda_1\mu_1l_1m_1}C^{\lambda_2\mu'_2}_{\lambda_2\mu_2l_2m_2}\:.
\end{equation}

We can finally compare this result with the 4-dimensional bispectrum components of Eq.~\eqref{eq:4D_bispectrum_components} 
and with the corresponding coupling tensor, $H$, of Eq.~\eqref{eq:tensor_H_SNAP}. From this comparison we can draw two 
conclusions. The first one confirms the findings of the last paragraph: indeed, the bispectrum components obtained here are invariant 
under any rotation of the 3-dimensional space. However, such rotations can always be represented as particular rotations in the 4-dimensional 
space. Since the tensor $C_{\bm \lambda\bm l \bm \mu \bm \mu'}$ cannot be generally obtained by means of a contraction of the tensor 
$H$ of Eq.~\ref{eq:tensor_H_SNAP}, the components obtained here cannot be derived from the ones of the 4-dimensional bispectrum 
of Eq.~\eqref{eq:4D_bispectrum_components}. 
This provides another proof of the fact that the SNAP formulation does not include all the possible scalars, since it is already lacks those 
related with standard 3-dimensional rotations. The second conclusion is that the compactness of the SNAP model is obtained by enforcing 
a non-physical invariance under rotation of the entire 4-dimensional space. Indeed, the three-dimensional bispectrum components obtained 
here, still necessitates of six indexes in the expansion, since they enforce rotational invariance with respect to standard 3-dimensional rotations. 
In contrast, the 4-dimensional bispectrum components are defined only in terms of three indexes, because of the additional invariance constraint
under rotations around the unphysical 4-dimensional axis. While it is true that the presence of the cut-off functions in 
Eq.~\eqref{eq:4D_bispectrum_explicit_SNAP} breaks this symmetry, it also implies that the cut-off functions provide enough descriptive power 
to express the dependence of the potentials $v_i^{(4)}$ on the atomic distances. This strong requirement is, generally, not satisfied.

\subsection{Connection with internal-coordinates representations}
In the last section we have derived the SNAP formalism and, with the aid of the general MultiSH expansion, we have also been 
able to investigate its shortcomings (the presence of uncovered portions of the rotationally-invariant space) and its advantages 
(such as the compactness of the representation). In this section we are going to investigate the connection between 
a MultiSHs representation and one written in terms of internal coordinates. Since this will be performed for the prediction of scalar 
quantities only, and in a multi-body expansion framework, we will exploit the construction presented in Section~\ref{sec:linear_ACE}. 
In particular, we will firstly derive a connection between the MTP formalism~\cite{MTP}, and then we will conclude by discussing 
the JLP formalism~\cite{JLP}. 

The assumptions for both MTP and JLP are the same as those introduced in Section~\ref{sec:linear_ACE}. 
However, there is a difference in the quantities expanded. In the MultiSHs expansion, we first expand over a general basis 
and then we project onto the scalar sub-space. When using internal coordinates, instead, we first select rotationally-invariant 
degrees of freedom, and then we perform the expansion. Given that any invariant constructed from the set of unit vectors $\{\ver r_{ji}\}$ 
can be written in terms of linear combinations of powers of the scalar products, $\{(\ver r_{ji}\cdot \ver r_{ki})\}$, only 
[see reference~\cite{Classical_group} for a treatment on the fundamental theorem of invariant theory], then, from 
Eq.~\eqref{eq:nu_cluster_expansion} we can write
\begin{equation}
        \varepsilon_i^{(\nu+1)} = \sum_{\bm j} v_i^{(\nu+1)}(\vb x_{\bm j i}) = \sum_{\bm j} v_i^{(\nu+1)}(x_{\bm j i}, s_{\bm j i})\:.
\end{equation}
Here, the $s_{\bm j i}$'s describe the set of all scalar products that can be constructed from the atomic unit vectors 
$\ver x_{\bm j i}$, namely
\begin{equation}
    s_{\bm j i} := ((\ver r_{j_1 i}\cdot \ver r_{j_2 i}),(\ver r_{j_1 i}\cdot \ver r_{j_3 i}), \ldots , (\ver r_{j_{\nu-1} i}\cdot \ver r_{j_\nu i}) )\:. \nonumber
\end{equation}
As clear by the fact that we can construct $\nu(\nu+1)/2$ unique scalar products out of $\nu$ unit vectors, considering all these 
scalar products produces a strongly redundant description already for relatively small $\nu$. For this reason, employing a 
representation in terms of scalar products is unfeasible already for $\nu\ge 6$. However, for smaller values, such representation 
is a valid alternative to the MultiSHs one. Moreover, since the two descriptions span the same space, there must exist a unitary 
matrix connecting the two. In other words, the following methods are, fundamentally, a change of basis from the MultiSHs framework.

\subsubsection{The Moments Tensor Potential}
In the MTP formalism, the expansion of $v_i^{(\nu+1)}$ is performed in terms of a polynomial expansion over the scalar products. 
Thus, essentially, this formalism is obtained by adopting the expansion
\begin{equation}\nonumber
    v_{i}^{(\nu+1)}(x_{\bm j i},s_{\bm j i}) = \sum_{\bm n \bm \alpha} a_{\bm n \bm \alpha} \mathcal{R}_{\bm n}(x_{\bm j i})\mathcal{M}_{\bm \alpha}(s_{\bm j i})\:,
\end{equation}
where we introduce the polynomial basis
\begin{align}
&\mathcal{M}_{\bm \alpha}(s_{\bm j i}):=(\ver r_{j_1 i}\cdot \ver r_{j_2 i})^{\alpha_{12}}(\ver r_{j_1 i}\cdot \ver r_{j_3 i})^{\alpha_{13}}\nonumber
\\&\qquad\qquad\qquad\qquad\qquad\times\ldots \times(\ver r_{j_{\nu-1} i}\cdot \ver r_{j_\nu i})^{\alpha_{\nu-1,\nu}} \nonumber\:,
\end{align}
and where $\bm \alpha$ is the matrix containing the exponents of the polynomial expansion (note that the sum on $\bm \alpha$ runs 
over all the possible exponents of the expansion). It is worth to stress that in this context the radial and the angular part are disentangled. 
Following the same steps of Section~\ref{sec:linear_ACE}, we can also cast the MTP into the simple form of 
Eq.~\eqref{eq:linear_scalar_model} as
\begin{align}
    &E^{\text{MTP}}_{\text{short}} = \sum_{\nu=0} E^{(\nu+1)}, \\
    &\text{with}\qquad E^{(\nu+1)} = \sum_{\bm n \bm \alpha }a_{\nu,\bm n \alpha} \sum_i \sum_{\bm j}\mathcal{R}_{\bm n}(x_{\bm j i})\mathcal{M}_{\bm \alpha}(s_{\bm j i}).\nonumber
\end{align}

We conclude this section by remarking again that, since the MultiSHs framework spans the entirety of the rotationally invariant 
space, then there must exist a connection between the MTP and a MultiSH-based one. In other words, it is possible to write
\begin{equation}\label{eq:MTP_to_multiSHs}
    \mathcal{M}_{\bm \alpha}(s_{\bm j i}) = \sum_{\bm l \bm L} U^{\bm \alpha}_{\bm l \bm L} \mY^{00}_{\bm l \bm L}(\ver x_{\bm j i})\:,
\end{equation}
with 
\begin{equation}
    U^{\bm \alpha}_{\bm l \bm L} = \int \dd \ver x_{\bm j i}\,  \mathcal{M}_{\bm \alpha}(s_{\bm j i})\mY^{00*}_{\bm l \bm L}(\ver x_{\bm j i})\:.
\end{equation}
Note that, not only we know that the scalar space ($\lambda=\mu=0$) is enough to represent $\mathcal{M}_{\bm \alpha}$, as expected 
when describing a scalar quantity, but also that the sum $\Sigma(\bm l)$ must be even (see Eq.~\eqref{eq:sum_Sigma}), since all the 
scalar products are invariant under inversion [See Section~\ref{sec:general_properties_exp_coeff}]. Indeed, an explicit evaluation of the 
integral above reveals that it is zero, whenever the sum $\Sigma(\bm l)$ is odd.

\subsubsection{Jacobi-Legendre Potential}
Our JLP \cite{JLP} follows the same overall approach of the MTP, but with a larger emphasis on the choice of the radial basis, and 
with an expansion in terms of the Legendre polynomials, $P_\gamma$ [see Ref.~\cite{abramowitz}], in place of the powers of scalar 
products (we use the unusual index $\gamma$ to differentiate with the approach in terms of MultiSHs). Indeed, it is known that any 
power of scalar products, can be written as a linear combination of Legendre polynomials as
\begin{equation}\label{eq:scalar_products_to_legendre_polynomials}
    (\ver r_{\bm ji}\cdot \ver r_{\bm ki})^\alpha = \sum_{l = \alpha, \alpha-2,\ldots} d_{\alpha \gamma} P_\gamma(\ver r_{\bm ji}\cdot \ver r_{\bm ki}),
\end{equation}
with the specific form of $d_{\alpha \gamma}$ given in Reference~\cite{abramowitz}. This means that the expansion proposed for the 
MTP, can yet again be re-casted as
\begin{equation}\label{eq:JLP_eq1}
     v_{i}^{(\nu+1)}(x_{\bm j i},s_{\bm j i}) = \sum_{\bm n \bm \gamma} a_{\bm n \bm \gamma} \mathcal{R}_{\bm n}(x_{\bm j i})\mathcal{P}_{\bm \gamma}(s_{\bm j i})\:,
\end{equation}
where
\begin{align}
    &\mathcal{P}_{\bm \gamma}(s_{\bm j i}):=P_{\gamma_{12}}(\ver r_{j_1 i}\cdot \ver r_{j_2 i})P_{\gamma_{13}}(\ver r_{j_1 i}\cdot \ver r_{j_3 i})\nonumber
\\&\qquad\qquad\qquad\qquad\qquad\times\ldots \times P_{\gamma_{\nu-1,\nu}}(\ver r_{j_{\nu-1} i}\cdot \ver r_{j_\nu i}) \nonumber\:.
\end{align}
Similarly the case of the matrix $\bm \alpha$, the matrix $\bm \gamma$ is the collection of all the indexes of the Legendre polynomials.

As previously mentioned, a great focus of the JLP lies in the optimization of the radial basis. We will briefly investigate the property of this 
construction, since it is independent from the angular part of the expansion and, thus, can be used also in any other formalism of choice, 
including the MultiSHs framework. 

As the name of the potential suggests, in the JLP the radial dependence is obtained by expanding over Jacobi polynomials, 
$P^{(\alpha,\beta)}$ (note that the $\alpha$ used here is different from the one used in relation to the MTP). In particular, for 
each fixed choice of $\alpha,\beta>-1$, the Jacobi polynomials form a complete and orthogonal set in the interval $[-1,1]$, with 
respect to the weight function $w^{\alpha\beta}(x) = (1-x)^\alpha(1+x)^\beta$. This allows one to explicitly optimize the radial basis, 
by optimizing $\alpha$ and $\beta$, without assuming any specific fixed form. The cosine map, defined as
\begin{equation}
    \sigma_{ji} = \sigma (r_{ji};r_\text{cut}) := \cos\left(\pi\dfrac{r_{ji}}{r_\text{cut}}\right)\:,\nonumber
\end{equation}
allows us to map each length $r_{ji}$ onto the interval $[-1,1]$, so that the expansion over the Jacobi polynomials can take place. 
An interesting property of this expansion is that, since the Legendre polynomials are obtained from the Jacobi polynomials by setting 
$\alpha=\beta=0$, it retains a degree of homogeneity in the basis used. We can also draw similarities with the SNAP potential, where 
the radial dependence is mapped onto a 4-dimensional hyper-axis [see Section~\ref{sec:SNAP}]. Indeed, we can see how using this 
map is almost identical to the $\cos(\theta_0)$ term of the Riemann map of Eq.~\eqref{eq:Riemann_map}, but for a different scaling 
factor. Namely, we could interpret this approach as a projection on an hyper-axis. With this construction in place, we have that Eq.~\eqref{eq:JLP_eq1} 
becomes 
\begin{equation}
     v_{i}^{(\nu+1)}(x_{\bm j i},s_{\bm j i}) = \sum_{\bm n \bm \gamma} a_{\bm n \bm \gamma} \mathcal{P}^{(\alpha,\beta)}_{\bm n\bm j i}\mathcal{P}_{\bm \gamma}(s_{\bm j i})\:,
\end{equation}
where we have used the short-hand notation
\begin{align}
    \mathcal{P}^{(\alpha,\beta)}_{\bm n\bm j i}:= P^{(\alpha,\beta)}_{n_1}(\sigma_{j_1 i})\times\cdots\times P^{(\alpha,\beta)}_{n_\nu}(\sigma_{j_\nu i})\:,
\end{align}
for the product of $\nu$ Jacobi polynomials. 
 
Another result of the JLP is the development of a straightforward constraining procedure for the $v_{i}^{(\nu+1)}$ terms, which 
impacts the properties of the basis and of the potentials itself. For example, contrary to most of the other methods that manually 
include a cut-off radius in the definition of the basis, the JLP is obtained by constraining directly the expansion coefficients $a_{\bm n\gamma}$. 
A specific derivation of this constraining procedure is beyond the scope of this work. Nevertheless, it is worth to note that this treatment 
has the two advantages. Firstly, it does not assume any specific form of the cut-off function, and secondly it has of a strict separation of 
the expression of the $v^{(\nu+1)}_i$, for different $\nu$. However, this last observation also implies that significant care must be taken, 
when considering unrestricted sums over all the possible atomic labels $\bm j$ since, in this case, self-interacting terms are not easily 
re-absorbed by the expansions at lower $\nu$. For a detailed treatment of the linearization procedure with this added constraints, we 
refer to the original work \cite{JLP}.

Now that the full basis has been defined, and following again the procedure outlined in Section~\ref{sec:linear_ACE}, the JLP model 
takes the compact form
\begin{align}
    &E^{\text{JLP}}_{\text{short}} = \sum_{\nu=0} E^{(\nu+1)}, \\
    &\text{with}\qquad E^{(\nu+1)} = \sum_{\bm n \bm \gamma }a_{\nu,\bm n\bm \gamma} \sum_i \sum_{\bm j}\mathcal{P}^{(\alpha,\beta)}_{\bm n\bm j i}\mathcal{P}_{\bm \gamma}(s_{\bm j i})\:.\nonumber
\end{align}

We conclude this final sub-section by mentioning that the connection between the JLP formalism and the MultiSHs is even more 
pronounced, because of the addition theorem for spherical harmonics \cite{Angular}
\begin{align}\label{eq:addition_theorem_bipo}
    P_\gamma(\ver r_{1}\cdot \ver r_{2}) &= \dfrac{4\pi}{2\gamma+1}\sum_{\mu=-\gamma}^\gamma (-1)^\mu Y^{\mu}_{\gamma}(\ver r_1)Y^{-\mu}_{\gamma}(\ver r_2) \nonumber\\
    &= (-1)^\gamma\dfrac{4\pi}{\sqrt{2\gamma+1}}\mY^{00}_{\gamma\gamma}(\ver r_1,\ver r_2)\:.
\end{align}
This expression shows the connection between the Legendre polynomials and the BipoSHs. By the same arguments presented 
in the MTP model, it is possible to express the Jacobi-Legendre basis in terms of the MultiSHs one by evaluating
\begin{equation}
    \mathcal{P}_{\bm \gamma}(s_{\bm j i}) = \sum_{\bm l \bm L}V^{\bm \gamma}_{\bm l \bm L} \mY^{00}_{\bm l \bm L}(\ver x_{\bm ji})\:,
\end{equation}
where 
\begin{equation}
    V^{\bm \alpha}_{\bm l \bm L} = \int \dd \ver x_{\bm j i}\,  \mathcal{P}_{\bm \gamma}(s_{\bm j i})\mY^{00*}_{\bm l \bm L}(\ver x_{\bm j i})\:.
\end{equation}
From Eq.~\eqref{eq:addition_theorem_bipo} we can deduce that the integral above can be casted as products of MultiSHs only, making 
its evaluation straightforward (although tedious). Moreover, since Eq.~\eqref{eq:scalar_products_to_legendre_polynomials} determines 
the tensor connecting the JLP representation with the MTP one, we are also able to determine the tensor $U^{\bm \alpha}_{\bm l\bm L}$ 
of Eq.~\eqref{eq:MTP_to_multiSHs}, connecting the MTP with the MultiSHs formalism, by means of $V^{\bm \alpha}_{\bm l\bm L}$. 
This allows to obtain analytical expressions that bridge the gap between a representation in terms of MultiSHs, and the analogous one 
in terms of internal coordinates.

\section{Linear models for covariant quantities}\label{sec:tensor_model_linear}
In the previous section we have discussed how to derive linear models for scalar quantities by means of multi-body expansion, 
with a focus on models for the short-ranged contribution to the energy, $E_{\text{short}}$. In doing so, we have constrained the 
expansion to admit only scalar coefficients. In this section we will investigate the consequences of abandoning such constraints. 
In other words, we will derive the general expression for linear models that target the harmonic components of a tensor $\bm T$. 
The assumption underlying this section will be, again, the short-ranged nature of the terms involved in this representation, the 
partitioning of the tensor in atomic contributions, $\bm T_i$, and the further partition on multi-body order terms $\bm \tau_i^{(\nu+1)}$. 
Explicitly, we will assume that the harmonic components of the tensor $\bm T$ can be written as
\begin{equation}
     T_{\lambda\mu} = \sum_i^\text{atoms}   T_{i,\lambda\mu} = \sum_i\sum_{\nu=0} \tau^{(\nu+1)}_{i,\lambda\mu}\:,
\end{equation}
where $\tau^{(\nu+1)}_{i,\lambda\mu}$ represents the local $(\nu+1)$-body term. Since the procedure is the same of the one of 
Section~\ref{sec:linear_ACE}, we can obtain an expression for these terms from Eq.~\eqref{eq:linear_model_MultiSHs_exp}. 
Indeed, the expansion in terms of MultiSHs is given by
\begin{equation}
    \tau^{(\nu+1)}_{i,\lambda\mu} = \sum_{\bm n \bm l \bm L} a^{\lambda}_{\nu,\bm n \bm l \bm L} u^{\lambda\mu*}_{i,\nu,\bm n \bm l \bm L}\:.
\end{equation}
We can now make two observations. Firstly, similarly to the $\lambda=\mu=0$ constraint of the scalar case, here we consider only 
the $(\lambda\mu)$ components of the MultiSHs expansion, since they are the ones that follow the required transformation rules. 
Moreover, because the correct rotational properties are already fully encoded into $u^{\lambda\mu*}_{i,\nu,\bm n \bm l \bm L}$, then 
the coefficients $a^{\lambda\mu}_{\nu,\bm n \bm l \bm L}$ cannot participate in the mixing of the magnetic number $\mu$. 
Explicitly, this implies that these coefficients cannot depend on $\mu$, namely 
$a^{\lambda\mu}_{\nu,\bm n \bm l \bm L}= a^{\lambda}_{\nu,\bm n \bm l \bm L}$. 

Finally, by defining $\bm T^{(\nu+1)}:= \sum_i \bm \tau_i^{(\nu+1)}$, we obtain the following general form of a linear model that targets 
the covariant components of the tensor $\bm T$
\begin{align}\label{eq:linear_model_tensors}
    & T_{\lambda\mu} = \sum_{\nu=0} T^{(\nu+1)}_{\lambda\mu}\:, \\
    &\text{with}\qquad T^{(\nu+1)}_{\lambda\mu} = \sum_{\bm n \bm l \bm L}a^\lambda_{\nu,\bm n \bm l \bm L} \sum_i u^{\lambda\mu*}_{i,\nu,\bm n \bm l \bm L}\:.\nonumber
\end{align}
This model encompasses the same basis used in reference~\cite{NICE} for the NICE model. Indeed, an example is provided by 
using the explicit definition of the expansion coefficients $u^{\lambda\mu*}_{i,\nu,\bm n \bm l \bm L}$ of Eq.~\eqref{eq:u_coeff_dirac_delta}, 
and the $\Gamma(\tau_1)$ tensor from Eq.~\eqref{eq:Gamma_n2_tau1}. In this case ($\nu =2$), we derive
\begin{align}
    &\sum_i u^{\lambda\mu*}_{i,2,n_1n_2l_1l_2} = \sum_i\sum_{j_1 j_2} R_{n_1l_1}(r_{j_1i})R_{n_2l_2}(r_{j_2i})\\
    &\qquad\qquad\qquad\quad\times\sum_{m_1m_2}C^{\lambda\mu}_{l_1m_1l_2m_2}Y_{l_1}^{m_1}(\ver r_{\bm j_1 i})Y_{l_2}^{m_2}(\ver r_{\bm j_2 i})\:.\nonumber
\end{align}
This is indeed the same contraction of the NICE basis for the $\ket{\rho^{\otimes 2}}$ case (while the radial part is generally represented 
from the point of view of an expansion of Gaussians), with the covariant behavior fully expressed by the contraction with the CG coefficient. 
Importantly, this model can be seen as a generalization of the scalar one of Eq.~\eqref{eq:linear_scalar_model}, since the former reduces 
to the latter when $\lambda=\mu=0$.

We conclude this investigation by mentioning that we can impose further constraints on the general linear model introduced above by 
investigating the nature of the target tensor, $\bm T$. Indeed, following the discussion on the parity of the coefficients 
$u^{\lambda\mu}_{\nu,\bm n \bm l \bm L}$, see Eq.~\eqref{eq:Multi_coefficients_parity}, we have that, when the sum 
$\Sigma(\bm l)+\lambda$ is even, the coefficients behave as the components of a tensor. Conversely, when $\Sigma(\bm l)+\lambda$ 
is odd, the coefficients behave as the components of pseudotensors. Thus, whenever $\bm T$ is a tensor or a pseudotensor, we can 
constraint the sum over $\bm l$ to values for which $\Sigma(\bm l)+\lambda$ is even or odd, respectively.

\section{Symmetries reduction}

In this conclusive section, we will investigate the consequences following the choice of different coupling schemes for the linear 
model of Eq.~\eqref{eq:linear_model_tensors}. Being the scalar model a particular case a tensorial one, it should be noted 
that what follows holds also for all the models examined in Sections~\eqref{sec:linear_ACE} and \eqref{sec:SNAP}. Here we will 
obtain results similar to the lexicographic ordering obtained in reference~\cite{ACE,ACE_dusson}, but generalized to the tensorial 
case. Moreover, the construction presented here will partially address the removal of the redundancies in linear models, investigated 
previously for the scalar \cite{Goffreport} and tensorial \cite{NICE} case.

Consistently with the message of this entire work, we will derive our results based on the property of the basis only, with no reference to the 
explicit form of the function $\rho^{\otimes \nu}(\vb x)$ except for the fact that it is symmetric. Moreover, the results obtained in this 
section will show that a similar ordering is not valid only for scalar models, but can be adopted also when targeting tensors.

We have already discussed how MultiSHs can be constructed by means of different coupling schemes (or trees), and that each 
coupling scheme produces a valid MultiSHs basis. Therefore there must always exist a change of basis between coupling schemes. 
Let us consider a fixed coupling scheme, $\tau_0$. With this coupling, the density $\rho^{\otimes \nu}$ is expanded as

\begin{equation}
    \rho^{\otimes \nu }(\vb x) = \sum_{\lambda\mu}\sum_{\bm n\bm l \bm L} u^{\lambda\mu}_{\bm n \bm l \bm L,\tau_0} \mathcal{R}_{\bm n \bm l}(x)\mY^{\lambda\mu}_{\bm l \bm L, \tau_0}(\ver x)\:.
\end{equation}
Now, we assume that the density $\rho^{\otimes \nu}$ is a symmetric function, which means that it is invariant under a permutation 
of its arguments. 
If we indicate with $\hat{\sigma}$ a general permutation, then we have $\rho^{\otimes \nu}(\hat{\sigma}\vb x) = \rho^{\otimes \nu}(\vb x)$.
We observe that the permutation $\hat{\sigma}$ does not act on the atomic positions, but on the ordering of the input of the 
functions $\rho^{\otimes \nu}(\vb x)$. In other words, these permutations lives in $S_\nu$, the space of permutation of $\nu$-elements, 
while the permutation acting on the atomic positions, which are not investigated here, belong to $S_{N}$, with $N$ being the number 
of atoms in the local environment. By expanding the function with permuted arguments, we write
\begin{align}
    \rho^{\otimes \nu }(\hat{\sigma}\vb x) = \sum_{\lambda\mu} \sum_{\bm n \bm L} u^{\lambda\mu}_{\bm n \bm l \bm L,\tau_0} \mathcal{R}_{\bm n \bm l}(\hat{\sigma}x) \mY^{\lambda\mu}_{\bm l \bm L, \tau_0}(\hat{\sigma}\ver x)\:.
\end{align}
Since the function $\mathcal{R}_{\bm n \bm l}$ is a simple product of radial functions, the permutation can be passed over the 
indexes. Indeed, for example, if $\hat{\sigma}$ swaps the first two arguments, we have
\begin{align}
    \mathcal{R}_{\bm n\bm l}(\hat{\sigma} x) &= \mathcal{R}_{\substack{n_1n_2\ldots n_\nu\\l_1l_2\ldots l_\nu}}(r_2, r_1, r_3,\ldots, r_\nu) =\nonumber\\
    &=\mathcal{R}_{\substack{n_2n_1\ldots n_\nu\\l_2l_1\ldots l_\nu}}(r_1, r_2, r_3,\ldots, r_\nu) = \mathcal{R}_{\hat{\sigma}\bm n,\hat{\sigma}\bm l}(x)\:,\nonumber
\end{align}
where, in the last equality we have used the definition of Eq.~\eqref{eq:def_product_radial_basis} to swap the indexes.

The situation is different for the MultiSHs, since they are not obtained by mean of simple products. However, we can observe that 
permuting the arguments of a MultiSH is equivalent to \emph{simultaneously} adopting a different coupling scheme and permuting 
the indexes of the $l$ channels. This can be seen graphically. For example, if $\hat{\sigma}$ swaps the first and third arguments, we have
\begin{equation}
    \includegraphics[width=.46\textwidth]{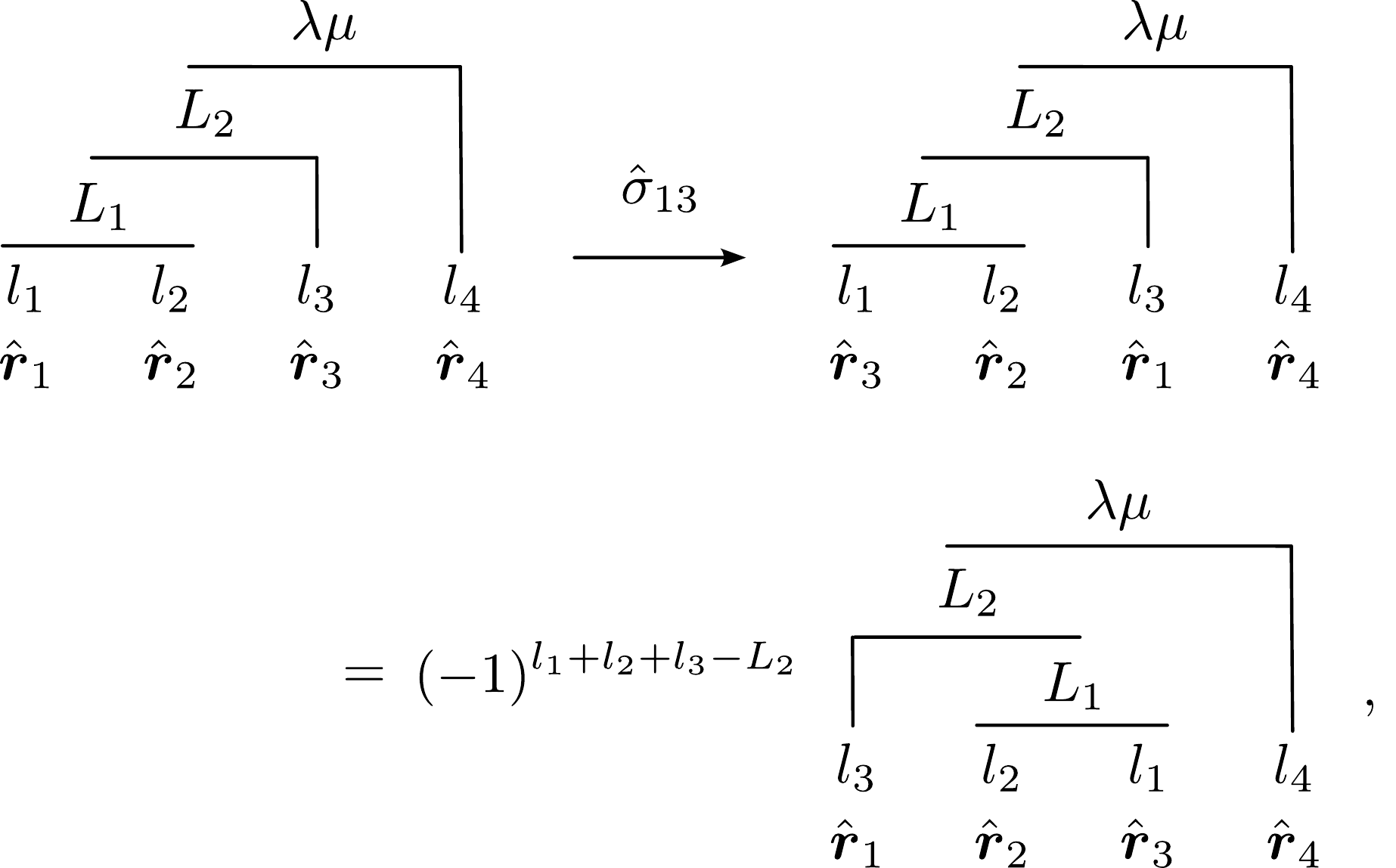}
    \nonumber
\end{equation}
where, for graphical reasons, we consider the $\nu=4$ case and where we have adopted the same coupling scheme of
Eq.~\eqref{eq:Gamma_generalized_graph}. The prefactor is induced by the change of order of the labels in the graph and 
is caused by the fact that the CG coefficients are not symmetric, as we already observed in Section~\eqref{sec:Bipo}. Importantly, 
we observe that the order of the unit vectors $\ver r_1,\ldots,\ver r_4$, has been restored to the one before the permutation. Since a 
similar construction can be done for any order and coupling scheme, then we have that a general permutation $\hat{\sigma}$ always 
induces a change of coupling scheme and of label ordering, namely
\begin{equation}\nonumber
    \mY^{\lambda\mu}_{\bm l \bm L, \tau_0}(\hat{\sigma}\ver x) = d_{\bm l\bm L} \mY^{\lambda\mu}_{\hat{\sigma}\bm l,\bm L,\tau(\hat{\sigma})}(\ver x)\:,
\end{equation}
where the prefactor $d_{\bm l\bm L}$ takes the value $+1$ or $-1$, and where $\tau(\hat{\sigma})$ is the new coupling tree induced 
by the permutation $\hat{\sigma}$. 

As we have already observed, two coupling schemes, $\tau_1$ and $\tau_2$ are always related to each other by a unitary matrix. We 
indicate with $W(\tau_1,\tau_2)$ the matrix transforming from the coupling $\tau_1$ to $\tau_2$. From this discussion, we deduce that 
there must exist a matrix $W(\tau(\hat{\sigma}), \tau_0)$, connecting the coupling scheme induced by the permutation $\hat{\sigma}$, here denoted with $\tau(\hat{\sigma})$,
with the initial coupling scheme $\tau_0$. Therefore, we must have
\begin{equation}\nonumber
    \mY^{\lambda\mu}_{\hat{\sigma}\bm l,\bm L,\tau(\hat{\sigma})}(\ver x) = \sum_{\bm L'} \left(W(\tau(\hat{\sigma}),\tau_0)\right)^{\lambda\mu}_{\hat{\sigma}\bm l, \bm L \bm L'}\mY^{\lambda\mu}_{\hat{\sigma}\bm l,\bm L',\tau_0}(\ver x)\:.
\end{equation}
We will not investigate the explicit expression of this general matrix, since it is irrelevant to our discussion. Instead we to refer 
to \cite{Angular} for the analytical form of the matrices that realize the connection between different coupling schemes. We just 
mention here that such matrices are proportional to $3nj$-Wigner symbols. 

If we now insert these results back into $\rho^{\otimes \nu}(\hat{\sigma}\vb x)$, we obtain
\begin{align}
    &\rho^{\otimes \nu }(\hat{\sigma}\vb x) = \sum_{\lambda\mu} \sum_{\bm n \bm L'} \left[\sum_{\bm L}d_{\bm l \bm L}W(\tau(\hat{\sigma}),\tau_0)^{\lambda\mu}_{\hat{\sigma}\bm l, \bm L \bm L'}u^{\lambda\mu}_{\bm n \bm l \bm L,\tau_0}\right]\nonumber\\
    &\qquad\qquad\qquad\qquad\qquad\qquad\times\mathcal{R}_{\hat{\sigma}\bm n,\hat{\sigma} \bm l}(x) \mY^{\lambda\mu}_{\hat{\sigma}\bm l, \bm L', \tau_0}(\ver x)\nonumber\:.
\end{align}
However, since $\rho^{\otimes \nu}$ is symmetric, then this expression is equivalent to the standard expansion 
of $\rho^{\otimes\nu}$. Therefore, it must hold that
\begin{equation}
    u^{\lambda\mu}_{\bm n\bm l\bm L,\tau_0}= \sum_{\bm L'}d_{\bm l \bm L'}W(\tau(\hat{\sigma}),\tau_0)^{\lambda\mu}_{\bm l, \bm L' \bm L}u^{\lambda\mu}_{\hat{\sigma}^{-1}\bm n,\hat{\sigma}^{-1}\bm l,\bm L',\tau_0}\nonumber\:,
\end{equation}
or, equivalently
\begin{equation}\label{eq:linear_relation_permutation_coeff}
     u^{\lambda\mu}_{\hat{\sigma}\bm n,\hat{\sigma}\bm l\bm L,\tau_0}= \sum_{\bm L'}d_{\bm l \bm L'}W(\tau(\hat{\sigma}^{-1}),\tau_0)^{\lambda\mu}_{\bm l, \bm L' \bm L}u^{\lambda\mu}_{\bm n\bm l\bm L',\tau_0}\:.
\end{equation}
We have already seen examples of these relations in Eqs.~\eqref{eq:symmetry_coeff} and \eqref{eq:Tripo_redundancy_reduction}, 
for the BipoSHs and TripoSHs, respectively.

Crucially, Eq.~\eqref{eq:linear_relation_permutation_coeff} shows that the coefficients obtained by applying a simultaneous permutation 
to the set of indexes $\bm n$ and $\bm l$ can be always written as a linear combination of the coefficients with unpermuted indexes. 
Moreover, the linear combination is realized by a summation over the intermediate channels, $\bm L$, only. In other words, the expansion 
coefficients obtained by permuting the indexes of known coefficients, do not carry any additional information. This observation has important 
consequences for the linear models of the form of Eq.~\eqref{eq:linear_model_tensors}. Indeed, given the linear relation of 
Eq.~\eqref{eq:linear_relation_permutation_coeff}, the full information carried from the expansion coefficients is encoded in terms satisfying 
a lexicographic order of the indexes $(n_\alpha,l_\alpha)$. Thus, the linear model for tensors becomes
\begin{align}
    &T_{\lambda\mu} = \sum_{\nu=0} T^{(\nu+1)}_{\lambda\mu}, \\
    &\text{with}\qquad
    T^{(\nu+1)}_{\lambda\mu} = \sum^{\text{ordered}}_{\bm n \bm l}\sum_{\bm L}a^\lambda_{\nu,\bm n \bm l \bm L} \sum_i u^{\lambda\mu*}_{i,\nu,\bm n \bm l \bm L}\:,\nonumber
\end{align}
where the first sum runs over the lexicographically ordered pairs $(n_1l_1)\ge(n_2l_2)\ge\ldots\ge (n_\nu l_\nu)$. 

The derivation proposed here leads us to a few observations. Firstly, we note again that this results is an extension of the one introduced 
in reference~\cite{ACE_dusson} for the case of scalar functions, but adapted to encompass also the case of linear-covariant models. 
The difference between this approach and the one used in previous works is that here we never use the explicit form of the expansion 
coefficients $u^{\lambda\mu*}_{i,\nu,\bm n \bm l \bm L}$. Instead, our derivation is based on the property of the radial and MultiSHs 
basis only. With this in mind, we are able to appreciate that the only necessary requirement is the symmetry of the function $\rho^{\otimes \nu}$.
Moreover, the results of this section aim to remove unnecessary redundancies in the expansion, in the same spirits of the observation 
done for the NICE framework \cite{NICE}, and in reference~\cite{Goffreport}.

However, this general construction does not address the redundancies over intermediate 
channels, $\bm L$, as clear from the summation of Eq.~\eqref{eq:linear_relation_permutation_coeff}. In fact, the explicit form of such 
redundancies depends on the chosen coupling scheme. Luckily, from this derivation, we deduce that if $\rho^{\otimes \nu}$ is not symmetric, 
namely it is not obtained by taking products of the same function $\rho(\vb r)$, but instead it is derived from products of $\nu$ different 
atomic densities, then the permutational symmetry will be broken. This can be achieved by considering different radial functions only. 
Indeed, in Section~\ref{sec:on_the_atomic_density} we have already proved that for a $\nu$-points density, obtained from the product 
of densities of the form of Eq.~\eqref{eq:intro_atomic_density}, taking different localization functions affects only the radial basis and 
not the MultiSHs one.

\subsubsection{Explicit construction}

To better clarify the role of the symmetry breaking induced by the localization functions, we proceed with a direct construction. 
As shown in Sec.~\ref{sec:on_the_atomic_density}, a localization function leads to an atomic density of the form
\begin{equation}
    \rho^{\alpha}_{i}(\vb r) = \sum_{nlm}c^\alpha_{i,nlm}R_{nl}(r)Y_l^m(\ver r)\:,
\end{equation}
where the expansion coefficients are
\begin{equation}
    c^\alpha_{i,nlm} = \sum^\text{atoms}_{j}q^\alpha_{nl}(r_{ji})Y_l^{m*}(\ver r_{ji})\:.
\end{equation}
Here, $\alpha$ indicates the localization function used (this could be also a continuous parameter, for example 
the width of Gaussians functions, or even a set of parameters). The functions $q^\alpha_{nl}(r_{ji})$ are obtained 
according to Eq.~\eqref{eq:radial_function_of_coefficients_expansion}. We can appreciate how the expansion coefficients, 
$c^\alpha_{i,nlm}$, are still permutationally invariant with respect to the atomic positions in the environment, since they 
contain a summation over all the atoms. From this, we can construct the $\nu$-density
\begin{equation}
    \rho^{\bm \alpha, \otimes \nu}_i(\vb r_1,\vb r_2,\cdots, \vb r_\nu) = \prod_{s=1}^\nu \rho_i^{\alpha_s}(\vb r_s),
\end{equation}
which shows that the inputs of the function $ \rho^{\otimes \nu}$ are fed to a different function $\rho_i^\alpha(\vb r_\alpha)$, 
indicated by the vector $\bm \alpha := (\alpha_1,\ldots, \alpha_\nu)$. If all the localization functions are different, then the 
function will not be symmetric as, for example, $\rho^{\otimes \nu}(\vb r_1,\vb r_2,\vb r_3,\ldots) \neq \rho^{\otimes \nu}(\vb r_2,\vb r_1, \vb r_3, \ldots)$.
The expansion coefficients are obtained directly from Eq.~\eqref{eq:Multi_coefficients_singlebody} (here reported for readability)
\begin{equation}\label{eq:expansion_coefficients_symmetry_breaking}
     u^{\bm \alpha,\lambda\mu}_{i,\bm n \bm l \bm L} = \sum_{\bm j} Q^{\bm \alpha}_{\bm n \bm l}(x_{\bm j i})\mY^{\lambda\mu *}_{\bm l \bm L}(\ver x_{\bm ji})\:,
\end{equation}
but where the functions $Q^{\bm \alpha}_{\bm n\bm l}$ now depend also on the localization functions, namely
\begin{equation}\label{eq:expansion_Q_symmetry_breaking}
    Q^{\bm \alpha}_{\bm n\bm l}(x_{\bm j i}) = q^{\alpha_1}_{n_1l_1}(x_{j_1 i})\cdot\ldots\cdot q^{\alpha_\nu}_{n_\nu l_\nu}(x_{j_\nu i}).
\end{equation}
The last two equations produce three consequences. Firstly, the localization function has an effect on the radial part only, while 
the angular part is untouched by this variation. Secondly, also the permutational invariance with respect to the atoms in the atomic 
environment is untouched by this procedure. Being it enforced at the level of the local density, it is automatically inherited by the 
$\nu$-density. Lastly, having functions $q^\alpha_{nl}$ that depend also on the angular number $l$ is crucial, since otherwise there 
would be exchange symmetries every time two or more of the radial numbers $n$ are equal. This is caused, again, by the fact that 
the MultiSHs remain the same regardless of the localization function of choice.

\subsubsection{Number of basis functions and performance}
The expansion coefficients of Eq.~\eqref{eq:expansion_coefficients_symmetry_breaking}, being obtained from 
symmetry-breaking localization functions, do not satisfy the same symmetries of Eq.~\eqref{eq:linear_relation_permutation_coeff}.
This implies that they are all independent, up to other hidden symmetries or numerical coincidences. Therefore, we cannot apply 
the lexicographical order anymore. In order to gauge the difference in the number of functions acquired, we can take the common 
example of a radial basis defined such that $n\in[0,n_\text{max}]$ and $l\in[0,n]$. For this case, the number of lexicographically 
ordered pairs has a $O(n_{\text{max}}^{2\nu}/\nu!)$ asymptotic scaling. On the contrary, the number of independent expansion 
coefficients with symmetry-breaking localization functions is of order $O(n_{\text{max}}^{2\nu})$, hence much larger already for 
relatively small $\nu$. Since in such representation the expansion coefficients are independent (loosely speaking they carry information), 
having more coefficients at lower order $\nu$ could be greatly beneficial. In fact, the cost of evaluating the expansion coefficients 
increases with both $n_\text{max}$ and $\nu$. Thus, for the same $n_\text{max}$ and $\nu$, we are able to introduce many more 
linearly independent terms in the expression, bringing more descriptivity in the representation, while keeping a lower order 
(computationally favorable) expansion.
We also mention that there are already instances in which different local densities are coupled, for example when considering 
the interaction between local densities which are relative to different species. In these cases, the local densities that constitute 
$\rho_i^{\otimes \nu}$ are not distinguished simply by different localization functions. Instead, they are obtained by different 
summation, as one can see from the explicit form of the expansion coefficients of the single density
\begin{equation}
    c^{z,\alpha}_{i,nlm} = \sum_{j\in z} q_{nl}^\alpha(r_{ji}) Y_l^{m*}(\ver r_{ji})\:,
\end{equation}
where the sum runs over the atoms in the local environment belonging to the species $z$ only. While one could keep using 
different localization functions (indicated by $\alpha$) for consistency reasons, the symmetry is automatically broken when 
the coupling coefficients are constructed over different atomic species $z$.

An important aspect of this discussion comes from Eqs.~\eqref{eq:expansion_coefficients_symmetry_breaking} 
and~\eqref{eq:expansion_Q_symmetry_breaking}, which show that the angular coupling induced by the MultiSHs 
is untouched from this symmetry breaking representation. 
Therefore, any implementation which is optimized to efficiently tackle the problem of coupling of angular momenta, will 
be just as effective in the case of different localization functions. Here we just mention that many of these approach use 
recursive formulas to evaluate the couplings (see for example Refs.~\cite{NICE,PACE,WignerKernel}) and therefore it 
is sufficient to incorporate a different radial function $q^\alpha_{nl}$ at each step of the iteration. Therefore, the most 
significant additional cost is to compute $\nu$ different radial basis and to construct the relative local densities. 
For everything else, the most common coupling calculators should be already effective.

\section{Conclusions}

In this work we have carried out an in-depth investigation and analysis of the most successful and well-know ML descriptors 
from the vantage point provided by choosing the MultiSHs as a multi-points angular basis. 

Typically, one approaches the investigation of the descriptors from the explicit construction of the expansion 
coefficients with respect to a radial basis and spherical harmonics. This leads to a bottom-up approach in deriving 
and enforcing the required transformation behavior under rotation or inversion of the atoms in the system. Instead, 
by following a top-down strategy that takes into account only the notion of the multi-point basis provided by the 
MultiSHs, here have we deduced all the required properties directly from the basis, with almost no notion of the 
nature of the expansion coefficients. 
In doing so, we have proved that one can deduce all the properties of interest by enforcing only two conditions, 
namely the orthogonality of the basis and its fixed behavior under rotations. From these properties, and with no 
underlying assumption on the explicit form of the expansion coefficients, we have been able to obtain their behavior 
under rotation, under conjugation and under inversion. Regarding the inversion, we have obtained a full 
characterization of their behavior as proper or pseudo objects.

As a main result, we have disentangled the nature of the coefficients from their transformation under these operations. 
On the one hand, this extends all the techniques and tools tailored on the atomic density to any other scalar function 
of choice, allowing for a straightforward applications to cases in which the underlying function is not necessarily the 
standard atomic density (for example, further applications could include our spin-powerspectrum~\cite{Domina}). 
On the other hand, we have shown how a basis-centric approach leads to a major simplification of the analytical 
calculations, such as for the case of the $\lambda$-SOAP kernel. Indeed, we have also shown that a compact and 
general inner product form of this kernel could be achieved by using only the orthogonality and the rotational behavior 
of the MultiSHs, in no more than a few lines of elementary algebraic manipulations.

As a further example of a formulation constructed around the MultiSHs, we have shown how the very definition of 
powerspectrum and bispectrum could be obtained directly from the MultiSHs formalism, with a built-in proof of 
their rotational invariance. In this derivation, we have pointed out that these properties (and their own definition) 
are linked only to the separability of the expanded function in products of single-center functions. This effectively 
unveils the real underlying hypothesis behind the definition of such descriptors. 

With the full framework in place and thoroughly investigated, we have then been able to derive the entire ACE 
formalism (up to any body order), by imposing a Dirac-delta form of the atomic density. Remarkably, this is the 
only instance throughout this work, where we have investigate a specific form of the expansion coefficients, despite 
having already derived all their transformation rules. As another example of the vantage point provided by the MultiSHs, 
we have pointed out the shortcomings of the SNAP (with its inability to explore the entire rotationally invariant space), 
while investigating the advantages provided by its compactness. This has led us to explore the straightforward extension of the MultiSHs formalism to a linear model for (covariant) tensorial quantities. 

Finally, we have concluded our investigations by analyzing the redundancies arising from descriptors based on symmetric 
functions. In particular, by exploiting the connection between different MultiSHs provided by the theory of re-coupling of 
angular momenta, we have provided a clear representation of these redundancies not only for the scalar case, but also 
for the descriptors of covariant quantities. Importantly, since we have identified the origin of these redundancies as 
associated to the symmetry of the expanded function, we have obtained another confirmation that a representation 
in terms of MultiSHs can be greatly informative. Furthermore, we have proved that the use of different atomic densities 
(obtained by employing different radial bases) can wash out these redundancies.

\begin{acknowledgments}

This work has been supported by Science Foundation Ireland through the Advanced Materials and BioEngineering 
Research (AMBER) (Grant: 12/RC/2278$_-$P2).

\end{acknowledgments}

\appendix
\section{From spherical harmonics to BipoSHs}\label{Appendix_sph_BipoSHs}
In this Appendix we will show how the property of Eq.~\eqref{eq:orthogonality_CG} leads to the change of representation 
from products of spherical harmonics to BipoSHs. If we assume that the two-points function $f(\ver r_1,\ver r_2)$ can be 
expanded in terms of spherical harmonics, then it will hold that
\begin{equation}
    f(\ver r_1, \ver r_2) = \sum_{l_1m_1l_2m_2} f_{l_1m_1l_2m_2} Y^{m_1}_{l_1}(\ver r_1)Y^{m_2}_{l_2}(\ver r_2)\:,
\end{equation}
for some coefficients $f_{l_1m_1l_2m_2}$. If we now employ Eq.~\eqref{eq:orthogonality_CG} for the orthogonality of 
two CG-coefficients, then we can write
\begin{align}
     &f(\ver r_1, \ver r_2) =\nonumber\\
     &= \sum_{\substack{l_1m_1m'_1\\l_2m_2m'_2}} f_{l_1m_1l_2m_2}\delta_{m_1m'_1}\delta_{m_2m'_2}Y^{m'_1}_{l_1}(\ver r_1)Y^{m'_2}_{l_2}(\ver r_2)=\nonumber\\
     &= \sum_{l_1l_2}\sum_{\lambda\mu}\left(\sum_{m_1m_2}C^{\lambda\mu}_{l_1m_1l_2m_2}f_{l_1m_1l_2m_2}\right)\nonumber\\
     &\qquad\qquad\qquad \times\left(\sum_{m'_1m'_2}C^{\lambda\mu}_{l_1m'_1l_2m'_2}Y^{m'_1}_{l_1}(\ver r_1)Y^{m'_2}_{l_2}(\ver r_2)\right)=\nonumber\\
     &= \sum_{l_1l_2}\sum_{\lambda\mu}u^{\lambda\mu}_{l_1l_2} \mY^{\lambda\mu}_{l_1l_2}(\ver r_1, \ver r_2)\nonumber\:,
\end{align}
In the last step we have introduced the BipoSHs, which allows us to identify the expansion coefficients $u^{\lambda\mu}_{l_1l_2}$ as
\begin{equation}
    u^{\lambda\mu}_{l_1l_2} =\sum_{m_1m_2}C^{\lambda\mu}_{l_1m_1l_2m_2}f_{l_1m_1l_2m_2}\:.
\end{equation}
This has the same structure of the BipoSHs themselves. Indeed, this derivation is a particular example of what stated in the more 
general Eq.~\eqref{eq:recipe_u_f}, the relation between the expansion in terms of spherical harmonics and the one in terms of MultiSHs. 
Moreover we have shown that the orthogonality of the CG coefficients of Eq.~\eqref{eq:orthogonality_CG} (generalized in 
Eq.~\eqref{eq:contraction_Gamma} for the $\Gamma$ tensors and failed by the $H$ tensors of Eq.~\eqref{eq:tensor_H_SNAP}) is 
crucial for performing the change of basis.

\section{Covariant behavior of the $\lambda$-SOAP kernel}\label{appendix_kernel_properties}

In this appendix we will prove that the kernel obtained in Eq.~\eqref{eq:multi_lambda_kernel} follows the correct transformations expected by a covariant kernel. 
In order to do so, we need 
to investigate the difference between the covariant and the contravariant nature of its components, alongside 
their relation with passive and active rotation. Our analysis begins by noticing that the rotation 
$\hat{R}$ written above is a \emph{passive} rotation (here we define a passive rotation as a rotation of the 
frame of reference, while an active rotation is a rotation of the atomic position).
Indeed, if we take the density around the $i$-th atom, $\rho^{\otimes \nu}_i$, then, the coefficients $u$ will 
depend on the atomic positions, $\{\vb r_{ji}\}$.
Therefore, a rotation $\hat{R}$, acting on $\vb x$, can be interpreted as a rotation that acts on the 
system's reference frame, here represented by the MultiSHs' basis. Because $\rho^{\otimes \nu}$ is a scalar 
function, this is equivalent to perform a rotation $\hat{R}^{-1}$ of the atomic positions. On the contrary, if we 
want to investigate an \emph{active} rotation, $\hat{R}_\text{active}$, on the atomic positions, we have to 
perform a rotation $\hat{R}^{-1}$ on the reference frame, namely $\hat{R}^{-1} = \hat{R}_\text{active}$. 
Thus, by using the following properties of the Wigner D-matrices
\begin{equation}
    D^\lambda_{\mu\mu'}(\hat{R}^{-1}) = D^{\lambda*}_{\mu'\mu}(\hat{R})\:,
    \end{equation}
we have that Eq.~\eqref{eq:transformation_rule_coeff_SH_passive} leads to
\begin{align}\label{eq:transformation_rule_coeff_SH_active}
     u^{\lambda\mu*}_{i,\substack{\bm n\bm l \bm L}}\xrightarrow{\hat{R}_\text{active}= \hat{R}^{-1}} &\sum_{\mu}D^{\lambda}_{\mu\mu'}(\hat{R}^{-1}) u_{i,\substack{\bm n\bm l \bm L}}^{\lambda\mu*}\nonumber\\
     &= \sum_{\mu}D^{\lambda*}_{\mu'\mu}(\hat{R}) u_{i,\substack{\bm n\bm l \bm L}}^{\lambda\mu*}\:,
\end{align}
which is exactly analogous to the transformation rule of Eq.~\eqref{eq:transf_rules_sph_harm} for the 
spherical harmonic $Y^m_l$. In other words, we can claim that the expansion coefficients transform 
contravariantly, while their complex conjugate transforms covariantly. This fact will be explored also in 
the next section, and confirmed again when we will investigate the explicit expression of the expansion 
coefficients, in Section~\ref{sec:on_the_atomic_density}.
Finally, if we define an active rotation on the atoms of $\rho^{\otimes \nu}$ as
\begin{equation}\label{eq:def_active_rot_density}
    \hat{R}\rho^{\otimes \nu}(\vb x; \{\vb r_{ji}\}) := \rho^{\otimes \nu}(\vb x; \{\hat{R}\vb r_{ji}\}) = \rho^{\otimes \nu}(\hat{R}^{-1}\vb x; \{\vb r_{ji}\})\:,
\end{equation}
where in the last equality we use the fact that $\rho_i$ is a scalar field, then we can derive the transformation 
rule for the $\lambda$-SOAP kernel as
\begin{align}
    &(K^{(\nu)}(\hat{R}_1\rho,\hat{R}_2\rho'))^\lambda_{\mu_1\mu_2} =\nonumber\\&\qquad=\sum_{\mu_1'\mu_2'}D^{\lambda}_{\mu_1\mu'_1}(\hat{R}_1)(K^{(\nu)}(\rho,\rho'))^\lambda_{\mu'_1\mu'_2}D^{\lambda*}_{\mu_2\mu'_2}(\hat{R}_2)\:.
\end{align}
This agrees with the derivation of reference~\cite{SAGPR} and proves that the general $\lambda$-SOAP 
kernel is suitable to describe \emph{covariant} quantities. In other words, we can state that such kernel is contravariant 
in its the first component and covariant in its the second component.

\section{Behaviour of the MultiSHs under parity transformations}\label{appendix:parity_MultiSH}
This appendix is devoted to prove that the MultiSHs behave as proper tensor when the sum $\sum(\bm l) + \lambda$ 
is even, and as a pseudotensor whenever the same sum is odd. We will prove this statement by exploiting a scalar 
contraction of the MultiSHs.

Given the MultiSH, $\mY^{\lambda\mu}_{\bm l \bm L }(\ver x)$, we can perform a scalar contraction with another spherical 
harmonics, by means of the recursion relation \eqref{eq:recursion_relation_multi}, projected onto the rotationally 
invariant space. Namely, let us consider the $\lambda'=\mu'=0$ components of the higher-order MultiSHs,
given by
\begin{equation}
    \mY^{00}_{\bm l \lambda \bm L \lambda}(\ver x,\ver r_{\nu+1}) = \dfrac{(-1)^{\lambda}}{\sqrt{2\lambda+1}}\sum_{\mu}(-1)^{\mu}\mY^{\lambda\mu}_{\bm l\bm L}(\ver x)Y^{\mu}_\lambda(\ver r_{\nu+1}),
\end{equation}
obtained from $$C^{00}_{l_1m_1l_2m_2}=\delta_{l_1l_2}\delta_{m_1-m_2}(-1)^{l_1-m_1}/\sqrt{2l_1+1}.$$ 
One can directly prove that we are into the rotationally invariant space, since the rotation property of 
Eq.~\eqref{eq:Multi_rotation} implies that
\begin{equation}
    \mY^{00}_{\bm l \lambda \bm L \lambda}(\hat{R}\ver x,\hat{R}\ver r_{\nu+1}) = \mY^{00}_{\bm l \lambda \bm L \lambda}(\ver x,\ver r_{\nu+1}), \nonumber
\end{equation}
obtained from $D^{0*}_{00}=1$. If we now perform an inversion of the frame of reference and use 
Eq.~\eqref{eq:inversion_MultiSHs}, we will obtain
\begin{equation}
     \mY^{00}_{\bm l \lambda \bm L \lambda}(-\ver x,-\ver r_{\nu+1}) = (-1)^{\Sigma(\bm l)+\lambda} \,\mY^{00}_{\bm l \lambda \bm L \lambda}(\ver x,\ver r_{\nu+1})\:.\nonumber
\end{equation}
Thus, when the sum $\Sigma(\bm l)+\lambda$ is even, we conclude that the expression above behaves as a scalar. 
In contrast, for an odd sum we have a pseudoscalar (because of the sign change under inversion). Note that, again, 
$\mY^{00}_{\bm l \lambda \bm L \lambda}$ is constructed by contracting $\mY^{\lambda\mu}_{\bm l \bm L }$ with a proper
tensor (the spherical harmonics). Therefore, whenever we have a pseudoscalar ($\Sigma(\bm l)+\lambda$ 
is odd) we must have that $\mY^{\lambda\mu}_{\bm l \bm L }$ is a pseudotensor. Instead, for a scalar ($\Sigma(\bm l)+\lambda$ is 
even) the multipolar spherical harmonics $\mY^{\lambda\mu}_{\bm l \bm L }$ must be a tensor, as expected.

\section{$\lambda-$SOAP Kernel for the group $O(3)$}\label{appendix_kernel_03}
In this Appendix, we will outline the derivation of a flavor of the $\lambda$-SOAP kernel suited to describe scalar and tensors, 
hence excluding objects behaving as pseudoscalars and pseudotensors. The procedure is essentially the same of the one shown in 
Section~\ref{sec:covariant_kernel_simple_derivation}, but with an extension of the Haar integration to any element of the $O(3)$ group, 
namely by including also parity transformations. This is achieved by defining the kernel as
\begin{align}
      (K^{(\nu)}(\rho,\rho'))^\lambda_{\mu_1\mu_2} :&=  \int \dd \vb x\, \rho^{\otimes \nu}(\vb x)\int \dd \hat{R}\, D^{\lambda}_{\mu_1\mu_2}(\hat{R})\,\nonumber\\
      &\times\hat{R}\bigg[(\rho')^{\otimes \nu}(\vb x)+(-1)^\lambda\hat{P}(\rho')^{\otimes \nu}(\vb x)\bigg]\:,
\end{align}
where $\hat{P}$ represents a parity transformation, as defined in Section~\ref{sec:general_properties_exp_coeff}. 
The exponent $(-1)^\lambda$ include the representation of an inversion for the space of angular momentum $\lambda$. Explicitly, the first addend in the square brackets is rotated and then integrated against the appropriate representation of the same rotation, provided by the Wigner $D$-matrix. In the same way, the second addend is transformed under the action of $\hat{R}\hat{P}$, namely a roto-inversion. In this way, the full term in square brackets covers an integration over the whole $O(3)$ group, including also inversions. As discussed in Sec.~\eqref{sec:general_properties_exp_coeff}, the parity of a proper tensor belonging to the space of angular momentum $\lambda$, is realized by a factor $(-1)^\lambda$. Therefore, this second term is integrated against the representation of the same $\hat{R}\hat{P}$ roto-inversion onto this space, which is given by $(-1)^\lambda D^\lambda_{\mu_1\mu_2}(\hat R)$.

Note that here 
we define the kernel in terms of \emph{active} rotations, $\hat{R}$, as described in Eq.~\eqref{eq:def_active_rot_density}. Following 
the same interpretation of the standard $\lambda$-SOAP kernel, this formulation compares the overlaps between the densities 
$\rho^{\otimes \nu}$ and $(\rho')^{\otimes \nu}$, for any orientation and inversion of the latter. 

We can now expand the $(\rho')^{\otimes \nu}$ in terms of the basis $\mathcal{R}_{\bm n \bm l}(x)\mY^{\lambda\mu}_{\bm l\bm L}(\ver x)$, 
and apply the parity transformation of Eq.~\eqref{eq:inversion_MultiSHs} to the MultiSHs. After this, we can follow the same calculation 
shown in Section~\ref{sec:covariant_kernel_simple_derivation} and eventually obtain
\begin{equation}
\begin{split}
    &(K^{(\nu)}(\rho,\rho'))^\lambda_{\mu_1\mu_2}\, \\
    &= \dfrac{8\pi^2}{2\lambda+1}\sum_{\bm n\bm l \bm L}\left(1+(-1)^{\Sigma(\bm l)+\lambda}\right)u^{\lambda \mu_1}_{\bm n \bm l \bm L}v^{\lambda \mu_2*}_{\bm n \bm l \bm L}\:.
\end{split}
\end{equation}
The formula above shows that the contributions vanish when $\Sigma(\bm l) +\lambda$ is odd, namely when the contraction is performed 
over the components that transform as pseudotensors. In other word, we can write the kernel as
\begin{equation}
    (K^{(\nu)}(\rho,\rho'))^\lambda_{\mu_1\mu_2}= \dfrac{16\pi^2}{2\lambda+1}\sum^{\text{proper}}_{\bm n\bm l \bm L}u^{\lambda \mu_1}_{\bm n \bm l \bm L}v^{\lambda \mu_2*}_{\bm n \bm l \bm L}\:,
\end{equation}
where the restriction on the sum indicates that only the proper tensorial components are considered, namely the ones for even $\Sigma(\bm l) +\lambda $.

\section{Expansion coefficients in terms of spherical harmonics}\label{sec:AppendC}

This Appendix is devoted to prove Eq.~\eqref{eq:radial_function_of_coefficients_expansion}, namely to 
explicitly show that the angular dependence of the functions $g_{nlm}(\vb r_{ji})$ is realized by spherical 
harmonics. In order to simplify the derivation, let us define the auxiliary function
\begin{equation}
    f_{nl}(\bm r_{ji}, \ver r) := \sum_m g_{nlm}(\vb r_{ji})Y_l^m(\ver r)\:,
\end{equation}
where the second argument is a unitary vector. As remarked in the main text, this function must behave 
as a scalar, since the density $\rho(\vb r)$ is a scalar field. Therefore, we can expand it over the 
scalar MultiSHs as
\begin{equation}
    f_{nl}(\bm r_{ji}, \ver r) = \sum_{l'}c_{nl,l'}(r_{ji})\mY_{l'l'}^{00}(\ver r_{ji}, \ver r)\:,
\end{equation}
where the functions $c_{nl,l'}(r_{ji})$ can be explicitly evaluated as
\begin{equation}
\begin{split}
    c_{nl,l'}(r_{ji}) &= \int \dd \ver r \,\dd \ver r_{ji} \, f_{nl}(\vb r_{ji}, \ver r) \mY^{00*}_{l'l'}(\ver r_{ji}, \ver r) =\\
    &= \delta_{ll'}\dfrac{(-1)^l}{\sqrt{2l+1}}\int \dd \ver r_{ji}\, \sum_{m} g_{nlm}(\ver r_{ji})Y^m_l(\ver r_{ji})\:,
\end{split}
\end{equation}
obtained by means of the integral
\begin{equation}
\begin{split}
    \int \dd \ver r \mY_{l'l'}^{00}(\ver r_{ji}, \ver r) Y^{m*}_{l}(\ver r)&= \delta_{ll'}C^{00}_{lml-m} Y^{-m}_l(\ver r_{ji})= \\
    &=\delta_{ll'}\dfrac{(-1)^{l}}{\sqrt{2l+1}}Y^{m*}_l(\ver r_{ji}) \:.
\end{split}
\end{equation}
This result has been obtained by using the explicitly the definition of the BipoSHs from Eq.~\eqref{eq:def_BipoSH}, 
the orthogonality of the spherical harmonics, the explicit value $C^{00}_{lml-m} = (-1)^{l-m}/\sqrt{2l+1}$ for the CG 
coefficients and the identity $(-1)^mY^{-m}_l(\ver r) = Y_l^{m*}(\ver r)$ for the complex conjugate of the spherical harmonics.

By exploiting again the orthogonality of the spherical harmonics, we can recover $ g_{nlm}(\vb r_{ji})$ 
from $ f_{nl}(\bm r_{ji}, \ver r)$ by means of the integral
\begin{equation}
\begin{split}
    g_{nlm}(\vb r_{ji}) &= \int \dd \ver r\,f_{nl}(\bm r_{ji}, \ver r)Y^{m*}_l(\ver r)\\
    &= \dfrac{(-1)^{l}}{\sqrt{2l+1}}c_{nl,l}(r_{ji})Y^{m*}_l(\ver r_{ji})\:,
\end{split}
\end{equation}
where we have used the expansion of the auxiliary function $ f_{nl}(\bm r_{ji}, \ver r)$. If we define 
$q_{nl}(r_{ji}) := (-1)^{l}c_{nl,l}/\sqrt{2l+1}$ and use the explicitly evaluated expression for $c_{nl,l}(r_{ji})$, 
we finally obtain
\begin{equation}
    q_{nl}(r_{ji}) = \dfrac{1}{2l+1}\int \dd \ver r_{ji} \sum_m g_{nlm}(\ver r_{ji}) Y^m_l(\ver r_{ji})\:,
\end{equation}
as required.

\bibliographystyle{unsrt}

\end{document}